\documentclass[prd, onecolumn, superscriptaddress, nofootinbib, notitlepage, floatfix, preprintnumbers,longbibliography]{revtex4-1}
\usepackage{xcolor}
\usepackage{graphicx}
\usepackage{wrapfig}
\usepackage{amsmath}
\usepackage{amsfonts}
\usepackage{amssymb}
\usepackage{multirow}
\usepackage{slashed}
\usepackage{physics}
\usepackage{soul}
\usepackage{comment}

\usepackage{ulem}
\usepackage[colorlinks=true, linkcolor=blue, citecolor=blue, urlcolor=blue]{hyperref}
\usepackage[paperwidth=210mm,paperheight=297mm,centering,hmargin=2cm,vmargin=2.5cm]{geometry}
\newcommand{\msbar}{{\overline{\mathrm{MS}}}}
\newcommand\befs{\begin{figure*}}
\newcommand\eefs[1]{\label{fig:#1}\end{figure*}}
\newcommand\bef{\begin{figure}}
\newcommand\eef[1]{\label{fig:#1}\end{figure}}
\newcommand\beq{\begin{equation}}
\newcommand\eeq[1]{\label{#1}\end{equation}}
\newcommand\beqa{\begin{eqnarray}}
\newcommand\eeqa[1]{\label{#1}\end{eqnarray}}
\newcommand\bet{\begin{table}}
\newcommand\eet[1]{\label{tb:#1}\end{table}}
\newcommand\bets{\begin{table*}}
\newcommand\eets[1]{\label{tb:#1}\end{table*}}
\newcommand\fgn[1]{Fig.\ \ref{fig:#1}}
\newcommand\eqn[1]{Eq.\ (\ref{#1})}

\newcommand\apx[1]{Appendix \ref{sec:#1}}
\newcommand\tbn[1]{Table \ref{tb:#1}}
\begin{document}
\title{
Parton physics of the large-$N_c$ mesons 
}

\author{Nikhil\ \surname{Karthik}}
\email{nkarthik.work@gmail.com}
\affiliation{Department of Physics, William and Mary, Williamsburg, Virginia, USA.}
\affiliation{Thomas Jefferson National Accelerator Facility, Newport News, Virginia, USA.}
\author{Rajamani\ \surname{Narayanan}}
\email{rajamani.narayanan@fiu.edu}
\affiliation{Department of Physics, Florida International University, Miami, FL 33199}
\begin{abstract}
We initiate the studies on the structural physics of the tower of
stable large-$N_c$ mesons through a first computation of the collinear
quark-structure of a large-$N_c$ pion using lattice Monte-Carlo methods.
We adapt the large-$N_c$ continuum reduction for the determination
of meson correlation functions involving the spatially-extended
quasi-PDF operators as a perfect strategy to concentrate only on
the short perturbative length scales.  We find the internal structures
of pion in the large-$N_c$ and $N_c=3$ theories to be quite similar.
Interestingly, we find hints that even the observed differences could arise to a
large extent via the different perturbative QCD evolution in the
two theories from similar initial conditions at low factorization
scales.
\end{abstract}

\preprint{JLAB-THY-22-3613}
\date{\today}
\maketitle

\section{Introduction}
Quantum Chromodynamics (QCD) in the limit of
large number of colors, $N_c$, at a fixed 't Hooft
coupling~\cite{tHooft:1973alw,tHooft:1974pnl} $\lambda = N_c
\alpha_s$, is greatly simplified by being a planar model in which
quarks are naturally quenched, and it is well known to be a realistic
QCD-like theory that approximately reproduces many features in the
real-world, such as the ratios of low-lying meson
masses~\cite{Perez:2020vbn,DeGrand:2016pur,Hernandez:2019qed,Bali:2013kia,Bali:2008an,DelDebbio:2007wk}.
The next frontier in QCD-physics is to understand the structural
aspects of hadrons in more detail, so as to relate the emergent
properties of hadrons, such as their masses and spins, 
to those of the short-distance quark-gluon
(parton) degrees of freedom and their interactions (e.g.,
see~\cite{AbdulKhalek:2022erw,Accardi:2012qut,Dudek:2012vr}).  In
this respect, the large-$N_c$ limit motivates and crystallizes
concepts in parton phenomenology, such as the linear Regge trajectories
(proven in two-dimensions~\cite{tHooft:1974pnl}), the dipole
approach to BFKL formalism~\cite{Mueller:1993rr}, and the concept
of quark-hadron duality~\cite{Poggio:1975af,Shifman:2000jv} to name
a few.

The large-$N_c$ baryons~\cite{Witten:1979kh} are ${\cal O}(N_c)$ heavier degrees
of freedom that can be described as a chiral
soliton~\cite{Skyrme:1962vh,Witten:1983tx,Adkins:1983ya,Diakonov:1987ty}.
Such an identification has lead to mean-field theory studies of the
parton distributions inside a nucleon (for initial works,
see~\cite{Diakonov:1996sr,Diakonov:1997vc,Weigel:1996ef,Gamberg:1998vg}.)
In contrast, the large-$N_c$ mesons are the leading lighter degrees of
freedom, and nonperturbative methods (e.g., lattice simulations)
are the only way to study them.  With the access to an infinite
tower of completely stable large-$N_c$ mesons of different $J^{PC}$,
it is an ideal realization of QCD that is conducive to investigate the partonic origin
of hadron physics. 
Development of such realistic models of mesons as
hard-scatterers is especially important due to the reinvigorated
experimental~\cite{Aguilar:2019teb,Dudek:2012vr,Adams:2018pwt} and
theoretical efforts towards the meson structures, especially of the
pion, the Goldstone mode of Chiral symmetry-breaking
(refer~\cite{Roberts:2021nhw} for a review,
and~\cite{Sufian:2019bol,Sufian:2020vzb,Izubuchi:2019lyk,Gao:2020ito,Gao:2021hvs,Karthik:2021qwz,Lin:2020ssv,Gao:2021dbh,Detmold:2021qln,Hua:2022kcm}
for recent numerical works).  Quite surprisingly, despite the
continued effort to understand large-$N_c$ QCD over the years, the
partonic nature of the large-$N_c$ mesons is to a large extent
unknown. To our knowledge, the study in Ref~\cite{RuizArriola:2006jge}
of the distribution amplitude of pion within a large-$N_c$ Regge
model is a singular work towards this direction.

\bef
\centering
\includegraphics[scale=0.35]{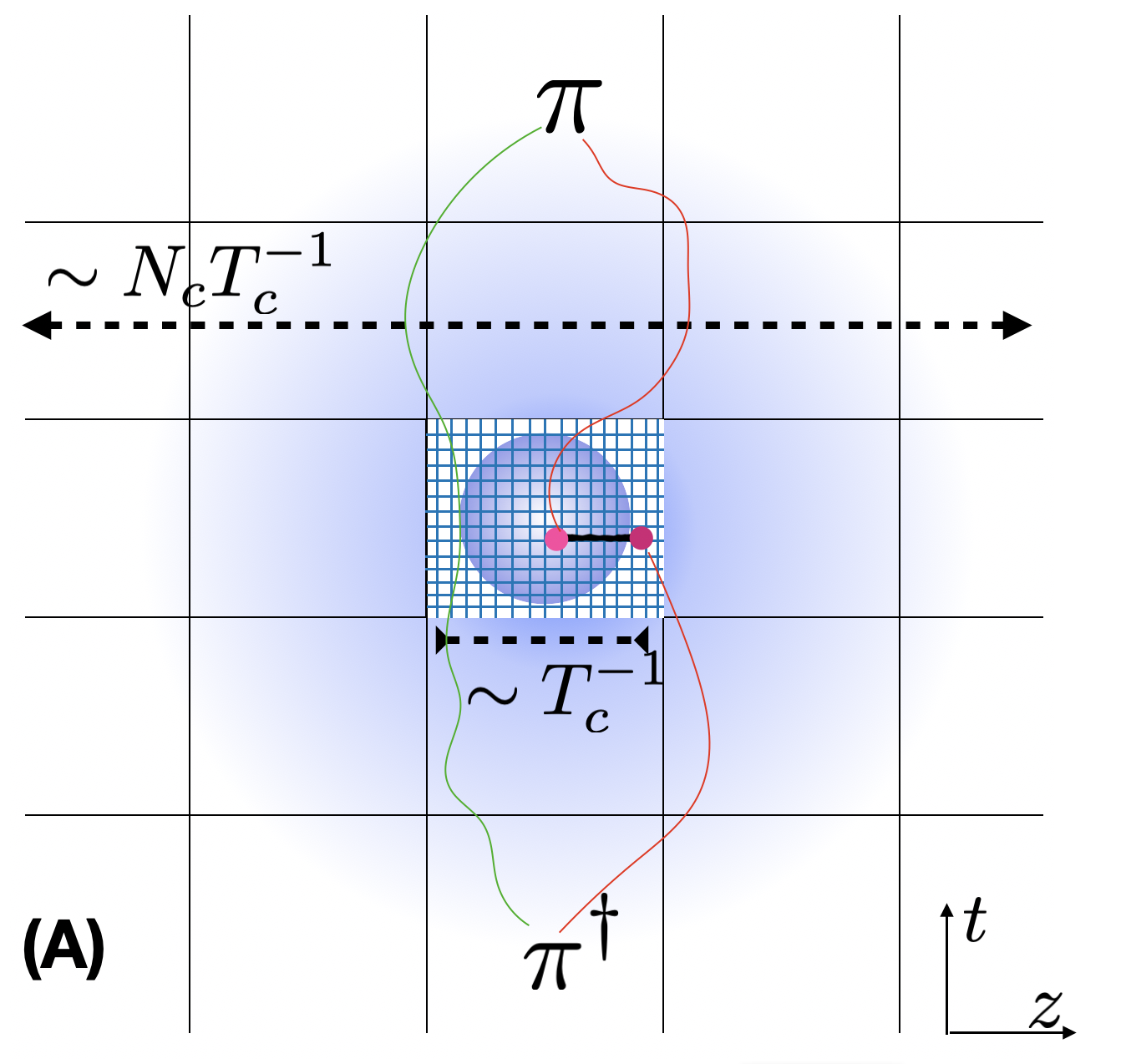}
\includegraphics[scale=0.45]{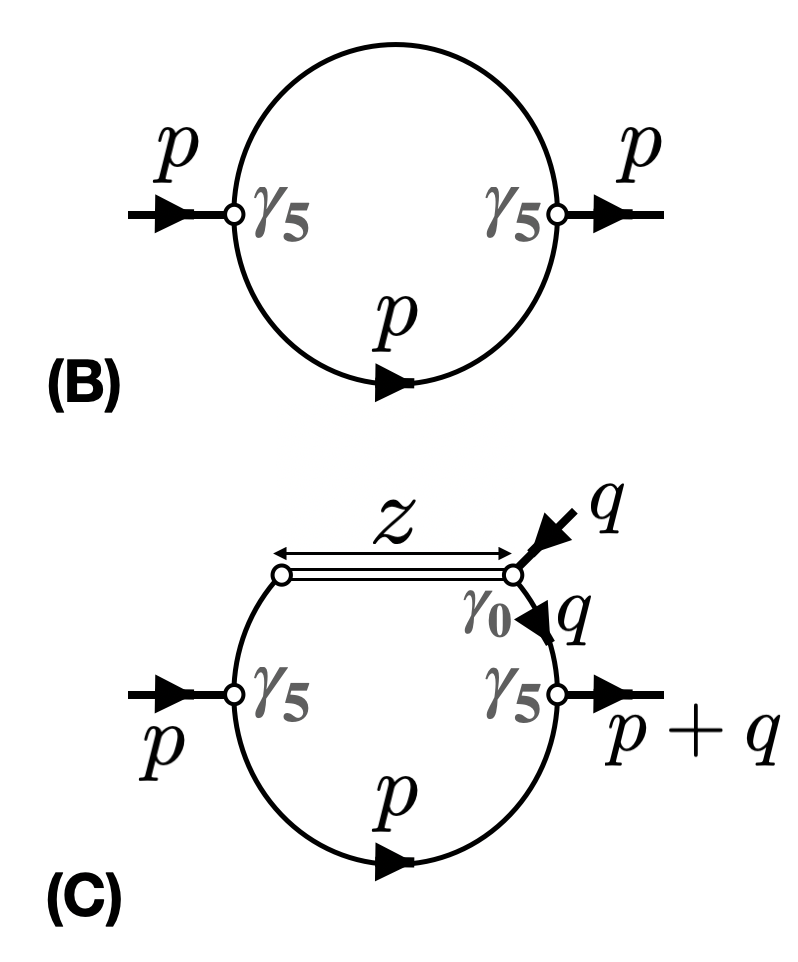}
\caption{(A) Schematic of large-$N_c$ continuum reduction for quasi-PDF
operator evaluated within a pion.  The gauge fields on~$\approx N_c
\ell$ sized box are obtained as replicas of gauge fields within a
$\ell$ sized box, with $\ell \approx T_c^{-1}$, the deconfinement
temperature. The quarks hopping on such crystalline configuration 
are labeled by their positions in periodic $\ell^4$ box and their Bloch momenta.
The correlation functions in the larger box can be
obtained using lattice implementation of momentum space Feynman diagrams
that use quark propagators in $\ell^4$ box.
(B and C) The momentum space Feynman diagrams implemented directly on
the lattice. The lines are quark propagators. The arrows show the
off-shell 4-momentum injected at the vertices.
The 2-point function of pion is shown in B.
The 3-point function of quasi-PDF operator (double line) with
pion creation and annihilation operators is shown in C.
}
\eef{cartoon}

Through the present work, we bridge this persisting gap in our
understanding of the canonical toy-model of QCD through a first
computation of quark distribution function of the large-$N_c$ pion, 
and thereby, lay the framework for comparative studies of internal structures of
different stable species of mesons.
As an important feature of the large-$N_c$ theory, we present the
large-$N_c$ continuum reduction as a novel tailor-made approach
for the operator product expansion (OPE) based
strategies~\cite{Braun:2007wv,Ji:2013dva,Radyushkin:2017cyf,Ma:2014jla} to
perform parton physics on the lattice.  We display the schematic
of the central idea of the calculation in \fgn{cartoon}(A), and we
elaborate on it in the following discussion. As an initial work in this 
direction, we keep the discussion simple by summarizing the main techniques and results in the main text, 
and by referring the reader to various appendices for the elaborate details.

\section{Basics of continuum reduction} 
Owing to the absence of a center-symmetry breaking deconfinement
phase transition in two Euclidean space-time dimensions, the
large-$N_c$ QCD$_2$ is well known to be reducible to a single-site
matrix model~\cite{Gross:1980he,Eguchi:1982nm}.  We can extend the
Eguchi-Kawai reduction~\cite{Eguchi:1982nm} to dimensions
$d>2$~\cite{Narayanan:2009xh,Narayanan:2007fb,Narayanan:2007ug,Narayanan:2003fc,Kiskis:2003rd},
provided we preserve the $U^d(1)$ center symmetry by reducing the theory
not to a point, but instead to a small box of volume $\ell^d$, with
$\ell \ge T_c^{-1}$, the inverse of deconfinement temperature.  The
powerful aspect of the large-$N_c$ reduction is that we can exactly
find the expectation values of gauge-invariant quantities (such as
a $w_1\times w_2$ Wilson loop) in $\mathbb{R}^d$ just from the
expectation values of the same quantity on the reduced $\ell^d$
periodic torus (even if $w_1,w_2>\ell$) through folding.  We can
regulate the reduced continuum theory on an $L^d$ periodic lattice
using a lattice coupling $b=\left(g^2N_c\right)^{-1}$ in the limit
$N_c\to\infty$ at fixed $b$, using $L$ greater than a critical
$L_c(b)$.  
The asymptotic scaling of $L_c(b)$ defined the critical size, $\ell_c = T_c^{-1}$~\cite{Kiskis:2003rd}.
As a corollary, we can unfold the torus by tessellating
$\mathbb{R}^d$ with the gauge configuration in $\ell^d$ box, resulting
in a path-integral over crystalline configurations. Consequently,
the quarks are labeled by position $x \in \ell^d$ and the Bloch
momentum $q$. Thereupon, we can write functions $F$ of a lattice
Dirac operator $\slashed{D}$ in $\mathbb{R}^d$, such as its propagator
$G$, in terms of functions $F^L$ of Dirac operator $\slashed{D}^L$
in $L^d$ periodic lattice, as 
\beq
F_{x,y}(U_\mu)  = \int \frac{d^d q}{(2\pi)^d} \ e^{i\frac{q\cdot(x-y)}{L}} F^L_{x,y}(U_\mu
e^{i\frac{q_\mu}{L}}).
\eeq{funcred}
We discuss the details behind such a construction in \apx{propreduc}.
By using such a relation, along with the
global $U^d(1)$ center symmetry, we can reduce all $n$-point functions
of quark bilinears in $\mathbb{R}^d$ to computations of $n$-point
functions on $L^d$ periodic lattice.  For example, 
as derived in \apx{twoptreduc},
we can write the
2-point function, $\tilde{C}_{\rm
2pt}(p)=\langle\pi(p)\pi^\dagger(p)\rangle$, for a pion ($\pi =
\bar d \gamma_5 u$) in momentum space as
\beq
\tilde{C}_{\rm 2pt}(p) =\left\langle \Tr\left [ \gamma_5 G^L(U_\mu e^{-i p_\mu}) \gamma_5 G^L\left(U_\mu\right)\right]\right\rangle,
\eeq{2ptfn}
where $p=(p_0,\mathbf{p})$ is the continuous-valued Euclidean four
momentum of the pion, the trace is over spin, color and the entire
$L^d$ lattice, and the ensemble average $\langle\ldots\rangle$ is
with respect to the pure gauge action. 
We show the Feynman diagram for \eqn{2ptfn} in \fgn{cartoon}(B).
From the spectral decomposition,
$\tilde{C}_{\rm 2pt}(p_0,\mathbf{p})=\sum_{i=0} 2 A_i
E_i(\mathbf{p})\left(p_0^2+E_i^2(\mathbf{p})\right)^{-1}$, we can
obtain the long-distance energy spectrum, $E_i(\mathbf{p})$, and 
amplitudes, $A_i$.
Alternatively, we can access the spectrum from the multi-exponential,
$A_i e^{- E_i(\mathbf{p}) t_s}$, decay of $C_{\rm 2pt}(t_s; \mathbf{p})
= \int \frac{dp_0}{2\pi}\tilde{C}_{\rm 2pt}(p_0,\mathbf{p}) e^{ip_0
t_s}$, in the Euclidean time $t_s$.  We see that the long-distance hadronic spectral physics is
trivialized by the ability to capture $|\mathbf{x}|, t_s\gtrsim \Lambda^{-1}_{\rm QCD}$
using only simulation of a box of size $\ell \approx \Lambda^{-1}_{\rm
QCD}$.

\section{Zooming in on parton scales with continuum reduction}
This large-$N_c$ continuum reduction leads to a key simplification
in lattice QCD computations of parton distributions.  Many recent
developments~\cite{Braun:2007wv,Ji:2013dva,Radyushkin:2017cyf,Ma:2014jla}
in the ab initio computations of the Bjorken $x$-dependent parton
distribution functions (PDFs), $f(x,\mu)$ at a $\msbar$ factorization scale
$\mu$, and related quantities, rely on the leading-twist expansion
of certain equal-time renormalized invariant amplitudes, ${\cal
M}(\nu, z^2)$ with $\nu=-z\cdot P$, involving an
operator-pair~\cite{Braun:2007wv,Ma:2014jla} or a bilocal extended
operator~\cite{Ji:2013dva,Radyushkin:2017cyf} with a spatial
separation $z_\mu =z_3 \delta_{\mu,3}$ that is evaluated within a
state $|P\rangle$ of an on-shell hadron moving with momentum $P =
\left(E(\mathbf{p}), \mathbf{p}\right)$.  Through lattice Monte
Carlo determination of ${\cal M}$, we can relate it to $f(x,\mu)$
through an OPE truncated at leading-twist
terms (see~\cite{Izubuchi:2018srq}),
\beq
{\cal M}(\nu,z^2)=\sum_{n=0} \frac{(i \nu)^n}{n!}C_n(\mu^2 z^2)\int_{-1}^1 x^n f(x,\mu)dx.
\eeq{tw2ope}
In the absence of higher-twist corrections, the Wilson coefficients
$C_n$ capture $\ln(-z^2\mu^2)$-type QCD contributions to ${\cal M}$
using perturbation theory and leads to $f(x,\mu)$ at a chosen scale
$\mu$. Thus, along with the necessity of non-zero $P_3$, the
short-distance $|z|$ is crucial for the validities of OPE, the
perturbation theory and for ignoring higher-twist terms.  On the
other hand, the leading-twist expansion is performed within hadronic
in- and out-states, and therefore, having control of the long-distance
aspects of QCD is equally important. Applying the above formalism
to the large-$N_c$ theory is much simpler -- 
we can capture the
long-distance hadronic states easily by the virtue of continuum
reduction, leaving only the relevant partonic scales for $z$ below the
inverse deconfinement transition temperature, $T^{-1}_c$, to be 
captured by Monte Carlo sampling of gauge fields within 
$T_c^{-1}$ extent.

With this realization, we extend the continuum reduction
approach to $n$-point functions involving an extended operator,
such as the $u$-quark quasi-PDF operator ${\cal O}(z; q) \equiv
\sum_{x} e^{i q \cdot x} \bar{u}_{x}\gamma_0 W_{x,x+z} u_{x+z}$,
for purely-spatial $z=(0,0,0,z_3)$, and $W_{x,x+z}$ is a straight Wilson-line
connecting $x$ to $x+z$. The spatial part $\mathbf{q}=0$ for the PDF we 
want to study.
Following our discussion of the 2-point function and the method of 
folding Wilson loops of any size on an $L^4$ lattice, we can 
similarly write the 3-point function $\tilde{C}_{\rm
3pt}(z,p,q)\equiv \left\langle \pi(p+q) {\cal O}(z;q)
\pi^\dagger(p)\right\rangle$, as
\beq
\tilde C_{\rm 3pt}=\sum_{x} \langle {\rm tr}\left(  \left[\gamma_0 W^L\right]_{x,x+z} \left[G^L \gamma_5 G^L\gamma_5 G^L\right]_{x+z,x}\right)\rangle,
\eeq{3ptfunc}
where the gauge-links $U_\mu$ entering the propagators from left
to right are multiplied by phases $1$, $e^{-i p_\mu}$ and $e^{iq_\mu}$
respectively.  The trace is over color and spin, and $W^L$ is the
folded Wilson line obtained by wrapping around the periodic lattice
if $|z_3| \ge L$.  
In \apx{threeptreduc}, we present a detailed 
derivation of the above equation. We show the Feynman diagram for \eqn{3ptfunc}
in \fgn{cartoon}(C).  Note that the quark-line disconnected piece in
$\tilde C_{\rm 3pt}$ is $N_c^{-1}$ suppressed, and therefore as
another large-$N_c$ advantage, we have ignored it in the above
equation. As in the 2-point function, we can obtain the
required bare quasi-PDF matrix element, 
\beq
2 P_0 h^B(z,P)\equiv \left\langle \pi; P|{\cal O}(z)|\pi; P\right\rangle, 
\eeq{defhb}
through the
spectral analysis of \eqn{3ptfunc} either in momentum space or in
the real-space $t_s$ after Fourier transforming $\tilde{C}$ with respect to $p_0$ to form $C_{\rm
3pt}(z,t_s,\mathbf{p}, q)$. A convenient choice $q=(0,\mathbf{0})$
gives the so-called summation method~\cite{Maiani:1987by}, wherein,
$C_{\rm 3pt}(z,t_s,\mathbf{p}, q=0)/C_{\rm 2pt}(t_s,\mathbf{p})=t_s
h^B(z,\mathbf{p})+{\rm constant}$, up to $O(e^{-(E_1-E_0) t_s})$
excited-state corrections.

\bef
\centering
\includegraphics[scale=1.2]{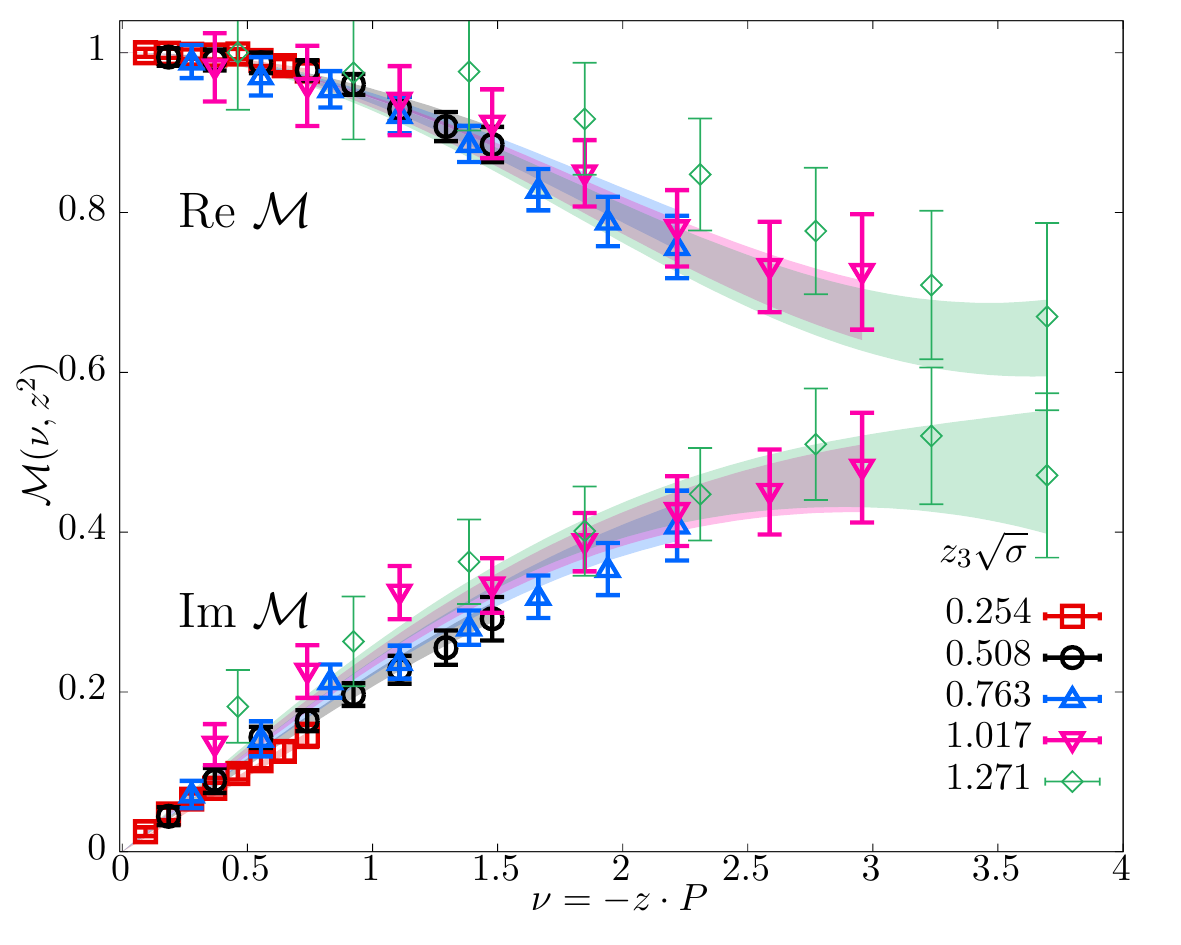}
\caption{The real and imaginary parts of the pseudo-ITD
of pion ${\cal M}(\nu,z^2)$ in the large-$N_c$ limit are shown.
The lattice data
from different values of quark-antiquark separation $z_3$
are shown using different colored symbols.
The bands are fits to
the leading-twist OPE with the large-$N_c$ NLO Wilson
coefficients at $\msbar$ scale $\mu=4.55\sqrt{\sigma}$.
The fit parameters are the Mellin moments.}
\eef{ritd}

\section{Computational details} 
We implemented the continuum reduction
approach to determine the $u$-quark PDF of the large-$N_c$ pion in $d=4$.
As a first exploratory study, we performed our computation at a
fixed simulation point at a large but finite value of $N_c=17$ on
an $L=8$ lattice using a coupling $b=0.355$, which is in the confined
phase~\cite{Kiskis:2003rd}.  The lattice spacing in units of string
tension~\cite{Kiskis:2009rf,Lucini:2003zr} is $\sqrt{\sigma}a=0.254(2)$.
Due to the finite large $N_c$, the $U^d(1)$ center symmetry reduces
to $Z^d_{N_c}$ discrete symmetry, and therefore, we quantized the lattice
momenta in units of $2\pi/(L N_c)$ and multiples thereof, to leave
the above results intact.  In this way, we effectively enlarged the
$8^4$ lattice into a $68\times 136^3$ lattice.  We used Wilson-Dirac
operator coupled to smeared gauge-links for $\slashed{D}^L$ and
tuned the quark mass to produce a pion of mass $m_\pi =0.86
\sqrt{\sigma}$. We stochastically computed $\tilde{C}_{\rm 2pt}(p)$
and $\tilde{C}_{\rm 3pt}(z,p,q=0)$ at all values of $p_0$ at each
given $\mathbf{p}=(0,0,P_3)$, using 15K-32K configurations, and
Fourier transformed them into functions of $t_s$.  We studied nine
different spatial momentum $P_3/\sqrt{\sigma}\in [0,5.82]$. 
We elaborate further on the lattice    
setup in \apx{calcdetails}. 
We determined $h^B$ using summation type fits to $C_{\rm 3pt}/C_{\rm
2pt}$ ratio.  
For further details on the spectral analysis of 
2-point and 3-point functions, the reader can refer to
\apx{twoptspect} and
\apx{threeptextrapol} respectively.
Since $h^B(z,P)$ is multiplicatively
renormalizable~\cite{Ishikawa:2017faj,Ji:2017oey}, we took the
renormalization group invariant ratio~\cite{Orginos:2017kos} of
quasi-PDF matrix elements at $P_3\ne 0$ with respect to $P_3=0$ to
form the pseudo Ioffe-time distribution (pseudo-ITD), ${\cal M}(\nu,
z^2)$.

\section{Collinear quark structure of the large-$N_c$ pion} 
We show
the real and imaginary parts of the $u$-quark pseudo-ITD, ${\cal
M}(\nu,z^2)$ as a function of $\nu$ in \fgn{ritd}. As seen from
\eqn{tw2ope}, the two are governed by $u-\bar u$ and $u+\bar u$
PDFs respectively.  In $SU(3)$ theory, $u+\bar u$ PDF mixes with
gluon PDF, however this mixing is $N_c^{-1}$ suppressed and hence
ignored here.  The data points are the result of our lattice
computation from different $(z_3, P_3)$ put together.  The
near-continuous set of momenta we were able to use, helped us pack
the range of $\nu$ with data points.  In the large-$N_c$ limit, the
string tension $\sqrt{\sigma}$ sets a fiducial scale that distinguishes
perturbative and nonperturbative length scales; therefore, we
restricted the data for ${\cal M}$ to only those up to the border-line
$\sqrt{\sigma} z_3\le 1.27$.  At the same time, a cautious use of
$|P_3|<a^{-1}$ only let us scan a range of $\nu<3.5$.  The near
universality of the data with respect to the scaling variable $\nu$
points to the viability of perturbative OPE methods in the large-$N_c$
theory. By fitting the lattice data using the leading-twist OPE in
\eqn{tw2ope}, we extracted the Mellin moments $\langle x^n
\rangle_{u\pm\bar u}\equiv \int_0^1 x^n f_{u\pm \bar u}(x,\mu)dx$
at a scale $\mu = 4.55 \sqrt{\sigma}$ using 1-loop
result~\cite{Izubuchi:2018srq} for the Wilson coefficients $C_n(\mu^2
z^2)$ in the large-$N_c$ limit; for this we used leading-order
value, $\lim_{N_c\to\infty}C_F(N_c)\alpha_s(\mu)=0.39$ using
$\Lambda_{\rm
\msbar}/\sqrt{\sigma}=0.503$~\cite{Allton:2008ty,Datta:2009ef}.  We
chose a scale $\mu\approx a^{-1}$ so that it is characteristic of
the typical small $z_3$ used in this work.
We gather the technical details for the OPE fits and on the
perturbative factors in \apx{implementation} and \apx{coupling} respectively.
We find for the first few moments
\beqa
&&\left[\langle x\rangle_{u+\bar u}, \langle x^3\rangle_{u+\bar u}\right]=\left[0.25(1), 0.10(2)\right]\ ({\rm via\ Im}{\cal M}),\cr
&&\left[\langle x^2\rangle_{u-\bar u}, \langle x^4\rangle_{u-\bar u}\right]=\left[0.13(2), 0.10(2)\right]\ ({\rm via\ Re}{\cal M}),
\eeqa{momentvals}
with correlated $\chi^2/{\rm df}\sim 39/26$ in the two cases. 
As a cross-check that perturbative OPE framework is working 
for the chosen range in $z_3$, we used the fixed-$z^2$ 
moments analysis~\cite{Karpie:2018zaz} as a diagnostic tool~\cite{Gao:2020ito,HadStruc:2021qdf} to detect any 
corrections -- as discussed in \apx{efficacy}, we found the 
method to work well within statistical errors.
In addition to the above Mellin moments analysis, we also
performed fits to the valence $u-\bar u$ data assuming a
phenomenologically motivated functional form~\cite{Barry:2021osv},
$f_{u-\bar u}(x;\alpha,\beta,s)={\cal N}x^{\alpha}(1-x)^\beta(1+s
x^2)$.  Since our access to the range of $\nu$ is limited in this
work, and the small-$x$ region is believed to be harder to access
on the lattice, we imposed a prior that $\alpha\in[-0.6,-0.4]$
motivated by the Regge phenomenology.  With the caveat of using an
Ansatz, we found the data to be best described by a large-$x$
exponent $\beta=0.7(3)$, similar to what is seen in recent lattice
$SU(3)$ QCD
results~\cite{Gao:2020ito,Sufian:2020vzb,Sufian:2019bol,Izubuchi:2019lyk}
as well as by global fits~\cite{Barry:2021osv,Barry:2018ort}. From
an indirect estimation of the valence momentum fraction, $\langle
x\rangle_{u-\bar u}=0.23(2)$ from the PDF Ansatz fit, we find it to be the
same as $\langle x\rangle_{u+\bar u}$ within errors; thus, there
might only be negligible amount of anti-$u$ in the
large-$N_c$ pion wavefunction.

\bef
\centering
\includegraphics[scale=1]{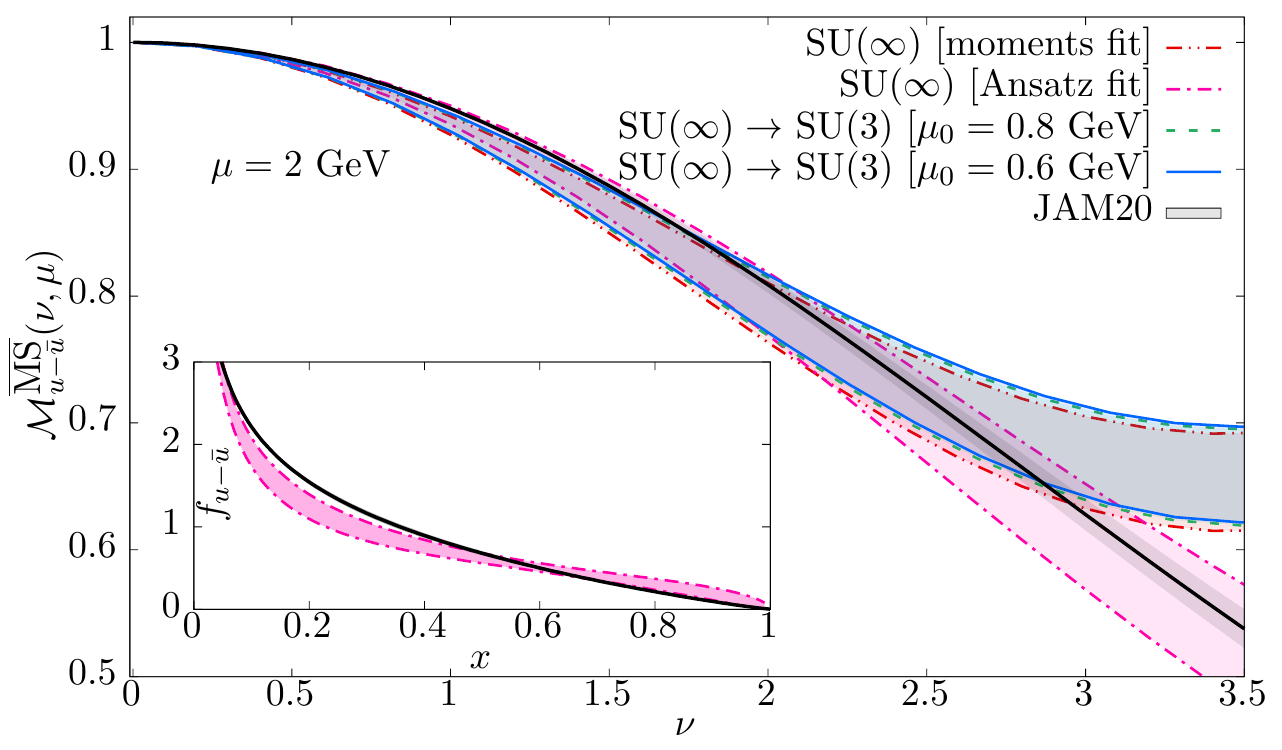}

\includegraphics[scale=1]{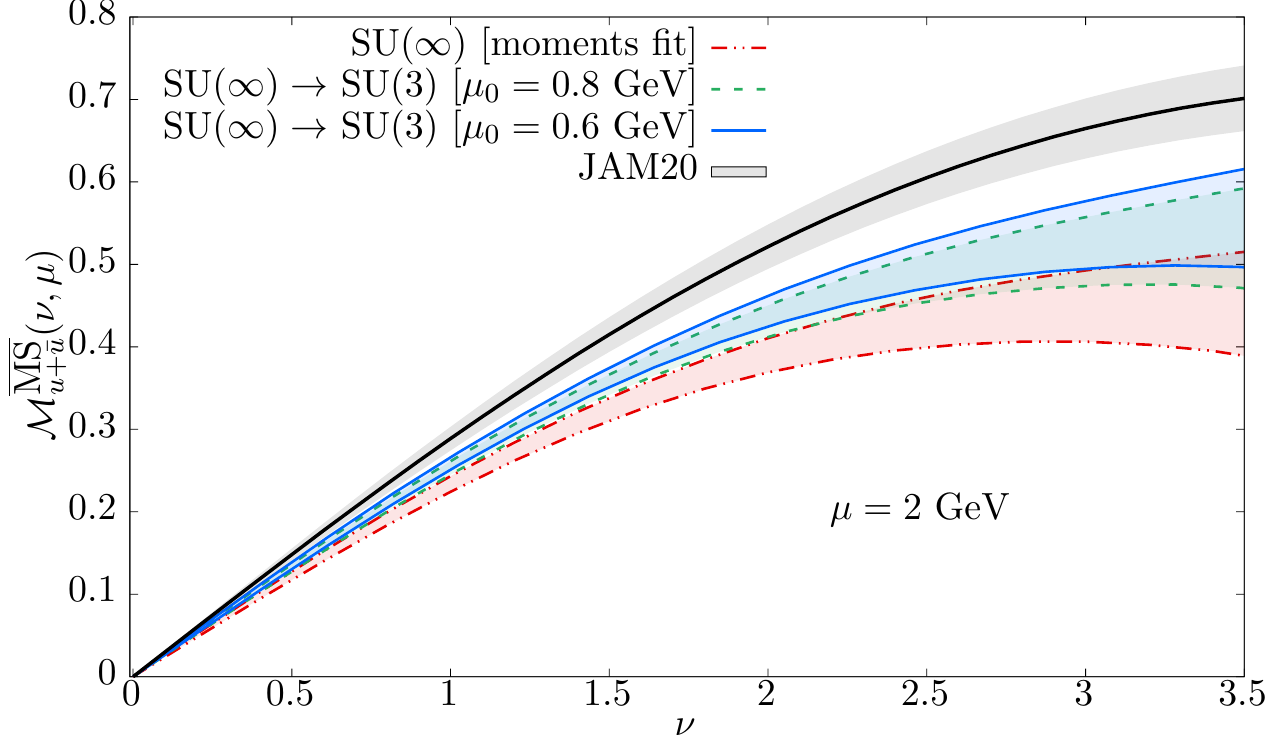}
\caption{
    The comparison of $\msbar$ ITDs in large-$N_c$ QCD (SU$(\infty)$)
    with the global fit results (JAM20) for the ITDs in SU(3) QCD
    at scale $\mu=2$ GeV.  The top and
    bottom panels are $u-\bar u$ and $u+\bar u$ ITDs respectively.  
    The red bands (moments fit) are
    results from OPE analysis by fitting Mellin moments. The green
    and blue bands are expectations for ITDs in SU(3) QCD based on
    the assumption of a nearly similar large-$N_c$ PDF at a lower
    factorization scale $\mu_0=0.8$ and 0.6 GeV respectively. The
    results for $u-\bar u$ ITD and PDF assuming an ansatz $f_{u-\bar
    u}(x,\alpha,\beta,s)$ are shown as purple bands (Ansatz fit) in the top-panel
    and its inset respectively.
}
\eef{ritdcompare}

\section{Phenomenology} 
In order to use large-$N_c$ theory as a model system to compare and
contrast the $SU(3)$ QCD with, we first set the GeV scale in the
$SU(\infty)$ world through a choice $\sqrt{\sigma}=0.44$ GeV that
is known~\cite{Teper:1997am,Perez:2020vbn,Bali:2008an} to result
in a low-energy meson spectrum that is numerically similar to the
real-world; this choice implies, $[a^{-1}, \mu,
m_\pi]=[1.73,2.00,0.38]$~GeV in our computation.  We use the $\msbar$
Ioffe-time distribution (ITD), ${\cal M}^{\msbar}_{u\pm\bar
u(}(\nu,\mu)$, which are the cosine (for $u-\bar u$) and sine (for
$u+\bar u$) Fourier transforms of PDFs from $x$
to $\nu$ space, to justifiably perform this comparison within the
range of $\nu$ spanned by our lattice data.  In \fgn{ritdcompare},
we compare ${\cal M}^{\msbar}_{u\pm\bar u}$ for the large-$N_c$
pion (red band), as inferred from the model-independent fits to
Mellin moments, with the JAM20 global fit result~\cite{Barry:2021osv}
(gray band) for the real-world pion at $\mu=2$ GeV.  We find a good
agreement between the two theories in the case of the valence $u-\bar
u$ ITD.  We suspect that the observed tendency for $SU(\infty)$
data to peel off at $\nu\approx 3$ could be a systematic effect due
to the absence of constraint from data beyond $\nu=3.5$, and in
fact, such a feature is absent in the ITD reconstructed from
$f_{u-\bar u}(x;\alpha,\beta,s)$ (purple band). In the inset of
\fgn{ritdcompare}, we also see a nearly similar $x$-dependencies of our Ansatz-based reconstruction
of $f_{u-\bar u}(x)$ (purple band) and the JAM20 result.
Thus, the valence structure of pion is likely to be weakly dependent
on $N_c$. It appears that features like the valence quarks that carry
$\approx 50\%$ of the pion momentum at few GeV resolutions,
could be typical in $SU(N_c)$ theories.

In the bottom panel of \fgn{ritdcompare}, we show a similar comparison
between $u+\bar u$ ITDs at $\mu=2$ GeV. Here, we see a visible
difference between the two theories. Based on a better agreement
seen in the valence sector, we ask if the difference seen in the
singlet $u+\bar u$ distribution could originate from the perturbative
radiative processes in the large-$N_c$ and $SU(3)$ QCD; as a main
difference, the $g \to q\bar q$ splitting is absent when $N_c\to\infty$.
In \apx{dglap}, we discuss the perturbative evolution aspects in 
the large-$N_c$ limit.
Working under a premise that the large-$N_c$ and SU(3) 
theories have similar $u+\bar
u$ PDFs at a low factorization scale $\mu_0$, we first evolved the
pairs, $[2\langle x^n\rangle_{u+\bar u},\langle x^n\rangle_g]$ at
$\mu=2$ GeV in the large-$N_c$ theory to a scale $\mu_0$ (= 0.8 to
0.6)~GeV, and evolved that result back to $\mu=2$ GeV using 3 flavor
SU(3) QCD DGLAP evolution. Since we have not explicitly calculated
the gluon moments for large-$N_c$ theory, we used the sum-rule
$\langle x\rangle_g = 1-2\langle x\rangle_{u+\bar u}$ and simply
set the other higher moments of the small-$x$ dominant gluon to be
negligible.  As we are looking only for qualitative tendencies, we
performed the evolution at leading-logarithmic order using the same
$\Lambda_{\msbar}$ in both theories.  We show the resulting ITDs
based on evolutions from $\mu_0=0.8$ and 0.6 GeV as the green and
the blue bands in \fgn{ritdcompare} lower panel. Remarkably,
the QCD evolution pulls the large-$N_c$ result closer to the
JAM20 result when successively smaller $\mu_0$ are used. Thus,
large-$N_c$ QCD presents itself as an interesting model system for
singlet parton physics where $g\to q\bar q$ splitting is switched
off, with all other splitting remaining intact.  Such a procedure
only lead to a negligible effect in valence $u-\bar u$ ITD as seen in \fgn{ritdcompare} top
panel.

\section{Discussion} 
We presented the large-$N_c$ mesons as an
interesting uncharted model-system for understanding partonic
physics, using the continuum reduction.  Our first lattice computation
of large-$N_c$ pion structure shows indeed that the structural
properties in the large-$N_c$ theory are likely to be similar to
our real-world, as has been seen in the meson spectrum; their
differences seem to be even more interesting as it gives us a version
of QCD where the sea is not radiatively proliferated with quark-antiquark
pairs, and hence could help understand the role of sea quarks in
real-world QCD.  Not to be mistaken, the method needs to be improved
by going to finer coupling, larger $N_c$, and also cross-checked with a complementary
twisted Eguchi-Kawai reduction~\cite{Gonzalez-Arroyo:2015bya}.  An
easy generalization of the method to QCD$_2$ might help in pruning
the Monte-Carlo methods by direct comparisons with analytical
results~\cite{tHooft:1974pnl,Bars:1977ud,Jia:2017uul,Jia:2018qee,Burkardt:2000uu}.
A large-$N_c$ advantage could be the exponential
suppression~\cite{Witten:1978bc,Teper:1979tq,Lucini:2001ej}, of
small instantons in large-$N_c$ limit that might suppress
instanton-induced power
corrections~\cite{Nason:1994xw,Shifman:1978bx,Andrei:1978xg}
to the OPE at typical short-distances reached in contemporary
lattice calculations.  It would be interesting to extend this work to
probe the differences in gluon structures of  the radial and
angular stable-excitations~\cite{Gao:2021hvs,Holl:2004fr} of the
ground-state mesons, and perform $x$-dependent spin physics of
stable higher-spin large-$N_c$ mesons, such as the $\rho$.

\begin{acknowledgments}
The authors thank R. Edwards, K. Orginos and J. Qiu for the valuable
discussions.  N.K. thanks P. Barry for helping with the JAM20 data.
The authors thank and acknowledge
the William \& Mary Research Computing for providing computational
resources and technical support that have contributed to the results
reported within this paper (\url{https://www.wm.edu/it/rc}); The
work was performed on the {\sl Femto} and {\sl Meltemi} computing
clusters at William \& Mary.
R.N. acknowledges partial support by the NSF under grant number
PHY-1913010.  
N.K. is supported by Jefferson Science Associates,
LLC under U.S.  DOE Contract \#DE-AC05-06OR23177 and in part by
U.S. DOE grant \#DE-FG02-04ER41302.  

\end{acknowledgments}

%\newpage
%\clearpage
%\pagestyle{empty}
%\onecolumngrid

%\begin{center}
%{\Large Supplementary Material}
%\end{center}

\appendix

\section{Details on continuum reduction}
We consider a $L^d$ periodic lattice. The gauge action in terms of plaquettes $U_p$ is
\beq
S_g = bN \sum_p \Tr (U_p + U_p^\dagger),
\eeq{wgact}
and gauge fields on an infinite $d$ dimensional lattice obey the periodic condition,
\beq
U_\mu(x) = U_\mu(x + L\hat \nu);\qquad \forall \quad \mu,\nu.
\eeq{gaugep}
As long as the lattice coupling $b < b_1(L)$ the theory is in the
confined phase and infinite volume results can be computed exactly
with finite lattice spacing effects. 
Below, we derive expressions for certain $n$-point functions of quark bilinears (mesons)
using the continuum reduction framework. We specify the $n$-point functions as $G^{(n)}$.
We derive results using continuum
reduction for a general dimension $d$, but we finally used only
$d=4$ in this work. 

\subsection{Quark propagator in infinite lattice from finite periodic
lattice: Bloch wavefunctions and $U^d(1)$ global symmetry}
\label{sec:propreduc}

Consider an operator $F(U)$ on an infinite lattice obtained by
copying the gauge fields from the $L^d$ lattice using periodicity.
As per Bloch's theorem, the eigenvalue problem takes the form
\beq
\sum_y^\infty F_{x,y} (U) q^i_y = \lambda_i (p) q^i_x(p);\qquad q^i_{x+nL} = e^{ip\cdot n} q^i_x(p),
\eeq{abloch}
where $n$ a tuple of integers.
Under a gauge transformation $g$, 
\beq
U_\mu^g( x) = g_{x} U_\mu( x) g^\dagger_{{x}+\hat\mu},\qquad 
F_{x,y}(U^g) = g_x F_{x,y}(U) g^\dagger_y.
\eeq{gtrans}
Of particular interest to us will be Abelian gauge transformations
of the form $g_x = e^{-i\frac{p\cdot x}{L}}$ on the infinite lattice.
Under these gauge transformations,
\beq
U_\mu^g(x) = U_\mu(x) e^{i\frac{p_\mu}{L}};\qquad(\text{written in short as\ } U e^{i \frac{p}{L}}), 
\eeq{aguage}
and we can rewrite the eigenvalue problem as
\beq
\sum_y^\infty F_{x,y} (Ue^{i\frac{p}{L}}) q^{ig}_y (p)= \lambda_i (p) q^{ig}_x(p);\qquad q^{ig}_x(p) = e^{-i\frac{p\cdot x}{L}} q^i_x(p);\qquad q^{ig}_{x+nL} (p)=  q^{ig}_x(p).
\eeq{gbloch}
One can use the periodicity of $q_x^{ig}(p)$ to further rewrite the
eigenvalue problem as
\beq
\sum_y^L F^L_{x,y}(Ue^{i\frac{p}{L}}) q_y^{ig}(p) = \lambda_i(p) q^{ig}_x(p);
\qquad F^L_{x,y}(U)\equiv \left[ \sum_{n=-\infty}^\infty F_{x,y+nL} (U)\right].
\eeq{pbloch}
and the induced operator $F^L$ on the finite periodic lattice satisfies
\beq
F^L_{x,y}(U) = F^L_{x,y+nL}(U) = F^L_{x+nL,y}(U)
\eeq{prop}
for any vector $n$ with integer entries.  The above eigenvalue
equation is for a finite size matrix on a finite periodic lattice.
We can write the operator and its inverse on the infinite lattice using their 
finite volume counterparts as
\beqa
F_{x,y}(U)  &=& \int \frac{d^d p}{(2\pi)^d} \sum_i \lambda_i(p) q_x^i (p)\left[q_y^i(p)\right]^\dagger
= \int \frac{d^d p}{(2\pi)^d} \ e^{i\frac{p\cdot(x-y)}{L}} F^L_{x,y}(Ue^{i\frac{p}{L}});\cr
F^{-1}_{x,y}(U)  &=& \int \frac{d^d p}{(2\pi)^d} \ e^{i\frac{p\cdot(x-y)}{L}} \left[ F^L\right]^{-1}_{x,y}(Ue^{i\frac{p}{L}}).
\eeqa{bloch}
One can extend the above relation to a product of operators:
\beqa
\sum^\infty_{z}A_{x,z}(Ue^{i\phi}) B_{z,y}(Ue^{i\chi})
&=&\sum^\infty_{z}\int \frac{d^dp }{(2\pi)^d} \frac{d^dq}{(2\pi)^d} e^{i\frac{p\cdot(x-z)}{L}} 
e^{i\frac{q\cdot(z-y)}{L}}
A_{x,z}^L\left(Ue^{i\phi}e^{i\frac{p}{L}}\right)B_{z,y}^L\left(Ue^{i\chi}e^{i\frac{q}{L}}\right)\cr
&=&\sum^L_{z}\sum_{k=-\infty}^\infty \int \frac{d^dp }{(2\pi)^d} \frac{d^dq}{(2\pi)^d} e^{i\frac{p\cdot(x-z-kL)}{L}} 
e^{i\frac{q\cdot(z+kL-y)}{L}}
A_{x,z}^L\left(Ue^{i\phi}e^{i\frac{p}{L}}\right)B_{z,y}^L\left(Ue^{i\chi}e^{i\frac{q}{L}}\right)\cr
&=& \int \frac{d^dp}{(2\pi)^d}
e^{i\frac{p_i\cdot(x-y)}{L}} \left[
\sum^L_{z} A_{x,z}^L\left(Ue^{i\phi}e^{i\frac{p}{L}}\right)B_{z,y}^L\left(Ue^{i\chi}e^{i\frac{p}{L}}\right)\right] .
\eeqa{prodinftop}

One application of the above reduction we will use involves the
quark propagator, $G_{x,y}(U)$, in a fixed gauge field background
and a smearing operator, $S^\phi_{x,y}(U)$, in a fixed gauge field
background where $\phi$ labels the type of smearing. Specific to
this work, $S(U_\mu)$ is the Wuppertal smearing kernel, and
$S^\phi_{x,y}(U_\mu)=S(e^{i\phi_\mu} U_\mu)$ for a phase $\phi=(0,\phi_1,\phi_2,\phi_3)$ with 
non-zero spatial components in general,
which we also set to 0.  Hence, what follows is for a more general
case than actually used in the present computation.  The smearing
operator is typically diagonal in spinor space. We will assume
$S^\phi(U)$ is Hermitian as is true in most cases that are typically functions of the 
covariant Lapacian operator. A form of reduction
we will need is
\beq
\sum^\infty_{x',x''}S^{\phi_1}_{x,x'} (U) G_{x',x''}(U) S^{\phi_2}_{x'',y}(U)
= \int \frac{d^dp}{(2\pi)^d}
e^{i\frac{p\cdot(x-y)}{L}} G^{L;\phi_1\phi_2}_{x,y} (U e^{i\frac{p}{L}}),
\eeq{inftofinp}
where we have used different smearing operators on either side of
the unsmeared quark propagator and we define the smeared propagator
on the finite periodic lattice by
\beq
G^{L;\phi_1\phi_2}_{x,y} (U ) =  \sum^L_{x',x''} S^{L\phi_1}_{x,x'}(U) G^{L}_{x',x''}(U) S^{L\phi_2}_{x'',y}(U)
\eeq{smearp}
which satisfies
\beq
G^{L;\phi_1\phi_2}_{x,y} (U ) = G^{L;\phi_1\phi_2}_{x,y+nL } (U )  = G^{L;\phi_1\phi_2}_{x+nL,y } (U ) 
\eeq{proppr}
for any vector $n$ with integer entries as expected of a propagator
on a periodic lattice.

 \subsection{Two point function of mesons}
\label{sec:twoptreduc}
Let 
\beq
M^\Gamma_{ij}(x) = \sum_{x',x''} \bar q^{(i)}_{x'} S^{\phi_i}_{x',x} (U) \Gamma  S^{\phi_j}_{x,x''}(U_\mu) q^{(j)}_{x''}
\eeq{mesonop}
be a gauge invariant meson operator located at $x$.  
Since $S_{x,y}\propto \delta_{x_0,y_0}$, the sums above are actually restricted to time-slice containing $x$ by construction, but 
written as a sum over the entire space-time.
The indices
$i,j$ provide the quark flavor indices and $\Gamma$ specifies the
type of fermion in a spinor space.  For the case of pion, considered
in the paper,
\beq
\pi(x) = M^{\gamma_5}_{du}(x)=\sum_{x',x''} \bar d_{x'} S_{x',x}(Ue^{i\phi}) \gamma_5 S_{x,x''}(Ue^{-i\phi}) u_{x''}.
\eeq{piop}
Using
\eqn{inftofinp}, the two point function of a meson of a type
$\Gamma_1$ with type $\Gamma_2$ in the infinite lattice is
\beq
G^{(2)}(x,y)=\left \langle M^{\Gamma_1}_{ij}(x) \left[M^{\Gamma_2}_{ij}\right]^\dagger(y)\right\rangle,
\eeq{m2ptdef}
which after Wick contraction yields
\beq
G^{(2)}(x,y;U) 
=  \int \frac{d^d q}{(2\pi)^d}   \frac{d^d q'}{(2\pi)^d} e^{i\frac{(q-q')\cdot(x-y)}{L}} 
{\tr } \left[ \Gamma_1 
 G^{L;\phi_2\phi_2}_{x,y } (U e^{i\frac{q}{L}})\Gamma_2
 G^{L;\phi_1\phi_1}_{y,x } (U e^{i\frac{q'}{L}}) 
 \right].
\eeq{m2ptfn}
If we write down the two point function in momentum space using
\beq
\tilde G^{(2)}(p',p;U) = \sum_{x,y}^\infty e^{i(p'\cdot x+p\cdot y)} G^{(2)}(x,y;U) 
\eeq{2ptft}
and split the infinite sum over $x$ and $y$ into blocks of finite
sums over finite periodic lattice (i.e., $x\to x + n {\rm\ mod\ } L$,
and replace sum to be over the periodic $x$ and $n$) and invoke the
periodicity property in \eqn{proppr}, we will arrive at condition $L p +q'-q=0$ along with 
momentum conservation $p+p'=0$. Using these, we can write the 2-point function as
\beq
 \tilde G^{(2)}(p',p;U) = \delta(p'+p)\int \frac{d^dq}{(2\pi)^d}
 \Tr \left[
     \Gamma_1 G^{L;\phi_2\phi_2} \left(U e^{i\frac{q}{L}} \right)
\Gamma_2 G^{L;\phi_1\phi_1} \left(U e^{i\frac{q}{L}+p'}\right) \right],
\eeq{m2ptfnm}
where $\Tr$ denotes the trace over the entire lattice and spin.
Invoking the $U^d(1)$ global symmetry present in the confined phase,
we can shift $U e^{i\frac{q}{L}} \to U$, we can write the propagator
in momentum space at a fixed gauge field background as
\beq
\tilde G^{(2)} (p',p; U_\mu) = \delta(p'+p) \Tr \left[
    \Gamma_1 G^{L\phi_2\phi_2} \left(U \right)
\Gamma_2 G^{L\phi_1\phi_1} \left(U e^{-ip}\right)
 \right].
\eeq{m2ptfnr}
For the pion, for which $\phi_1 = -\phi_2=\phi$ so as to preserve isospin symmetry during 
quark-smearing, we defined the 2-point function in the main text as
\beq
\tilde C_{\rm 2pt} (p) = \Tr \left[
    \gamma_5 G^{L;-\phi,-\phi} \left(U \right)
\gamma_5 G^{L;\phi,\phi} \left(U e^{-ip}\right)
 \right],
\eeq{m2ptfnrmaintext}
with explicit smearing factors included in the detailed expression.

\subsection{QuasiPDF-pion-pion three-point function}
\label{sec:threeptreduc}

Let the fermion bilinear connected by a spatial Wilson line from
$w$ to $w+z$ for $z=(0,0,0,z_3)$, i.e., the quasi-PDF operator, is
given by
\beq
{\cal O}(w;z) =  \bar u_{w} \gamma_0 W_{w;w+z} u_{w} 
\eeq{wlineop}
Our focus will be on the three point function
\beq
G^{(3)}(x,y,w;z)\equiv \langle \pi(x) {\cal O}(w;z) \pi^\dagger(y) \rangle.
\eeq{3ptfn}
Strictly in the large $N_c$ limit, we can ignore the quark-line disconnected diagrams and 
write
\beqa
G^{(3)}(x,y,w;z)
&=&  \sum_{x',x'',y',y''} {\rm tr}\Biggl(\gamma_5 S_{x,x'}(U e^{-i\phi}) G_{x',w} (U) \gamma_0 
  W_{w,w+z} (U)
G_{w+z,x''} (U) \cr
&&\qquad\qquad S_{x'',y}(Ue^{-i\phi})\gamma_5 
  S_{y,y'}(Ue^{i\phi})  G_{y',y''} (U)   S_{y'',x}(Ue^{i\phi})\Biggr).
\eeqa{3ptfns}
Replacing infinite lattice propagators by propagators on periodic lattice as we have done before,
\beqa
G^{(3)}(x,y,w;z)
&=&  \int \frac{d^4 q}{(2\pi)^4} \frac{d^4 q'}{(2\pi)^4} \frac{d^4 q''}{(2\pi)^4} 
e^{-i q\cdot (w-x)/L} e^{-i q'\cdot (y-w-z)/L } e^{-i q''\cdot (x-y)/L}\cr&&  {\rm tr}\Biggl(\gamma_5 G^{L;-\phi,\emptyset}_{x,w}(Ue^{i q/L}) \gamma_0 W_{w,w+z}(U) 
 G^{L;\emptyset;-\phi}_{w+z,y}(Ue^{iq'/L}) 
\gamma_5 G^{L;\phi,\phi}_{y,x}(U e^{iq''/L})\Biggr),
\eeqa{3ptfnsp}
where we have used the following notation,
\beqa
G^{L;-\phi,\emptyset}_{x,y}(U) &=& \sum_{x'} S^L_{x,x'}(Ue^{-i\phi}) G^L_{x',y}(U)\cr
G^{L;\emptyset,-\phi}_{x,y}(U) &=& \sum_{x'} G^L_{x,x'}(U) S^L_{x',y}(Ue^{-i\phi})\cr
G^{L;\phi,\phi}_{x,y}(U) &=& \sum_{x',x''} S^L_{x,x'}(Ue^{i\phi}) G^L_{x',x''}(U) S^L_{x'',y}(Ue^{i\phi}).
\eeqa{smprops}
Fourier transforming over  $(x,w,y)\to (p',Q,p)$ on the infinite lattice keeping $z$ fixed,
\beqa
&&\tilde G^{(3)}(p',Q,p;z,U) \cr
&=& \sum_{x,w,y}^\infty \int \frac{d^4 q}{(2\pi)^4} \frac{d^4 q'}{(2\pi)^4} \frac{d^4 q''}{(2\pi)^4} 
e^{-i q\cdot (w-x)/L}  
 e^{-i q'\cdot (y-w-z)/L } e^{-i q''\cdot (x-y)/L} 
 e^{i(p'\cdot x + Q\cdot w  + p\cdot y)} \cr
&& \qquad\qquad {\rm tr}\Biggl(\gamma_5 G^{L;-\phi,\emptyset}_{x,w}(Ue^{iq/L}) \gamma_0 W_{w,w+z}(U) 
G^{L;\emptyset,-\phi}_{w+z,y}(Ue^{iq'/L}) \gamma_5 G^{L;\phi,\phi}_{y,x}(U e^{iq''/L})\Biggr).
\eeqa{3ptfnsp1}
We can split the infinite sum over $x,y,w$ into blocks of
sums over finite periodic lattice, invoke the periodicity property
in \eqn{proppr} and use the folded property of Wilson lines. This
will result in
\beq
 q-q''+p' L =0;\qquad q'-q+Q L=0;  \qquad q''-q'+p L=0,
 \eeq{momrel}
 which includes the momentum conservation, $p'+ Q + p=0$.
 We arrive at
\beqa
&&\tilde G^{(3)}(p',Q,p;z,U)
= \delta(p'+Q+p) \int \frac{d^4 q'}{(2\pi)^4}  \cr&& \sum_{x,w,y}^L 
{\rm tr}\left(\gamma_5 G^{L;-\phi,\emptyset}_{x,w}(Ue^{i\left(\frac{q'}{L}+Q\right)}) \gamma_0 W_{w,w+z}(Ue^{i\frac{q'}{L}}) G^{L;\emptyset,-\phi}_{w+z,y}(Ue^{i \frac{q'}{L}}) \gamma_5 G^{L;\phi,\phi}_{y,x}(U e^{i\left(\frac{q'}{L}-p\right)})\right).
\eeqa{3ptfnsp2}
Using $U^d(1)$ symmetry,
\beqa
&&\tilde G^{(3)}(p',Q,p;z,U)=\cr
&=&\delta(p'+Q+p) \sum_{x,w,y}^L {\rm tr}\left(  \gamma_0 W_{w,w+z}(U) G^{L;\emptyset,-\phi}_{w+z,y}(U) \gamma_5 G^{L;\phi,\phi}_{y,x}(U e^{-ip})\gamma_5 G^{L;-\phi,\emptyset}_{x,w}(Ue^{iQ})\right).
\eeqa{3ptfunsp3}
We referred to the above equation in the main text, now with an explicit specification of quark smearing factors, as
\beq
\tilde{C}_{\rm 3pt}(z,p,Q) =  \sum_{x,w,y}^L {\rm tr}\left(  \gamma_0 W_{w,w+z}(U) G^{L;\emptyset,-\phi}_{w+z,y}(U) \gamma_5 G^{L;\phi,\phi}_{y,x}(U e^{-ip})\gamma_5 G^{L;-\phi,\emptyset}_{x,w}(Ue^{iQ})\right).
\eeq{maintext3pt}
We used $Q=0$ in this work.

\section{Details of the lattice calculation}
\label{sec:calcdetails}
\bet
\begin{tabular}{|c|c|c|c|}
\hline
    $n_3$ & $P_3 a$ & $P_3/\sqrt{\sigma}$ & configurations \cr
\hline
    0 & 0 & 0 & 15353  \cr
    2 & 0.092 & 0.363 & 16320  \cr
    4 & 0.185 & 0.727 & 11520  \cr
    6 & 0.277 & 1.090 & 19200  \cr
    8 & 0.370 & 1.454 & 30720  \cr
    10 & 0.462 & 1.817 & 27552 \cr
    12 & 0.554 & 2.181 & 30720 \cr
    14 & 0.647 & 2.544 & 30720 \cr
    16 & 0.739 & 2.910 & 30504 \cr
\hline
\end{tabular}
\caption{The table lists the momenta $P_3 = \left(\frac{2\pi}{L_s
N_c}\right) n_3$, and the amount of statistics at each momentum.
The statistics comes from two sources; namely, independent number
of gauge field configurations (second column) and the number of
$Z_2$ stochastic vectors in each configuration.  We used 3 $Z_2$
random vectors which are diluted in chirality and in even-odd lattice
sites, which comes out to 12 set of inversions over the components
of the noise vectors. We fixed this for all momenta. To convert $P_3$ to GeV,
$P_z = 0.081 n_3$ GeV.  All the momenta used in this work are below
the lattice $a^{-1}$ scale.  }
\eet{details}

In the present work, we used a fixed large value of $N_c=17$, since
it is the smallest value of $N_c$ beyond which the $1/N_c$ corrections
are typically found to small in previous works.  We used $L^4$
lattices in this paper with $L=8$. We used the standard single plaquette Wilson gauge action
 and set the lattice coupling $b=0.355$, such that
it is close to being the largest $b$ possible on $L=8$ and keep the
lattice gauge theory is in the confined phase (phase 0c). The critical 
$L_c(b)=6.6$ for the value of $b$ we used. Since
fermion loops are $1/N_c$ suppressed, the quenched lattice computation
of fermionic quantities is exact in the large-$N_c$ limit. 
Each update of the gauge fields on the entire lattice was made up of $\frac{N_c(N_c-1)}{2}$ SU(2) heat-bath updates on every link followed by one SU(N) over-relaxation update on every link~\cite{Kiskis:2003rd}. We performed $100$ such updates between measurements to avoid autocorrelation.
To
make sure the configurations thermalized to the 0c phase, we
successively decreased the value of $b$ from a higher value of
$b=0.365$.  By monitoring the gap in the Polyakov loop eigenvalues~\cite{Kiskis:2003rd}
in all four directions, we ensured that the configurations were in
the correct phase. We computed the pion-pion
two-point and pion-quasiPDF-pion three-point functions (Equations 2 and 4 in the main text) 
on every configuration
at 9 different values of momentum,
\beq
P_3 a = \frac{2\pi n_3}{N_c L} = \frac{2\pi n_3}{136},
\eeq{momvals}
for $n_3=0,2,4,\ldots,16$. We used gradually more number of configuration, $N_{\rm cfg}$ ($\sim$ 32K)
at the higher momenta compared to the lower ones ($N_{\rm cfg}\sim$ 12K). We
have collected the details of the statistics in \tbn{details}.

We evaluated the two-point and three-point functions (Equations 2 and 4 in the main text) stochastically.
Namely, for the two-point function, the stochastic estimator using noise vectors $\xi$,
\beq
\tilde{C}_{\rm 2pt}(p)=\overline{\xi^\dagger \gamma_5 G^L(U) \gamma_5 G^L\left(Ue^{ip}\right) \xi} = \overline{\chi^\dagger(0) \phi(p)},
\eeq{2ptstochastic}
with $\phi(p)\equiv G^L\left(Ue^{iq}\right) \xi$ and $\chi(0)\equiv \gamma_5 [G^L(U)]^\dagger \gamma_5 \xi$. The 
combined noise and ensemble average is 
\beq
\overline{\xi^\dagger A \xi} = \frac{1}{N_{\rm vec} N_{\rm cfg}} \sum_{i=1}^{N_{\rm vec}\times N_{\rm cfg}} \xi_i^\dagger A \xi_i.
\eeq{stochav}
We used $N_{\rm vec}=3$ number of $\mathbb{Z}_2$ noise vectors for
$\xi$; that is $\xi^{\alpha,a,x}=\frac{1\pm i}{\sqrt{2}}$ for spin,
color and position indices $\alpha,a,x$ respectively. We further
diluted the noise-vectors over even-odd lattice sites and over the
two chiral projections. For the three-point function, we used the stochastic estimator as
\beq
\tilde{C}_{\rm 3pt}(z,p,q) =\sum_{x} \overline{ \xi^\dagger_x \gamma_0 W_{x,x+z} \phi_{x+z}(p)};\qquad \phi(p)=G^L(U)\gamma_5 G^L(Ue^{-i p})\gamma_5 G^L(Ue^{iq})\xi.
\eeq{3ptstochastic}
In this work, we only used $q=0$ above.
We used 2-steps of Stout smearing for the gauge-links that are used to construct the Wilson line $W$.

We used Wilson-Dirac operator $\slashed{D}^L(m_w)$ to compute
the propagators $G^L = [\slashed{D}^L]^{-1}$.  We improved the Dirac
operator by using gauge-links that are smeared by two steps of the
large-$N_c$ version of the Stout smearing~\cite{Morningstar:2003gk}. With smearing,
we expect the zero quark mass to be in the region of the Wilson
mass $m_w=[-0.38, -0.39]$.  We tuned to $m_w=-0.36$ to realize a
pion mass that was feasible given the computational resource available
to us.  We implemented the Wilson-Dirac inversion using BiCG-Stab
algorithm~\cite{Frommer:1994vn}.

We used smeared quark sources in the construction of two-point and
three-point functions using a smearing kernel $S(U;N_{\rm wup},\delta)$ for the
Wuppertal smearing~\cite{Gusken:1989ad}. We implemented Wuppertal smearing using
$(N_{\rm wup}, \delta)=(40,0.6)$, which we chose to be optimal
through a set of initial tuning runs. We kept the radius of Wuppertal smearing fixed 
at all pion momenta. A puzzling experience during the 
tuning process at non-zero momenta was the negligible effect of phased momentum smearing~\cite{Bali:2016lva}, $S(e^{i\phi} U)$,
which typically improves the signal to noise ratio at higher momenta in the SU(3) QCD at some value of $\phi$; 
we did not find any such improvement within statistical errors 
during the tuning phase in which we used only about $\sim 1000$ 
configurations. Therefore, we simply used unphased ($\phi=0$) Wuppertal kernel
for quark smearing at all momenta.

\section{Spectral content of pion two-point functions}
\label{sec:twoptspect}

\subsection{Construction}

We determined the two-point function $\tilde{C}_{\rm 2pt}(p)$ with
$p=(p_0,0,0,P_3)$ for $p_0 = 2\pi n_0/L^{\rm eff}_0$ using $n_0\in[0,\frac{L^{\rm eff}_0}{2}]$ and 
the effective temporal extent of $L^{\rm eff}_0=68 = N_c L/2$. We
could have used an effective temporal extent of up to $136$, but used a
smaller one to make the computation easier, and an effective temporal
extent of 68 is quite comparable to what is being used in present
structural computations in SU(3) QCD. One possibility to investigate
the spectral content in the 2-point function is to fit the data at
different fixed spatial $P_3$ to
\beq
\tilde{C}_{\rm 2pt}(p_0,P_3) = B + \sum_{i=0}^{N_{\rm st}-1} \frac{|A_i|^2 \sinh(a E_i(P_3))}{\cosh(a E_i(P_3)-\cos(a p_0))};\qquad A_i=\langle E_i|\pi^\dagger|0\rangle,
\eeq{momspect}
where one can truncate the momentum space spectral decomposition
at $N_{\rm st}$ number of states. It is to be remembered that, even
if higher excited states might not contribute to the $p_0$ dependence
of the correlator in the range of smaller $p_0$, they can still
contribute a momentum independent constant to the above correlator.
Therefore, we corrected such a truncated series by a constant term, $B$,
to account for such effects of all other higher excited states.

In this paper, we used the computed momentum space 2-point functions to 
Fourier transform them into real space, so that we can perform a rather 
traditional lattice QCD analysis via effective masses and multi-exponential
fits. That is, the real space correlator is
\beq
C_{\rm 2pt}(t_s, P_3) = \sum_{n_0=0}^{L^{\rm eff}_t-1} \tilde{C}_{\rm 2pt}(p_0,P_3) e^{i p_0\frac{t_s}{a}};\quad \tilde{C}_{\rm 2pt}(L^{\rm eff}_0-n_0,P_3) = \tilde{C}_{\rm 2pt}(n_0,P_3).
\eeq{realc2pt}
Note that we have used the momenta $p$ and their integer quanta $n$
interchangeably as arguments above, and we will do so in the
rest of the text without any obvious confusion.
After the above Fourier transformation, we performed the usual $N_{\rm st}$-state fits to study their spectral content,
\beq
C_{\rm 2pt}(t_s, P_3) = \sum_{i=0}^{N_{\rm st}-1} |A_i|^2 \left(e^{-E_i \frac{t_s}{a}}+e^{-E_i (L^{\rm eff}_0-\frac{t_s}{a})}\right),
\eeq{nstatefit}
and obtained their effective masses by solving $\frac{\cosh(E_i
(L^{\rm eff}_0/2-(t_s+a)/a))}{\cosh(E_i (L^{\rm eff}_0/2-t_s/a))}
= \frac{C_{\rm 2pt}(t_s+a, P_3)}{C_{\rm 2pt}(t_s, P_3)}$.

\subsection{An issue with long-tailed distributions in the zero-momentum case}
\bef
\centering
\includegraphics[scale=0.75]{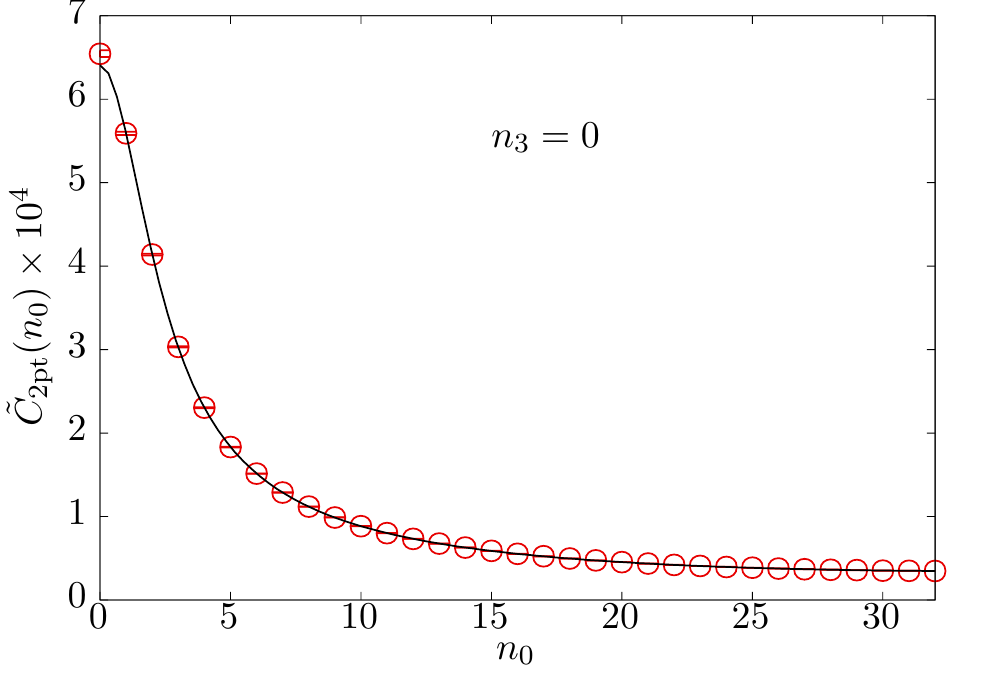}
\includegraphics[scale=0.75]{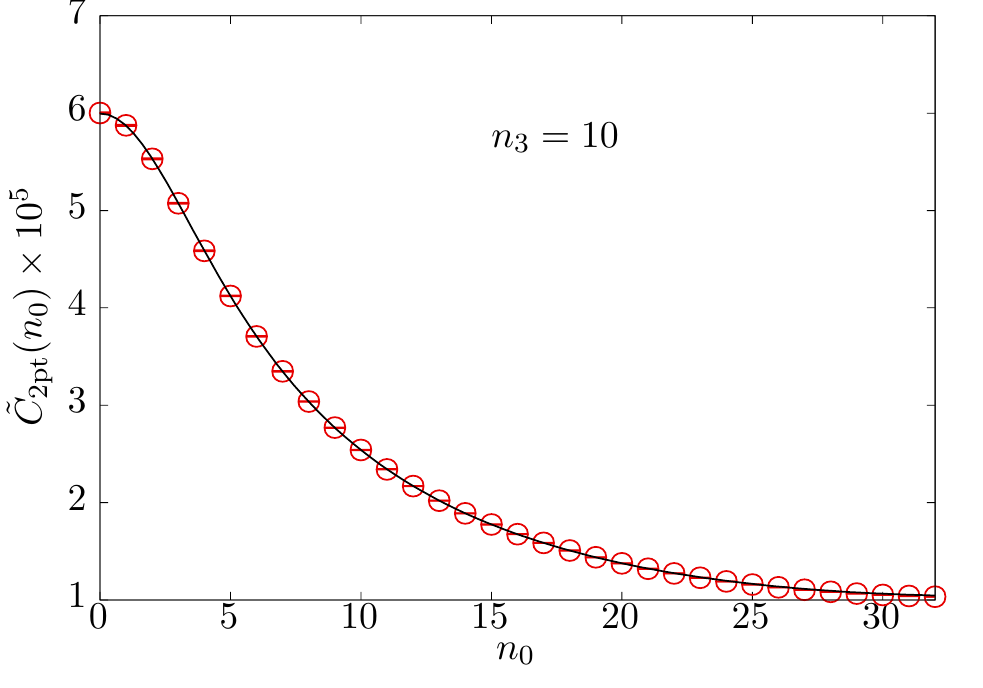}

\includegraphics[scale=0.75]{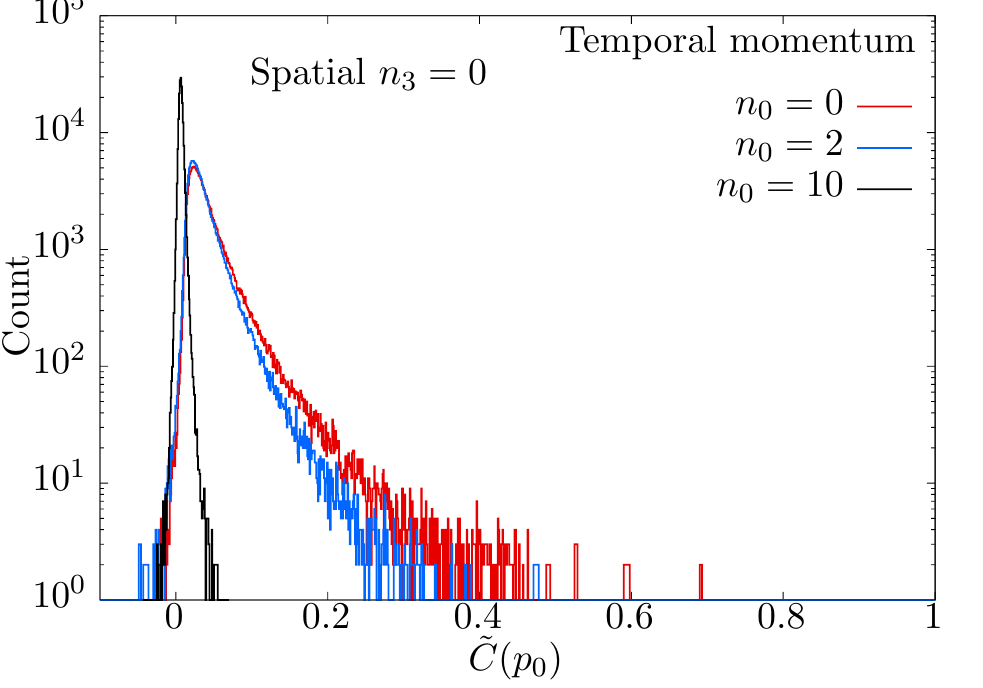}
\includegraphics[scale=0.75]{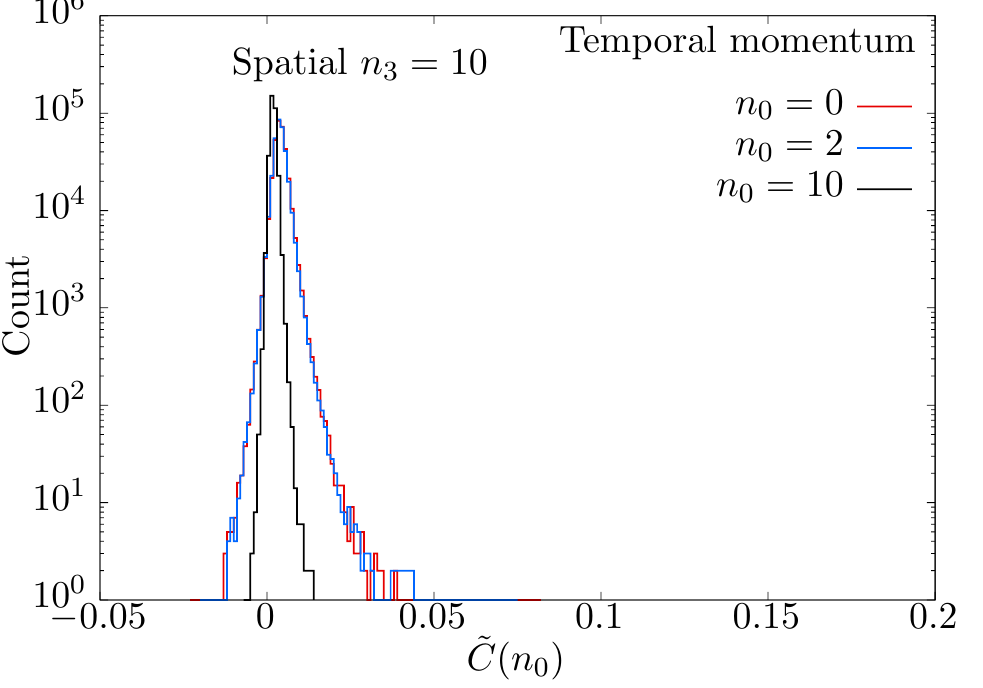}
\caption{(top panels) Sample pion-pion two point functions
$\tilde{C}_{\rm 2pt}(n_0; n_3)$ at spatial momenta $n_3=0$ (left)
and $n_3=10$ (right) as a function of temporal momentum $n_0$. The
black curves are fits to the momentum space correlator to \eqn{momspect} with $N_{\rm st}=3$ to data
from $n_0 \in [1,20]$.  For $n_3=10$, the fit automatically passes
through the $n_0=0$ data point. For $n_3=0$, the $n_0=0$ data point
is slightly above the expectation. This causes a problem for $n_3=0$
real space correlator construction and shows up as a pathological
state with near zero mass. The possible issue is purely numerical
and has to do with $n_3=0,n_0=0$ being the pion susceptibility that
is hard to evaluate stochastically due to the long tailed nature
of its Monte Carlo histogram. (Bottom panels) Such Monte Carlo
histograms for $\tilde{C}_{\rm 2pt}(n_0,n_3)$ are shown in the bottom
two panels for $n_3=0$ and $n_3=10$.  Indeed the $n_3=0, n_0=0$
case is long tailed, and it might require even larger
statistics to evaluate it robustly. The histograms immediately get
narrower at non-zero $n_3$ and $n_0$ and thereby, does not cause any issues.
}
\eef{fitpx}
First, we discuss the case of zero spatial momentum $P_3=0$ which
we found to be challenging within the stochastic approach of
constructing trace along with the Fourier transform to real space.
In the top panels of \fgn{fitpx}, we show the momentum space
correlator $\tilde{C}(p_0,P_3)$ as a function of temporal momentum
$p_0$; the top-left and top-right panels show the results at $P_3=0$
and $P_3/(2\pi/136)=n_3 = 10$ respectively. The black curves are the
best fit curves using \eqn{momspect} truncated at
$N_{\rm st}=3$ and fitted over a range of $n_0\in [1,20]$. The fits
work well with $\chi^2/{\rm dof}\approx 1$.  However, we note that
the curve for $P_3=0$ when extrapolated to $n_0=0$ is slightly, but
in a statistically significant manner, below the actual stochastically
evaluated data point for $\tilde{C}(p_0=0,P_3=0)$. Such a problem
existed only at $P_3=0$ case, and at other non-zero $P_3$ (such as
$n_3=10$ case on the top-right) the fitted curve automatically
passed through $n_0=0$ data point as well.  While the problem is
easy to fix by avoiding the $n_0=0$ data point while performing
fits, it causes problem when reconstructing real-space correlators
via Fourier transform; namely, if $n_0=0$ is not evaluated very
accurately, then its effect is to add a spurious low-mass state
into the real-space correlator.  The origin of the problem is easy
to understand. In the bottom left and bottom right panels of
\fgn{fitpx}, we show the Monte Carlo histogram of the stochastic
estimator in \eqn{2ptstochastic} for the two spatial momenta. For
each of them, we have shown the histograms at three values of $n_0$.
We see that for $n_0=0, n_3=0$, which is nothing but pion susceptibility,
the distribution is very long-tailed, and hence, it is likely that
the difficulty we are finding is due to the inability to robustly
estimate the mean and the statistical error of such a long-tailed
distribution. At non-zero $n_0$, the distribution gets narrower.
Also, the distribution at $n_0=0$ gets narrower at non-zero $n_3$,
and hence we were able to reconstruct real-space correlators well.
Having understood the problem, we found the following procedure to
correct the $n_0=0$ data points to solve the issue; we took the
lowest three non-zero $n_0=1,2,3$ data points that is dominated by
the ground-state $E_0$, and solved a system of equations,
\beq
\frac{A}{\cos(a p_0)-\cosh(a E_0)} + B = \tilde{C}_{\rm 2pt}(p_0,P_3),
\eeq{solvesystem}
to find the unknown parameters $A, B, E_0$. Using them, we corrected the
$n_0=0$ point with the estimated value $\frac{A}{1-\cosh(aE_0)}+B$.
Using this corrected $n_0=0$ data point, we used \eqn{realc2pt} to
perform the Fourier transform and obtained the correlator as a function
of $t_s$. With this procedure, the spurious low mode disappeared
at $P_3=0$. At non-zero $P_3$, such a procedure did not have any
significant effect at all.

\bef
\centering
\includegraphics[scale=0.75]{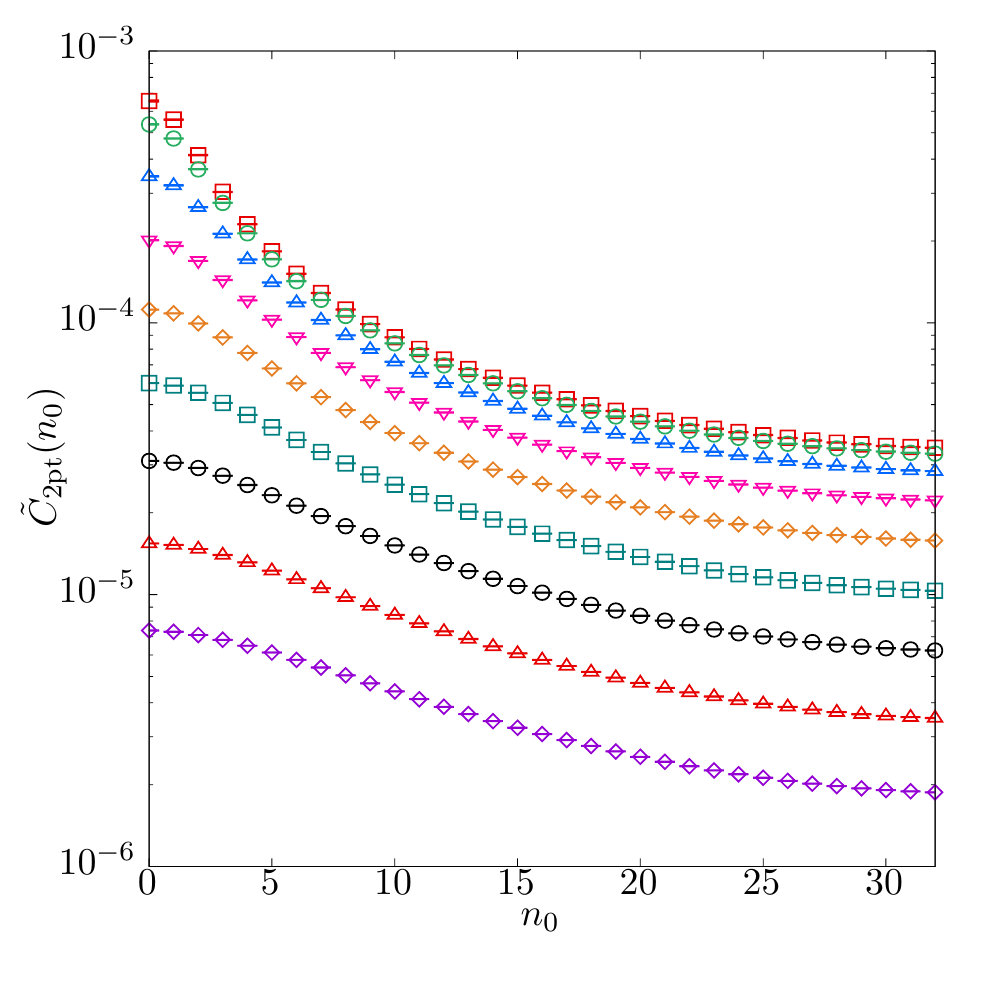}
\includegraphics[scale=0.75]{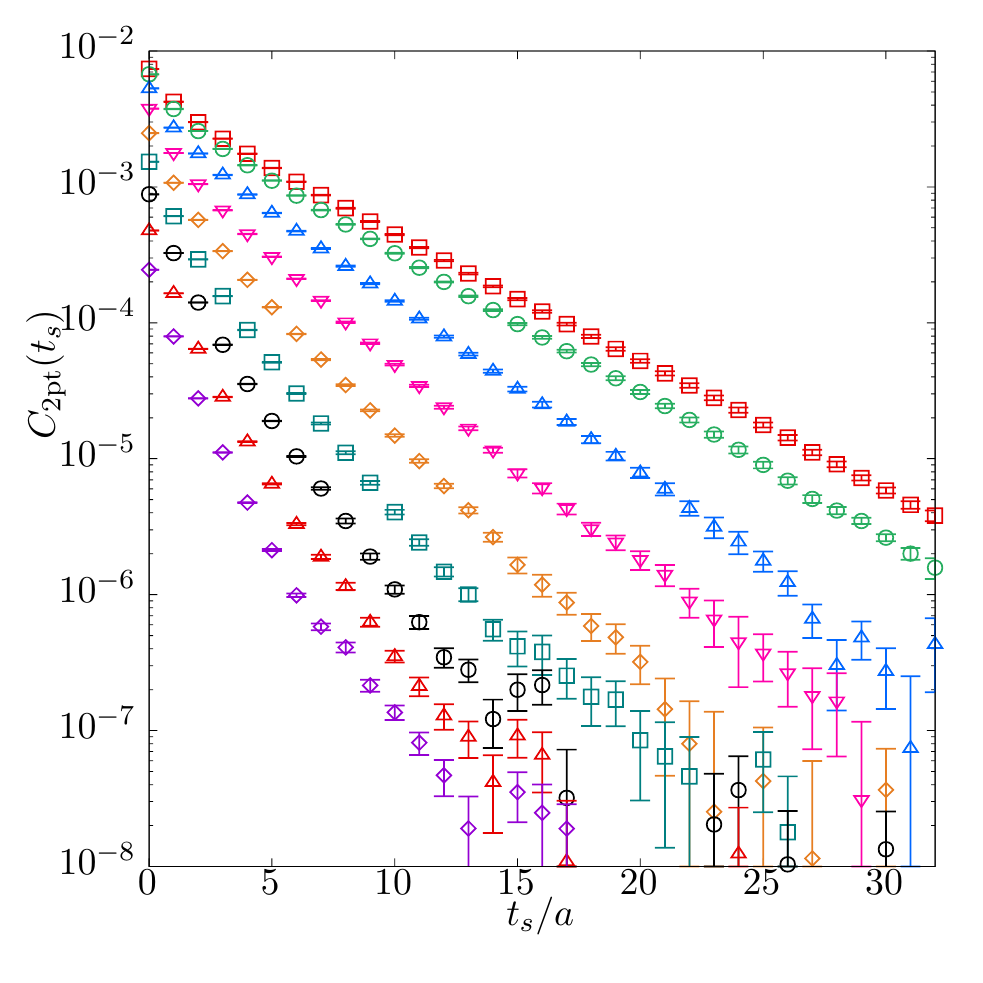}
\caption{Pion two point function correlator as a function of 
temporal Euclidean momentum $n_0$ (left) and as a function of
temporal separation $t_s/a$ (right) as constructed from the momentum 
space correlations via Fourier transformation. The different symbols from 
top to bottom in the two panels are the data points at different 
spatial momentum along the $z$-direction, $n_3$, from 
$n_3=0$ to 16 in steps of 2.}
\eef{correl}

\bef
\centering
\includegraphics[scale=0.85]{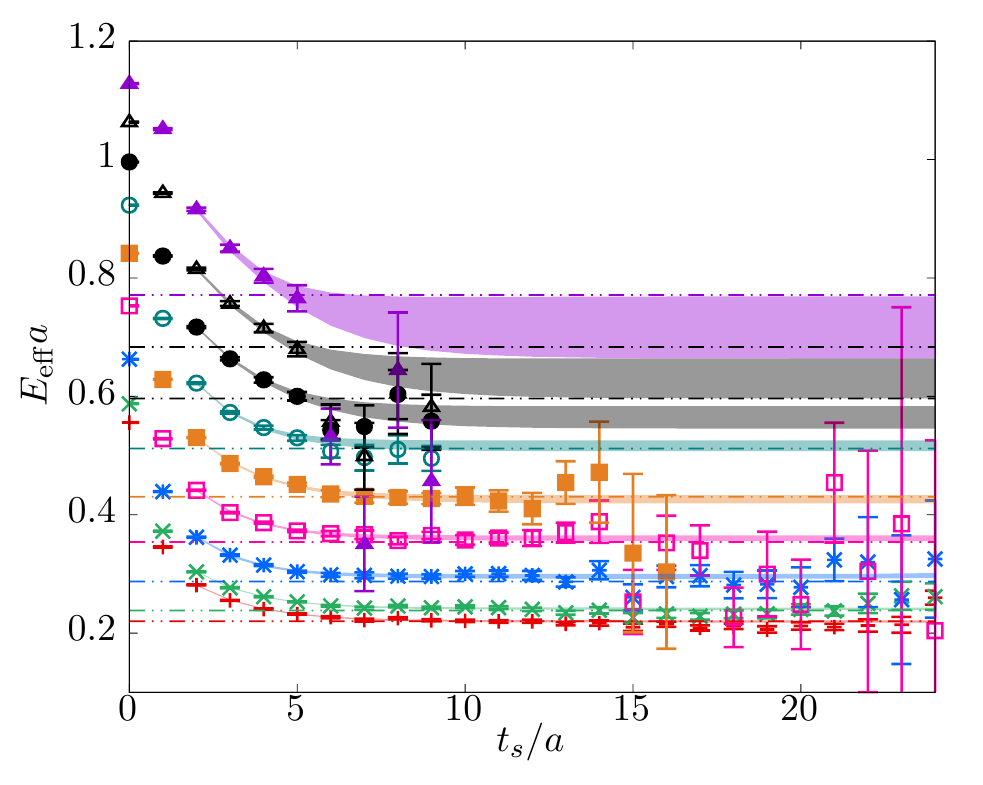}
\caption{The effective mass as determined from the real space 
two-point function $C_{\rm 2pt}(t_s,P_3)$. The different symbols are 
the results at different spatial momenta $P_3=2\pi n_3/136$
from $n_3=0,2,\ldots,16$, from bottom to top. The bands are 
two-state fits to correlators.
The horizontal lines are expectations from continuum single-particle
dispersion.
}
\eef{effmass}

\subsection{Analysis of 2-point functions in real-space}

In the left panel of \fgn{correl}, we show our ``raw" data for
$\tilde{C}_{\rm 2pt}(p_0,P_3)$ that we directly computed on the lattice as a
function of $p_0$ for the 9 different spatial $P_3$. As such, we
find our determination of $\tilde{C}_{\rm 2pt}(p_0,P_3)$ to be smoothly varying
in both $p_0$ and $P_3$, up to an issue noted above for $P_3=0$.
The well-determined nature of $\tilde{C}$ in Fourier space is
deceiving, as the long-distance exponential fall-off in the Fourier
transformed $C_{\rm 2pt}(t_s)$ comes from delicate cancellations between
different $\tilde{C}_{\rm 2pt}(p_0)$, resulting in noise at larger $t_s$. This can be seen in the right panel
of \fgn{correl}, where we show such an inferred real-space two-point
function, $C_{\rm 2pt}(t_s,P_3)$ using \eqn{realc2pt}. We have displayed the
results at various spatial momenta as a function of $t_s/a$.

In the determination of the quasi-PDF matrix element, the spectral
data of two-point function does not enter in the summation type
analysis we performed (as we discuss in the next section). We present
our analysis of the spectral content of the two-point function now
for the sake of completion. In \fgn{effmass}, we show the effective
mass $E_{\rm eff}(t_s)$ determined from $C_{\rm 2pt}(t_s,P_3)$. We were
able to perform stable two-state fits ($N_{\rm st}=2$ in \eqn{nstatefit})
to $C_{\rm 2pt}(t_s)$. For this, we used a fit range $t_s\in [2a,15a]$
for $n_3=0,2,4,6,8$, $t_s\in[2a,10a]$ for $n_3=10,12$, and
$t_s\in[2a,6a]$ for $n_3=14,16$.  We used a smaller minimum of $2a$
so as to be sensitive to excited states, and at the same time make
the fits stable. We changed the maximum range of $t_s$ so as to
avoid the noisier data points, as well as those that are not
well-determined after all the intricate cancellations in the Fourier
transform from $\tilde{C}_{\rm 2pt}(p_0)$ resulting in orders of
magnitude smaller values for $C_{\rm 2pt}(t_s)$ as seen in \fgn{correl}
(for example, the $t_s>6a$ data points in \fgn{effmass} for $n_3=14,16$
that are suddenly pulled to smaller values than expected, and it
is clear that they are not well-determined numerically and might
need more precise data for $\tilde{C}$.) In this way, we found the
pion mass in our calculation to be $m_\pi a = 0.219(2)$ in lattice
units. We show the resulting effective mass curves from the two-state
fits at different momentum $n_3$ as the bands in \fgn{effmass}.
For comparison, the expected values for $E(P_3)$ from one-particle
dispersion relation based on the value of $m_\pi a =0.219$ are shown
as the dot-dashed horizontal lines in the figure. We see that the
resulting ground-state energies agree with the continuum dispersion
within errors. As a curious observation that is unrelated to the
results in the paper, we found the first excited state energy at
$n_3=0$ to be $a E_1(P_3=0)=0.79(4)$ from the two-state fits. Using
a string-tension value of $\sqrt{\sigma}=440$ MeV, we find this
value to be about 1.3 GeV which seems to agree quite nicely with
the pion radial excitation pole-mass in SU(3) QCD.  Thus, the
usage of string tension to set the large-$N_c$ GeV scale has its
advantage as noted in the main text.  However, we only found a poor
agreement of the momentum dependence of $E_1(P_3)$ with a single
particle dispersion curve and hence we cannot rule out the possibility
of the agreement with pion(1300) at $P_3=0$ to be a numerical
coincidence in our calculation, and the $E_1$ could simply be
effectively capturing the tower of excited states.

\bef
\centering
\includegraphics[scale=0.4]{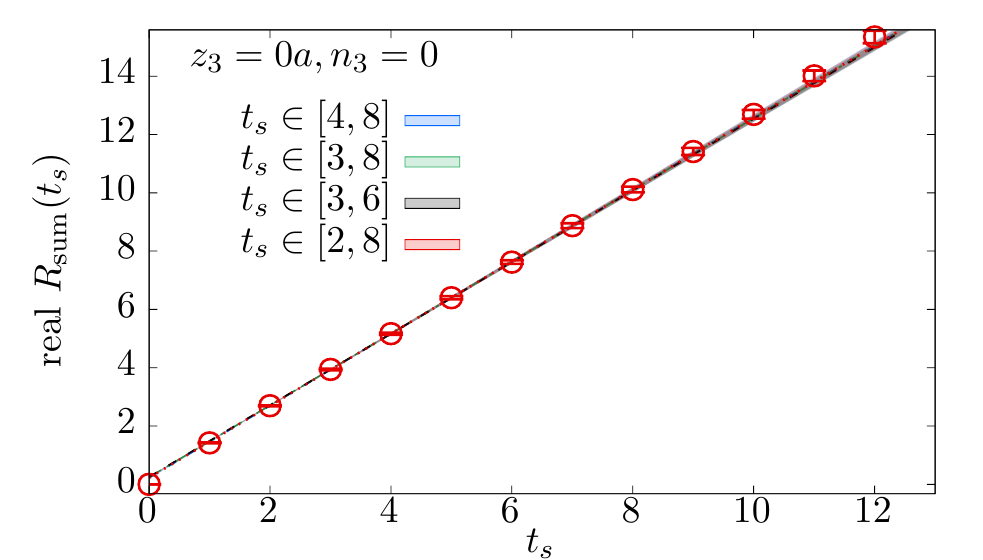}
\includegraphics[scale=0.4]{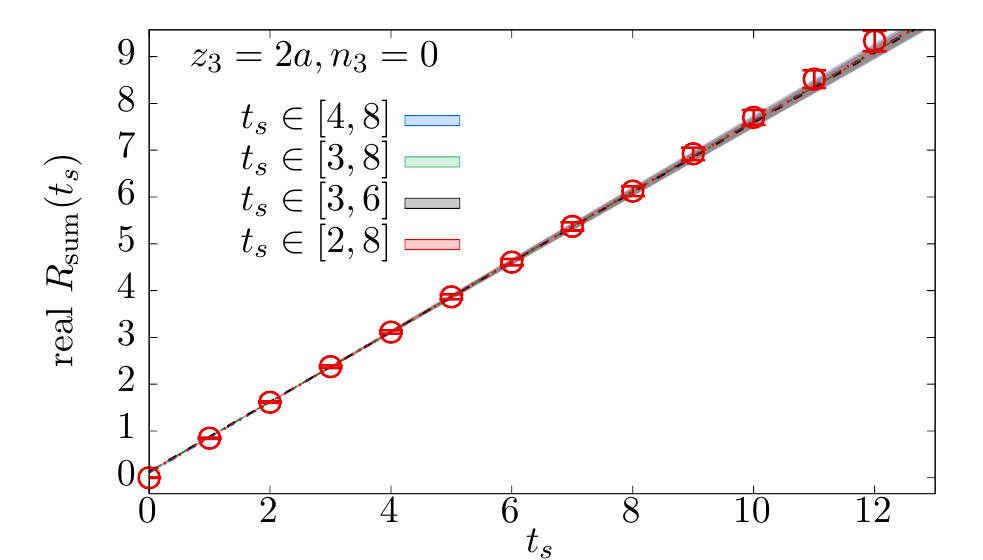}
\includegraphics[scale=0.4]{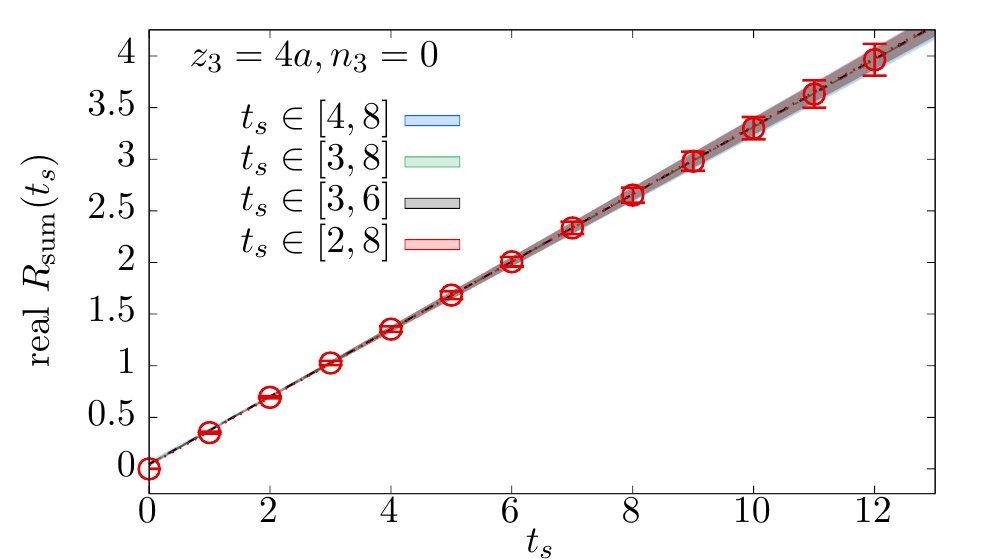}
\includegraphics[scale=0.4]{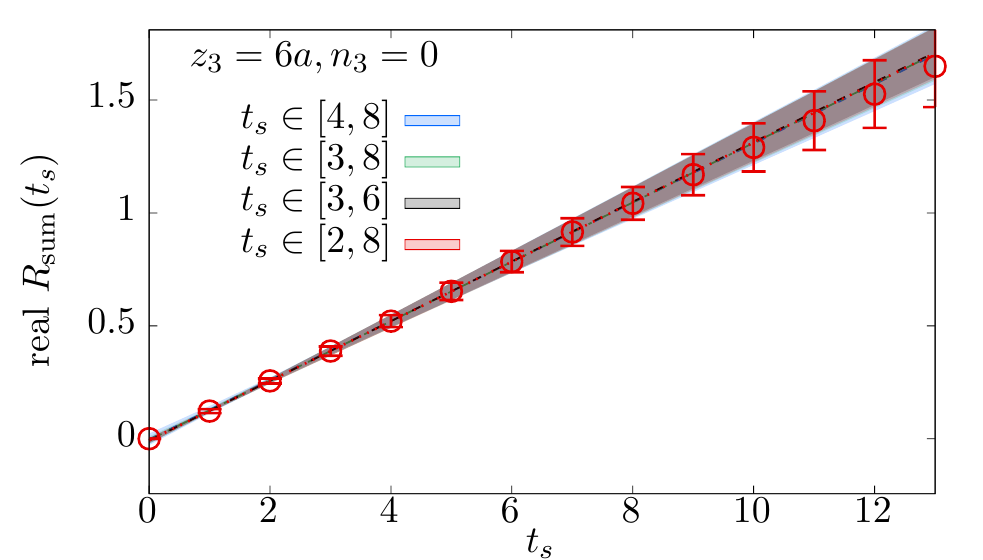}

\includegraphics[scale=0.4]{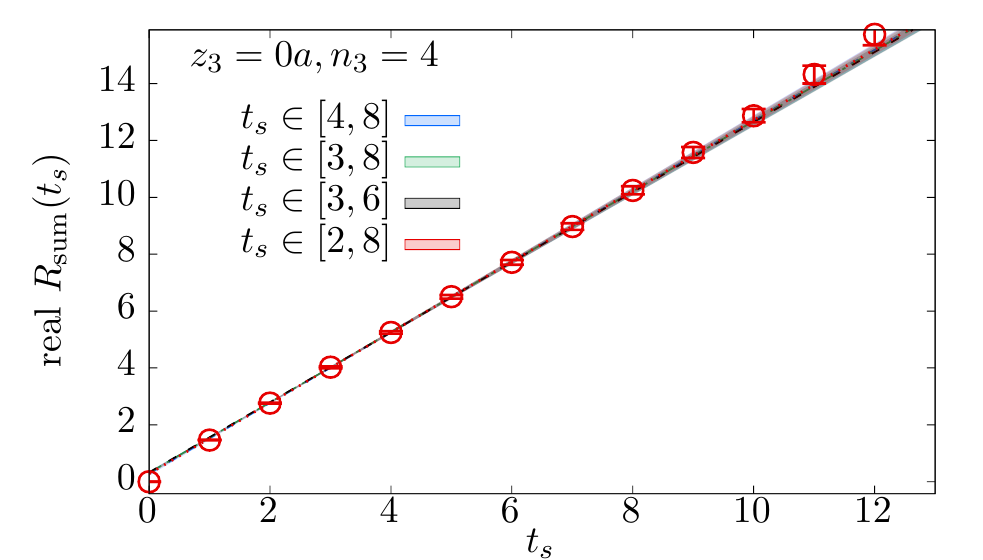}
\includegraphics[scale=0.4]{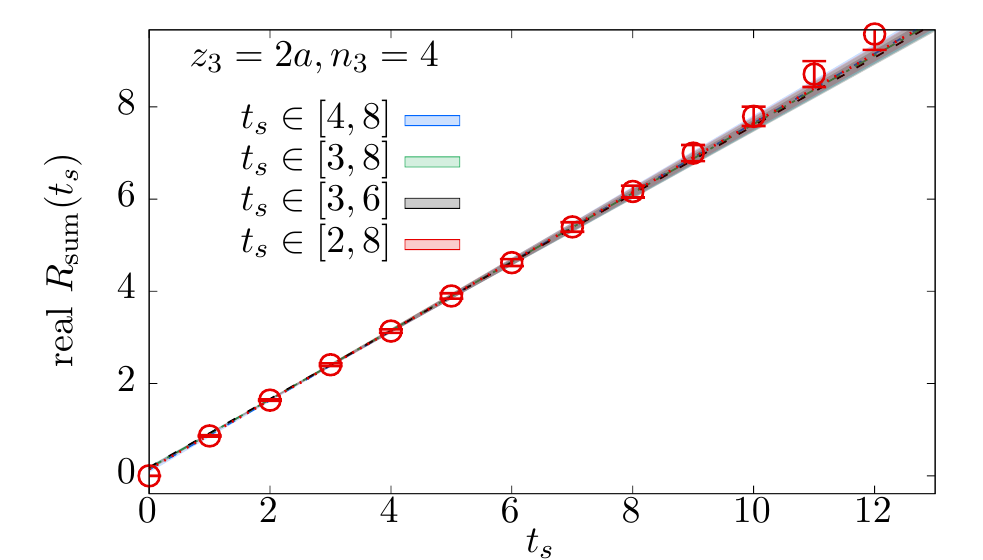}
\includegraphics[scale=0.4]{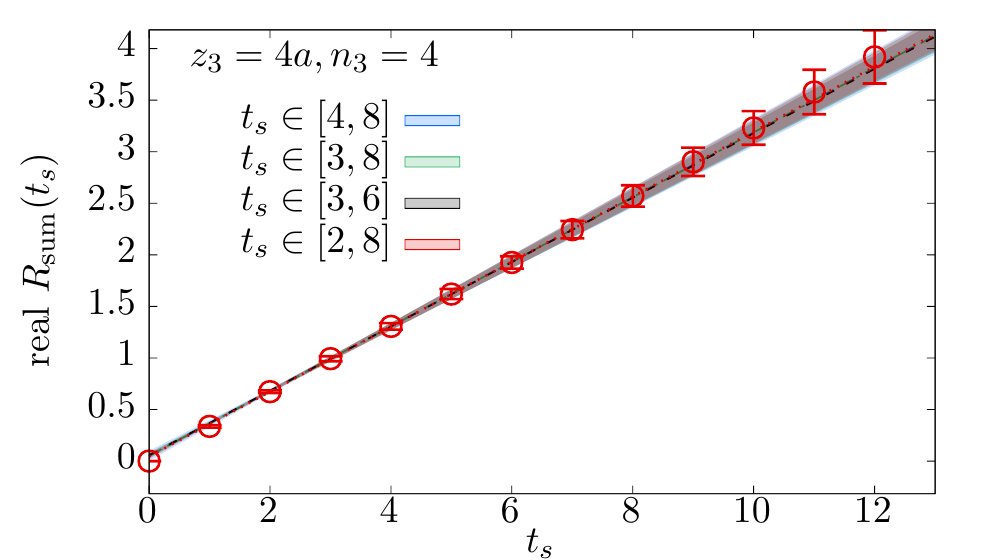}
\includegraphics[scale=0.4]{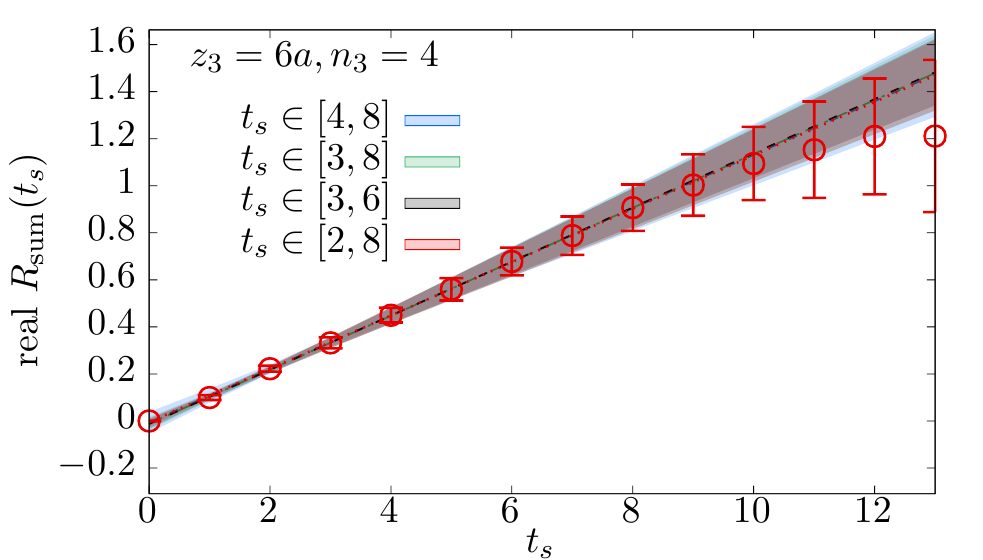}

\includegraphics[scale=0.4]{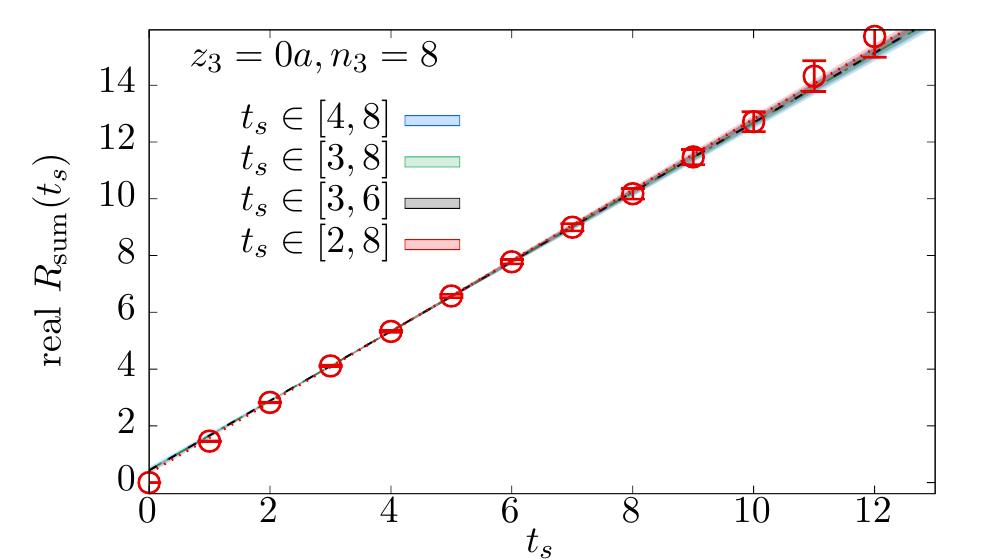}
\includegraphics[scale=0.4]{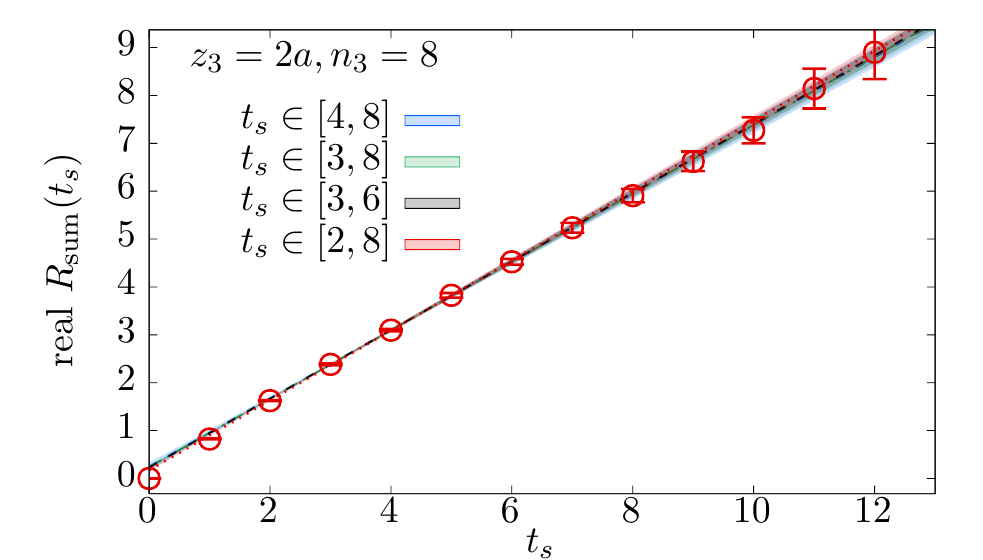}
\includegraphics[scale=0.4]{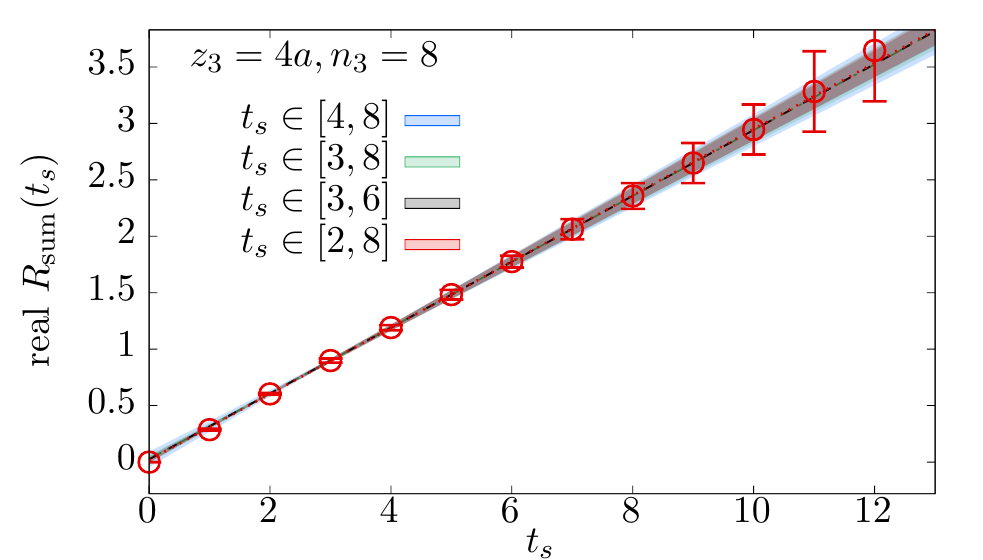}
\includegraphics[scale=0.4]{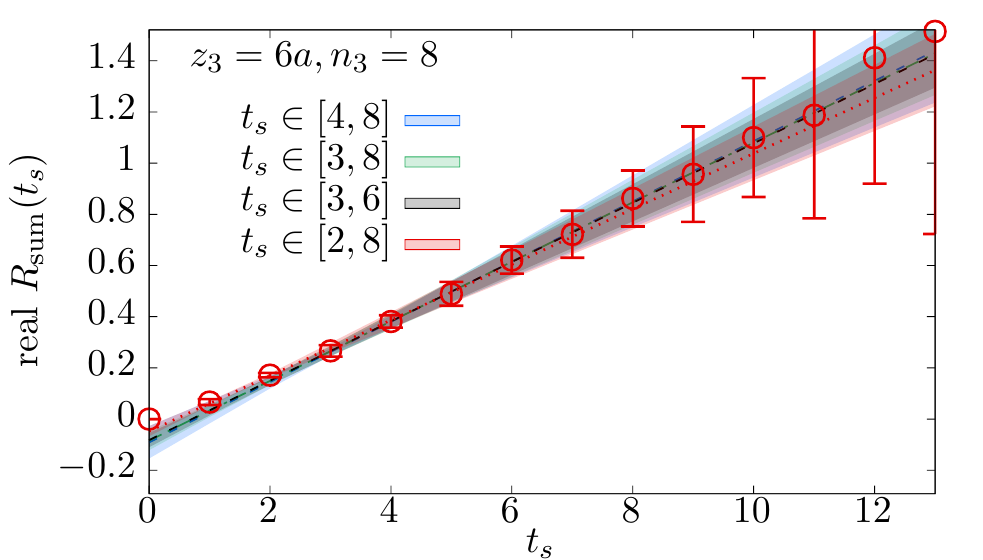}

\includegraphics[scale=0.4]{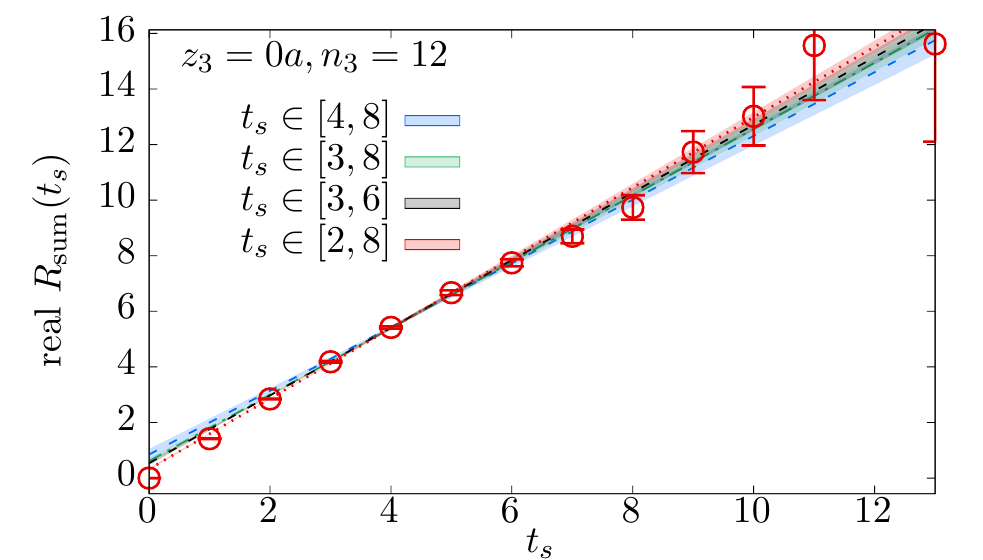}
\includegraphics[scale=0.4]{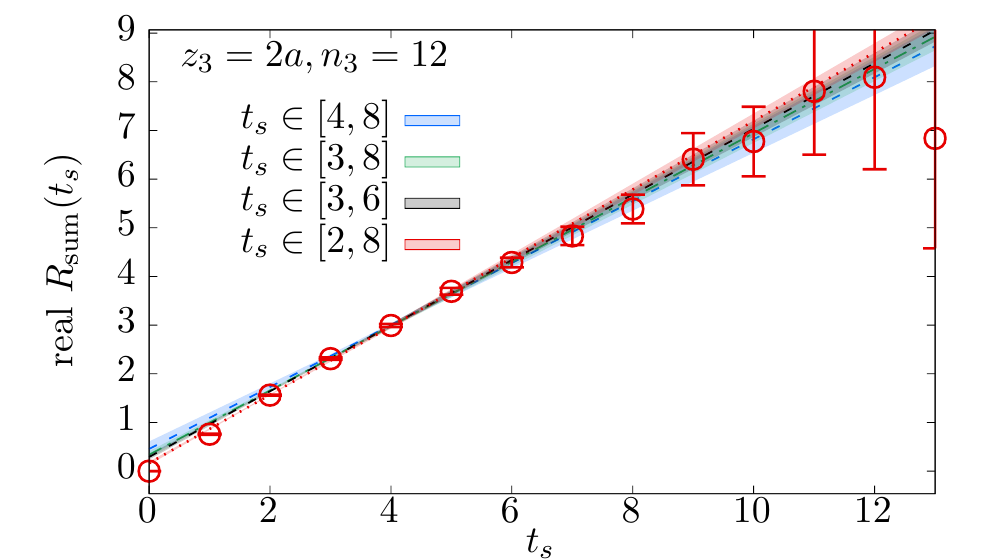}
\includegraphics[scale=0.4]{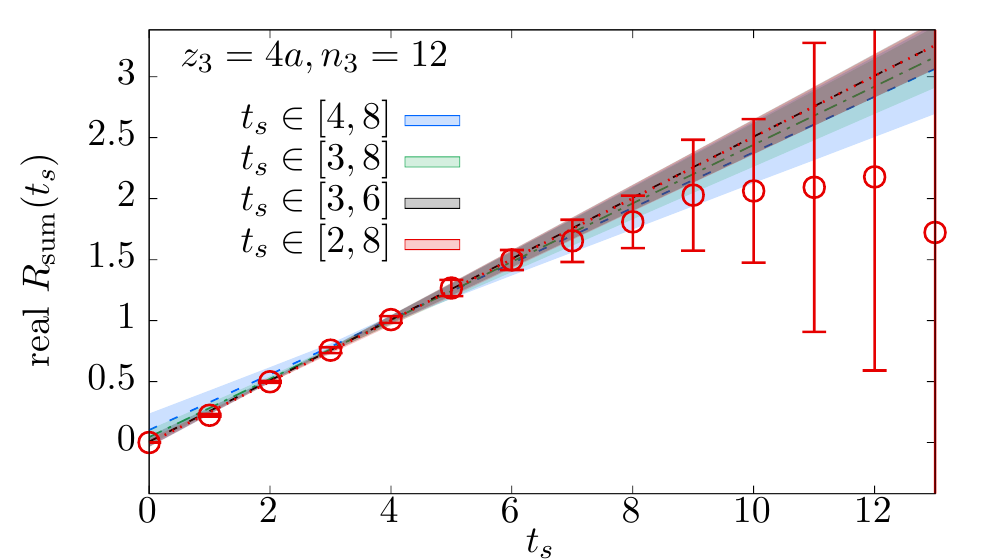}
\includegraphics[scale=0.4]{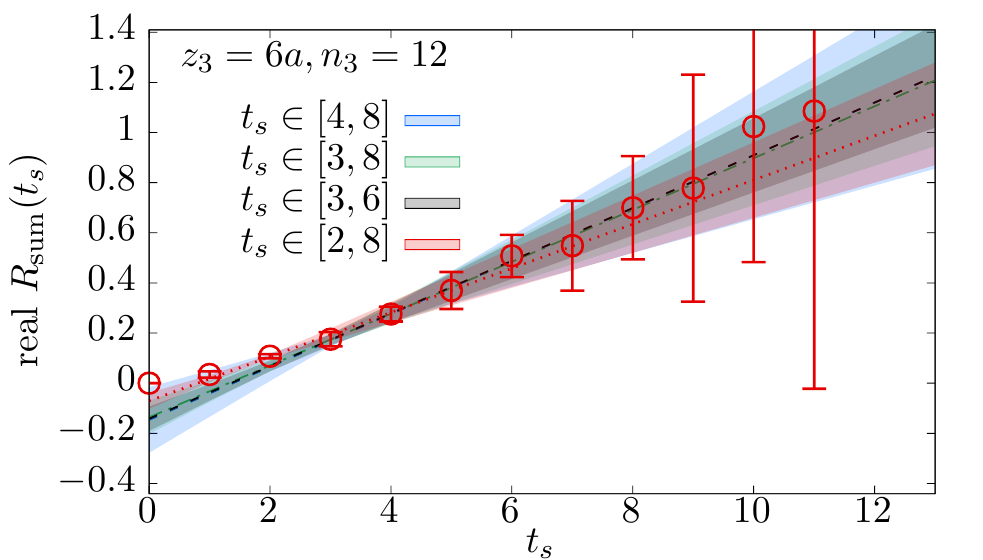}

\includegraphics[scale=0.4]{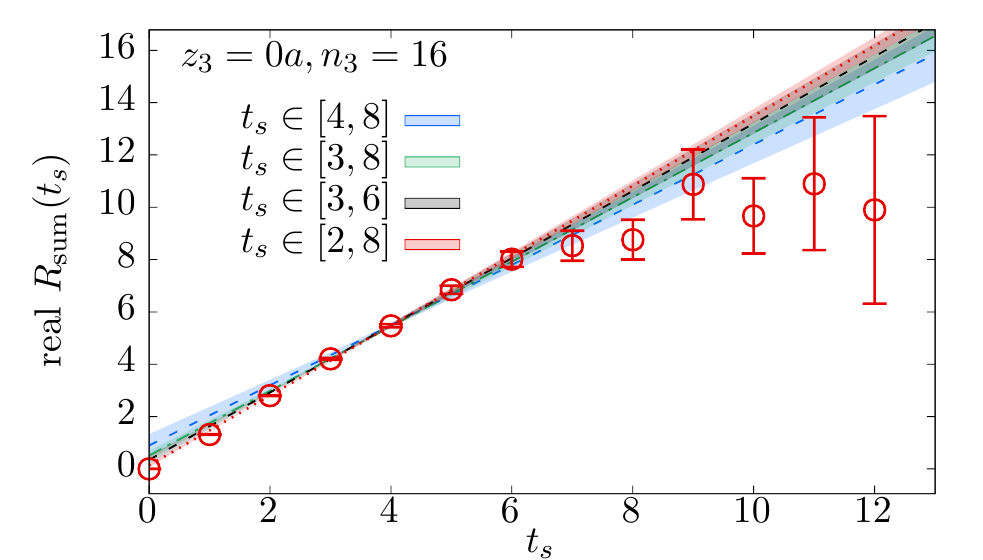}
\includegraphics[scale=0.4]{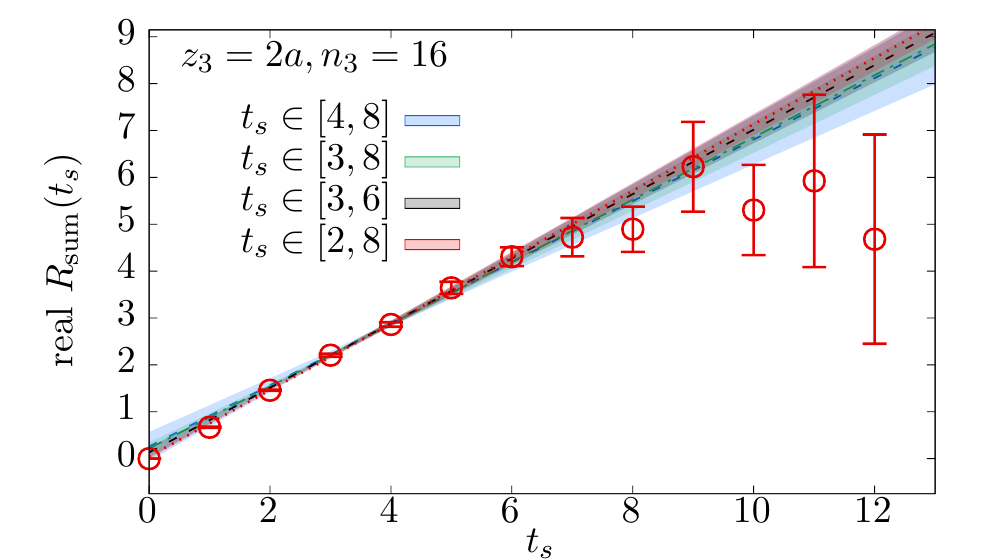}
\includegraphics[scale=0.4]{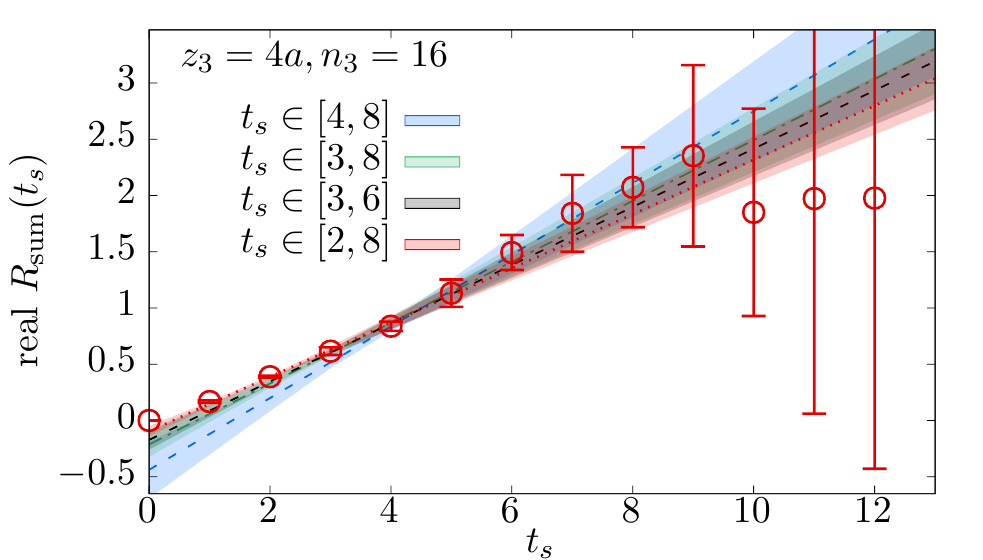}
\includegraphics[scale=0.4]{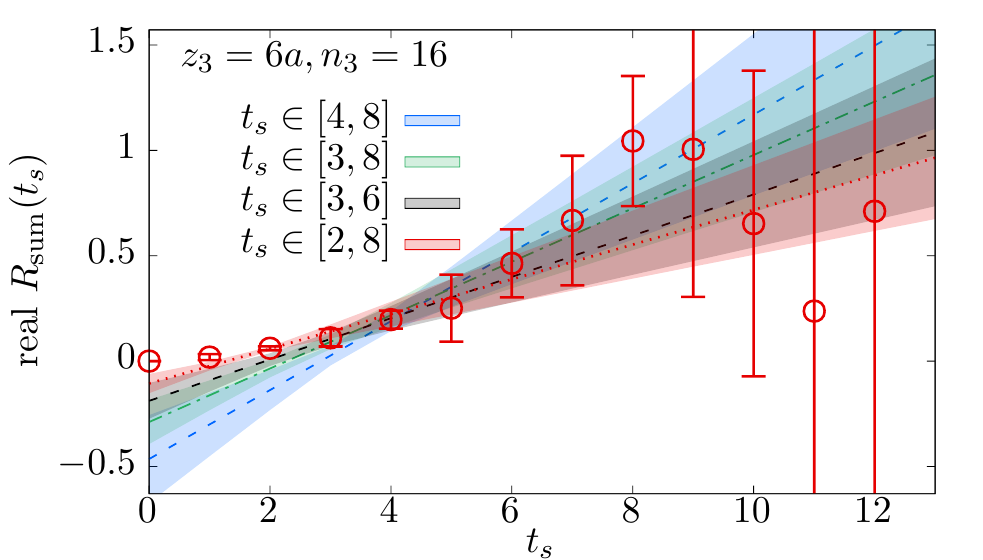}

\caption{Extraction of the real-part of the ground state 
bare quasi-PDF matrix element, ${\rm Re} h^B(z_3,n_3)$, via summation 
method. The bands are straight line fits, $h^B t_s + C$ to the data over 
different ranges of $t_s$ that is specified in the legend. The different panels show the data and the fits
from different $n_3$ (rows) and different $z_3/a$ (columns).
}
\eef{resumfit}

\bef
\centering
\includegraphics[scale=0.4]{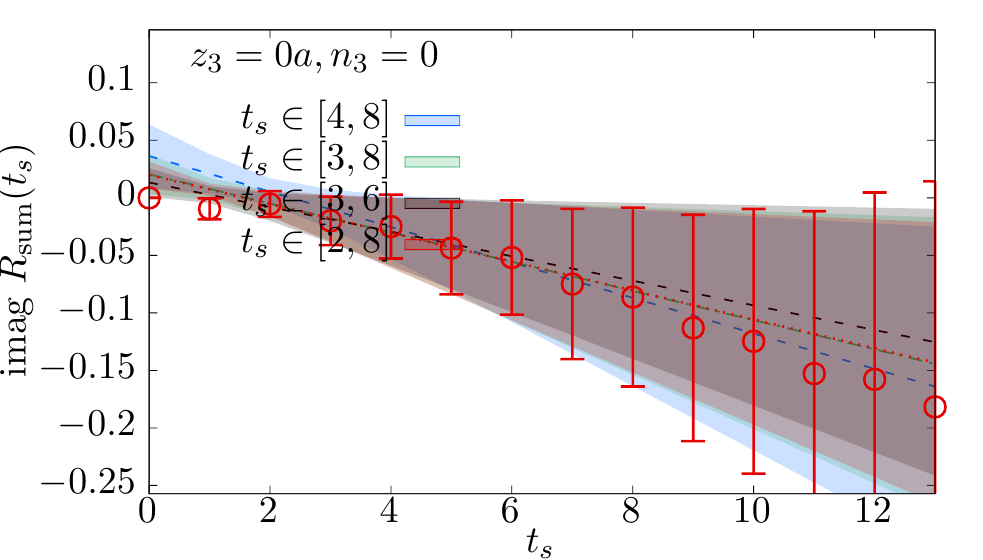}
\includegraphics[scale=0.4]{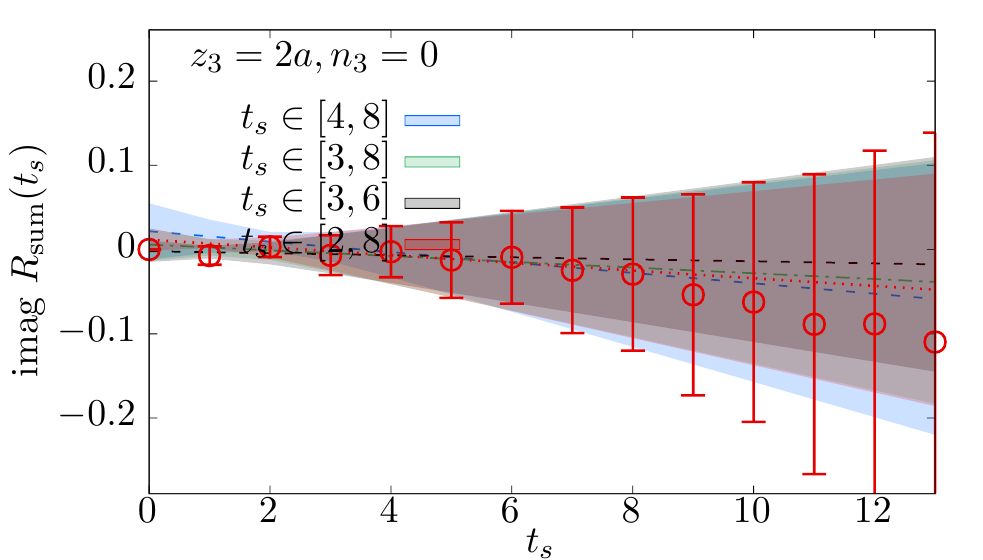}
\includegraphics[scale=0.4]{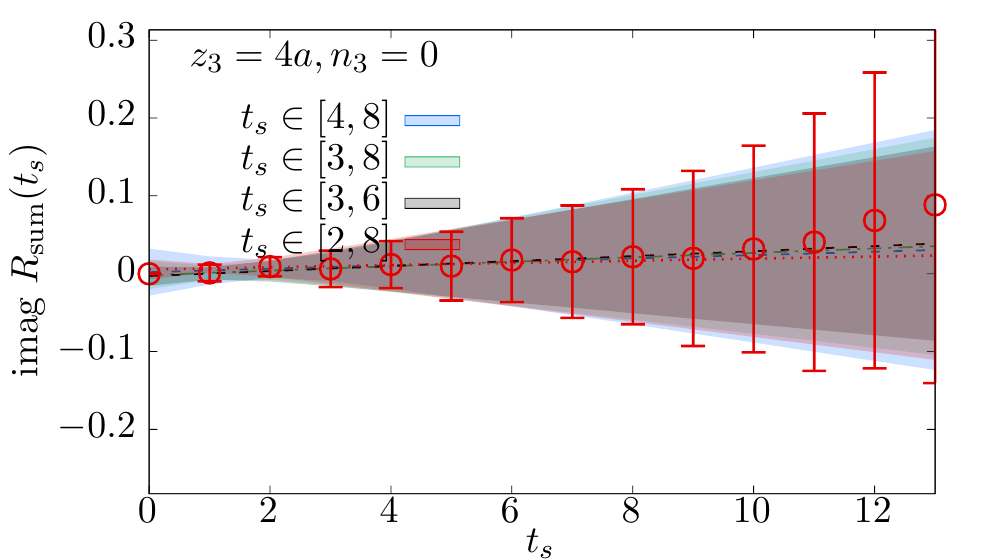}
\includegraphics[scale=0.4]{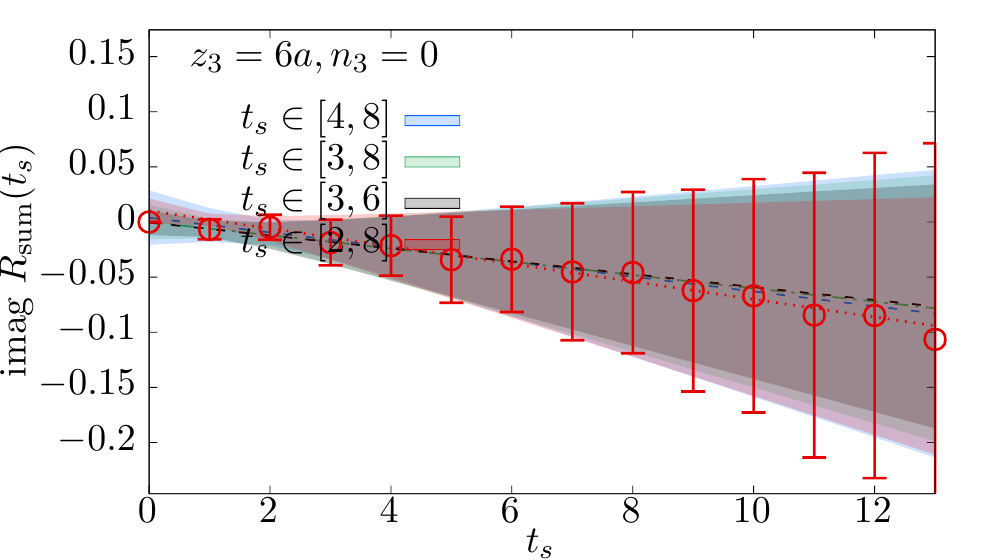}

\includegraphics[scale=0.4]{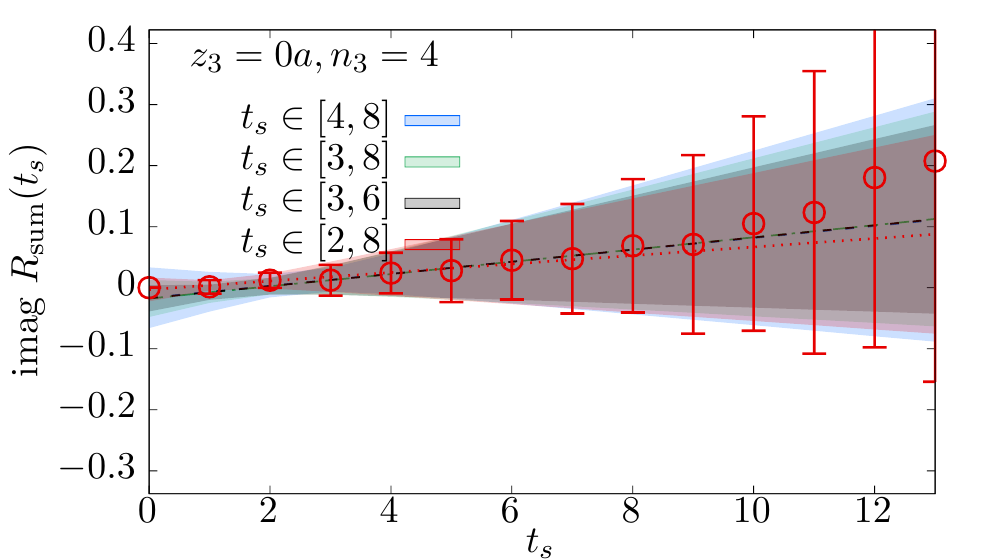}
\includegraphics[scale=0.4]{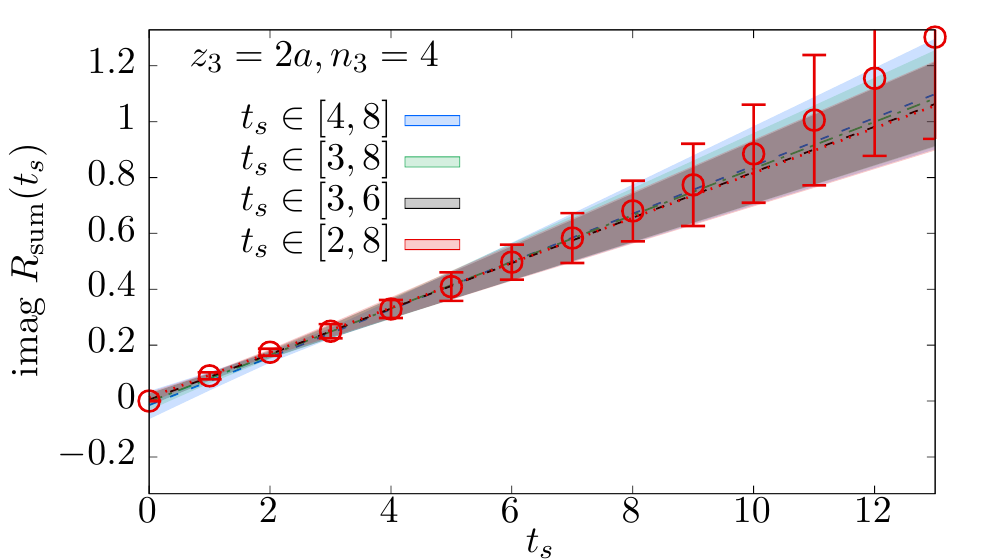}
\includegraphics[scale=0.4]{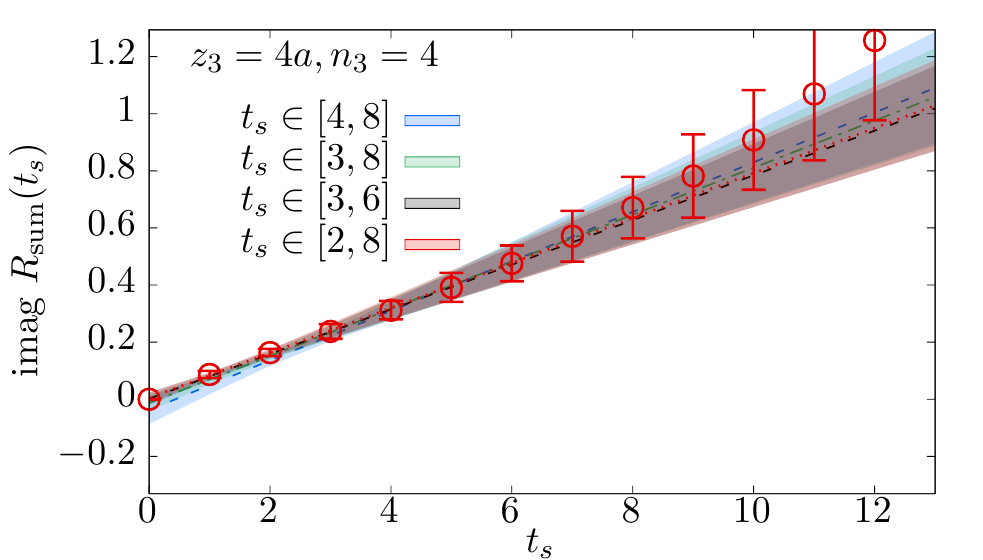}
\includegraphics[scale=0.4]{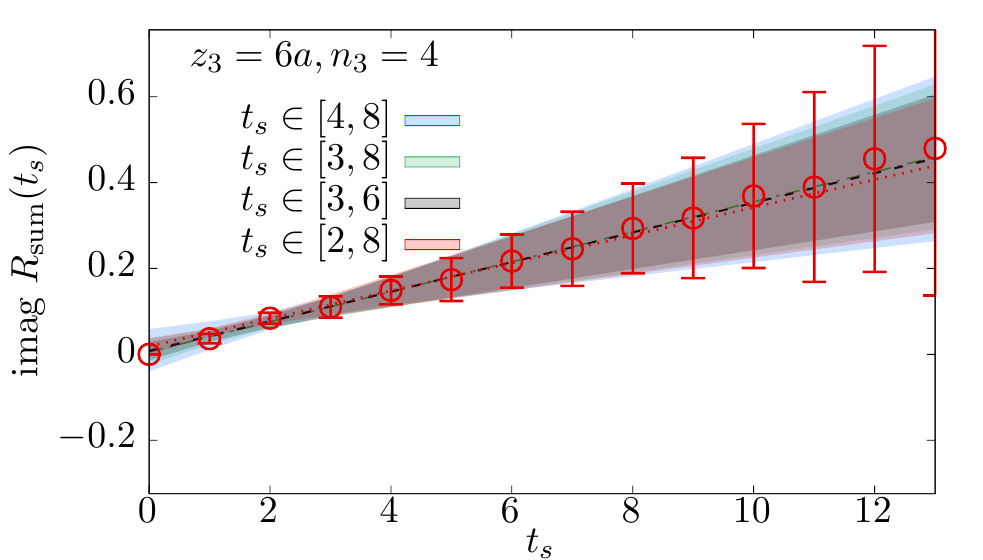}

\includegraphics[scale=0.4]{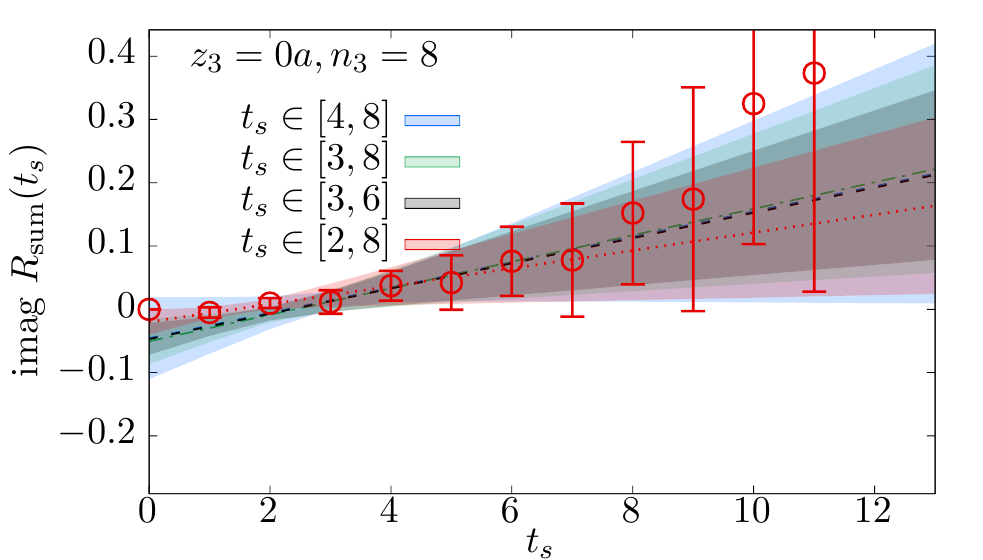}
\includegraphics[scale=0.4]{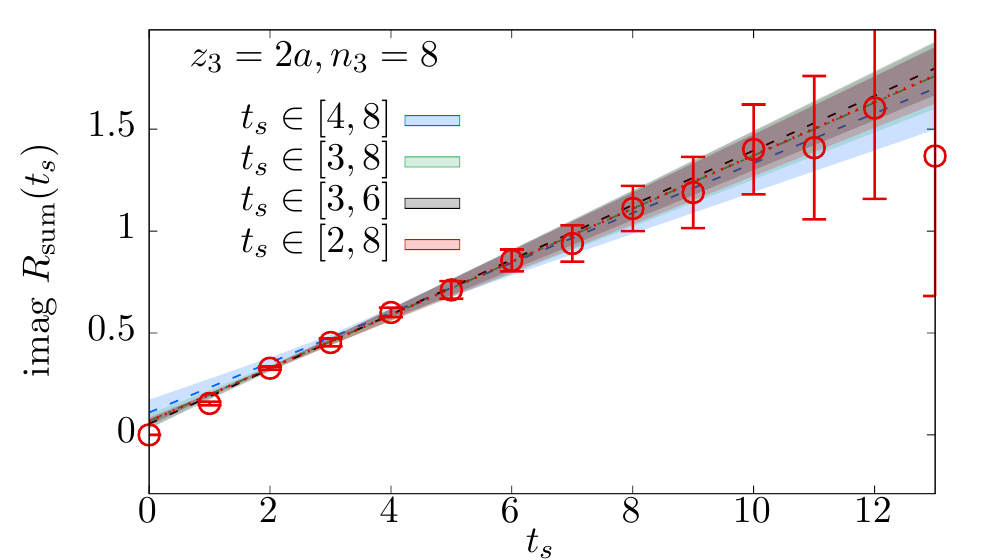}
\includegraphics[scale=0.4]{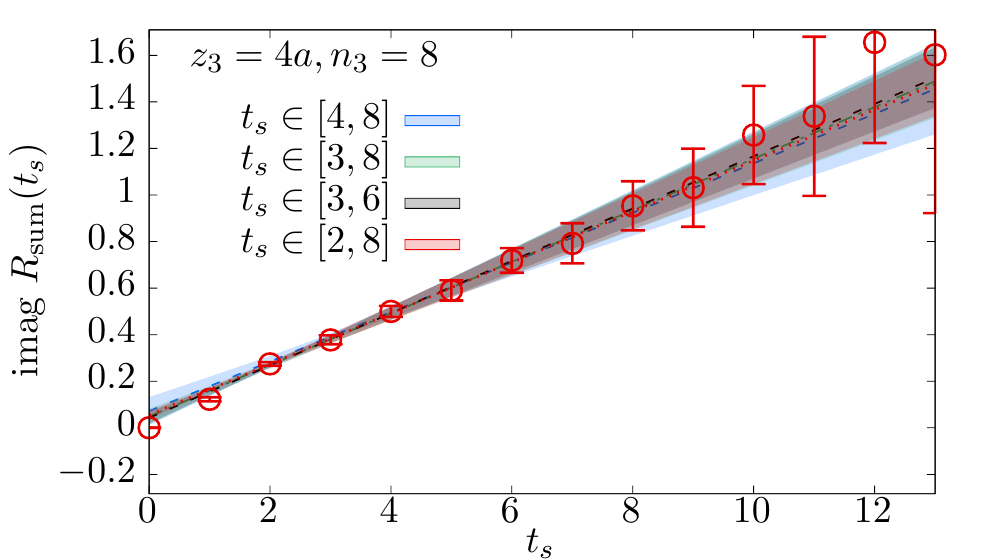}
\includegraphics[scale=0.4]{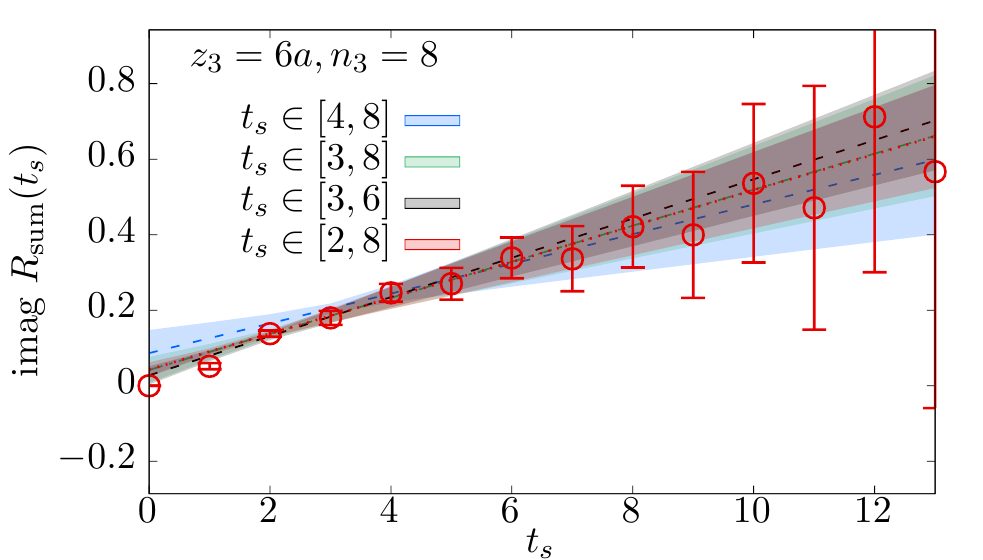}

\includegraphics[scale=0.4]{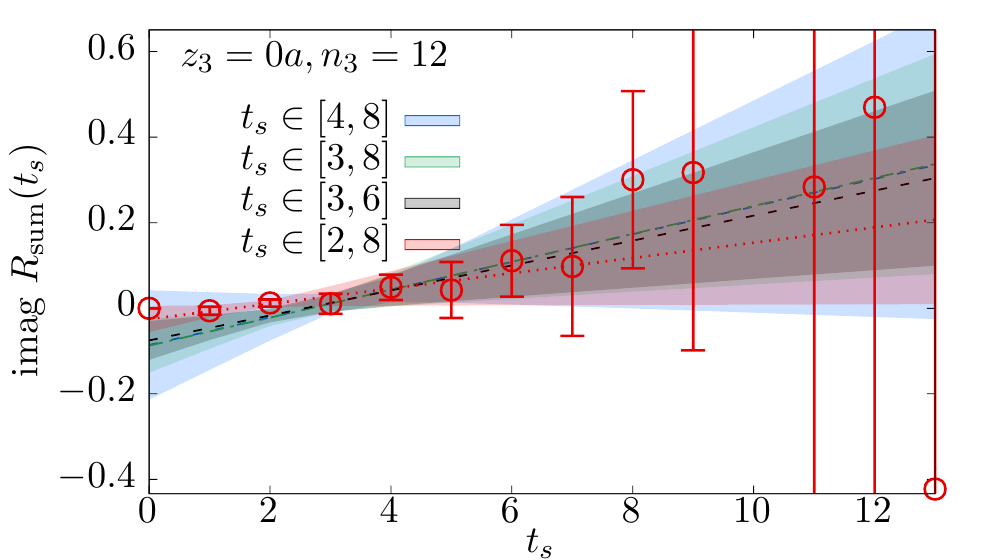}
\includegraphics[scale=0.4]{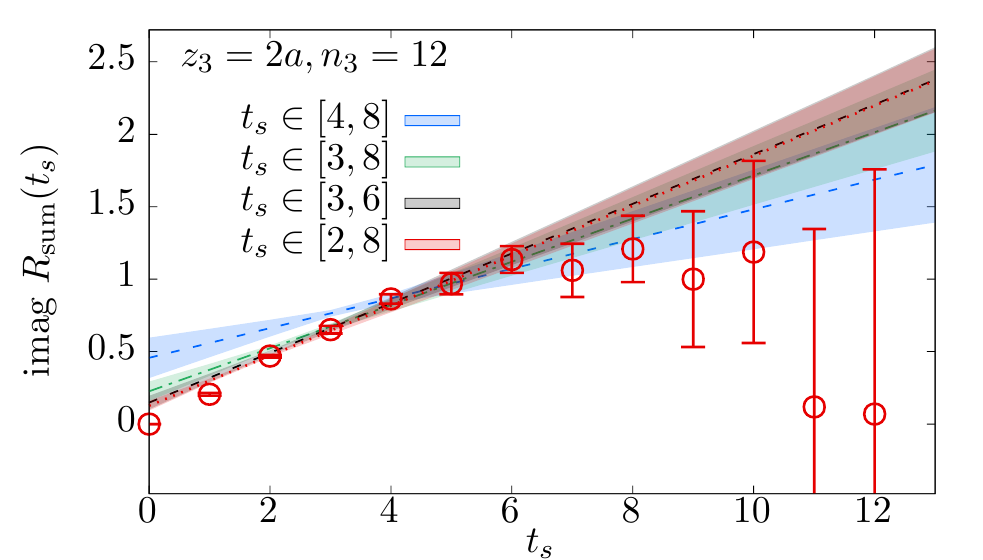}
\includegraphics[scale=0.4]{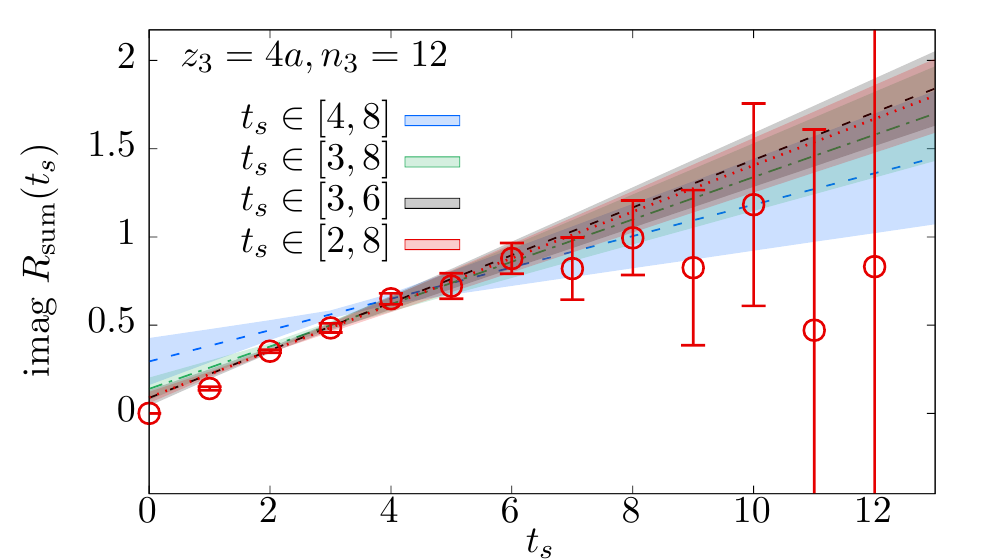}
\includegraphics[scale=0.4]{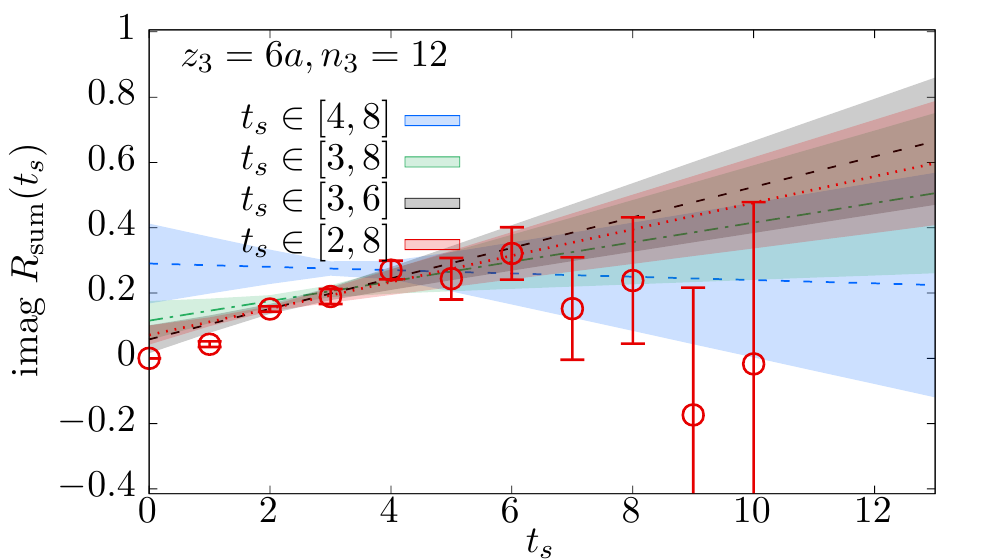}

\includegraphics[scale=0.4]{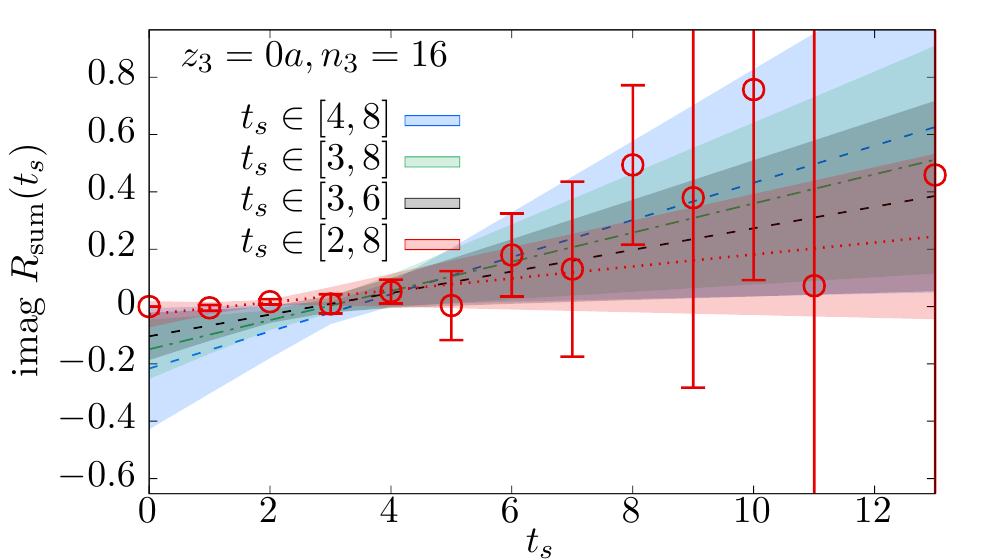}
\includegraphics[scale=0.4]{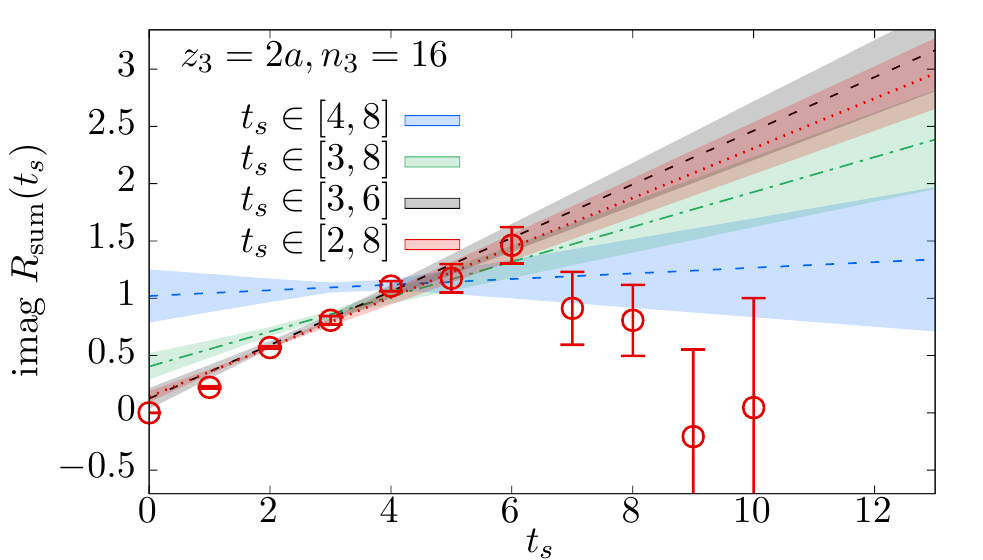}
\includegraphics[scale=0.4]{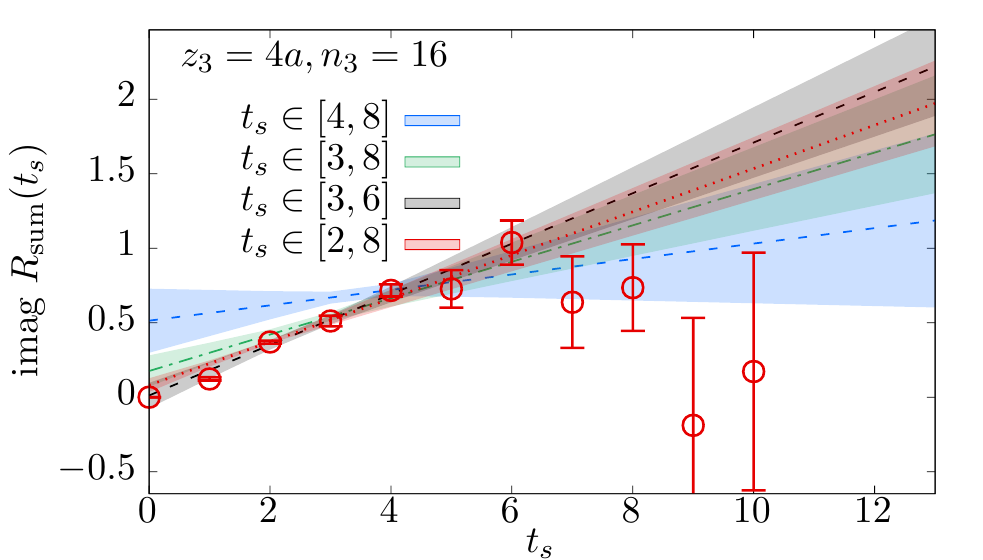}
\includegraphics[scale=0.4]{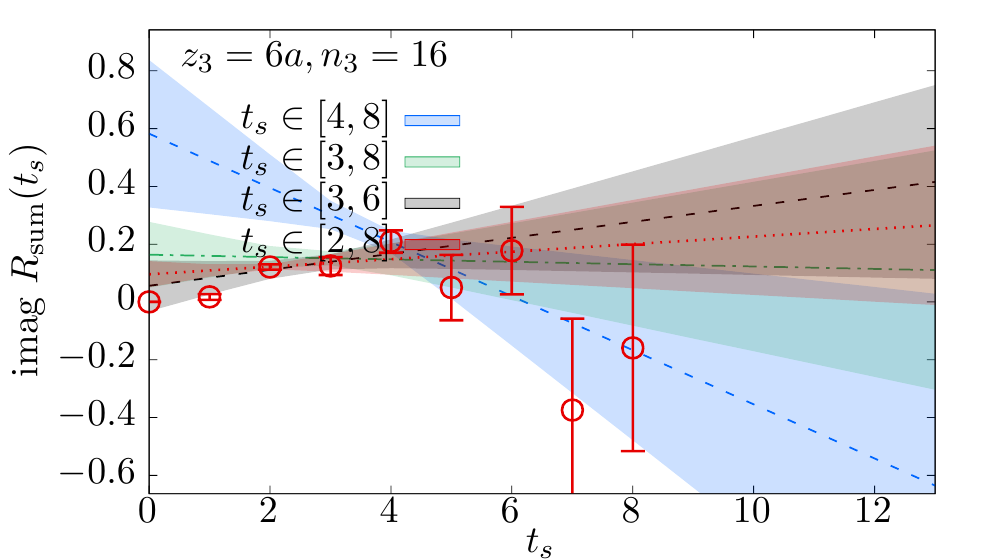}

\caption{
Extraction of the imaginary-part of the ground state 
bare quasi-PDF matrix element, ${\rm Im} h^B(z_3,n_3)$, via summation 
method. The bands are straight line fits, $h^B t_s + B$ to the data over 
different ranges of $t_s$.
The different panels show the data and the fits
from different $n_3$ (rows) and different $z_3/a$ (columns).}
\eef{imsumfit}

\bef
\centering
\includegraphics[scale=0.5]{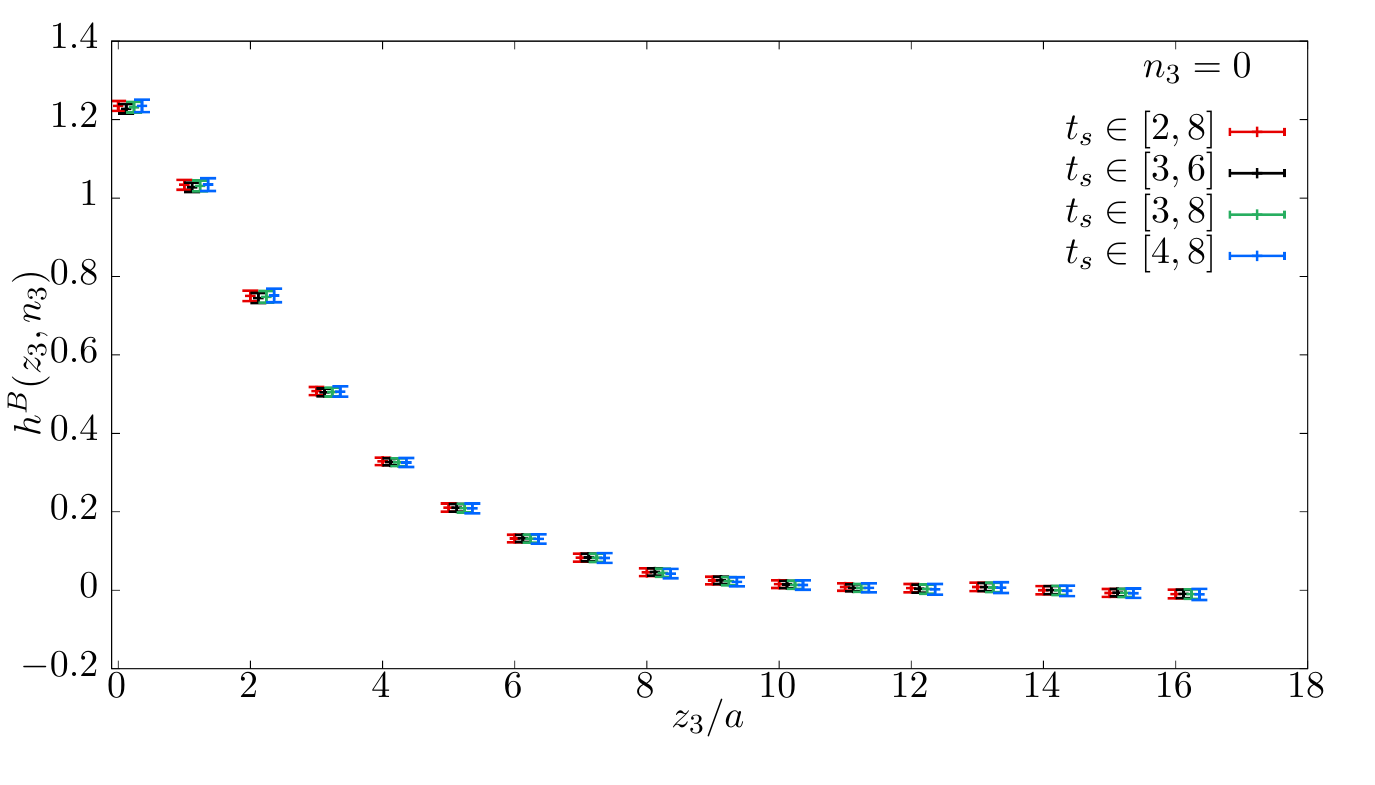}
\includegraphics[scale=0.5]{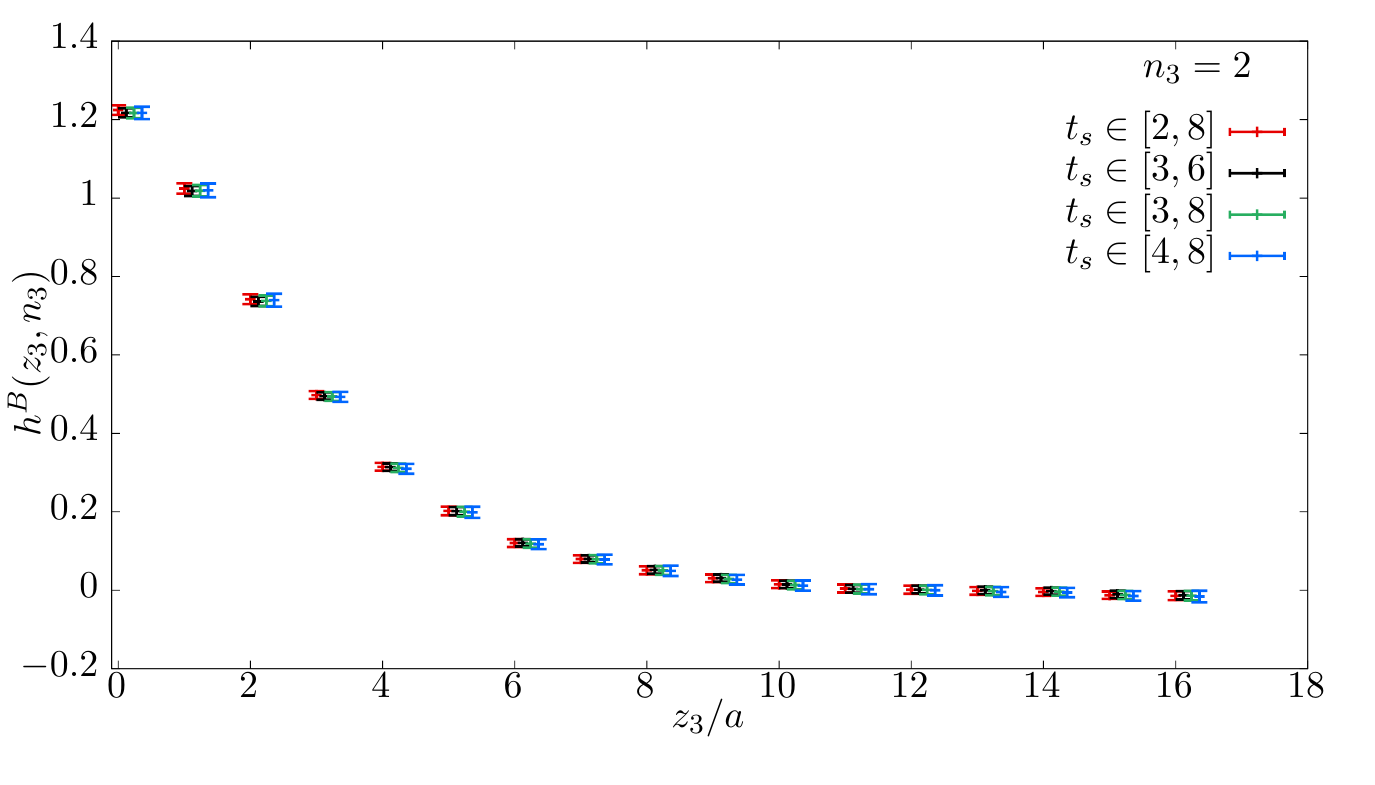}

\includegraphics[scale=0.5]{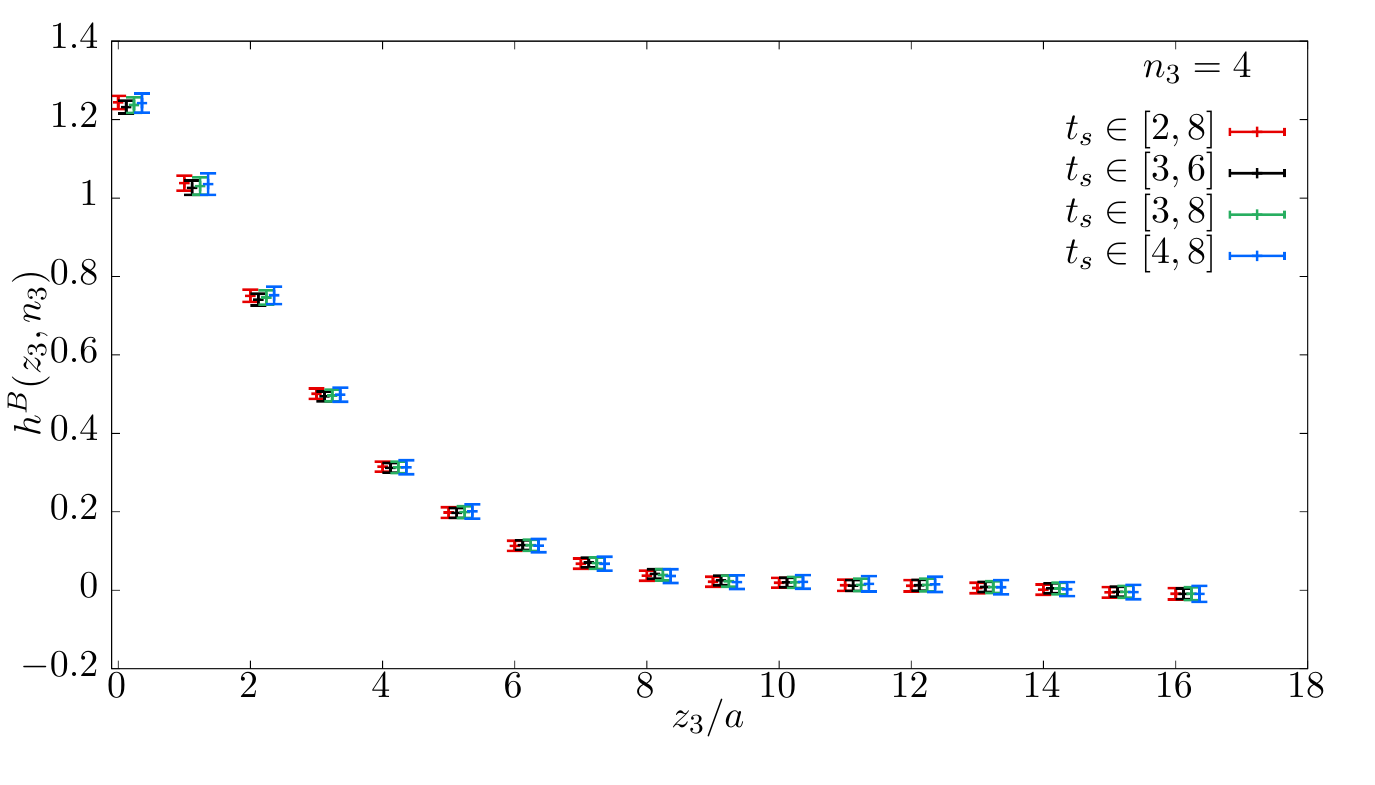}
\includegraphics[scale=0.5]{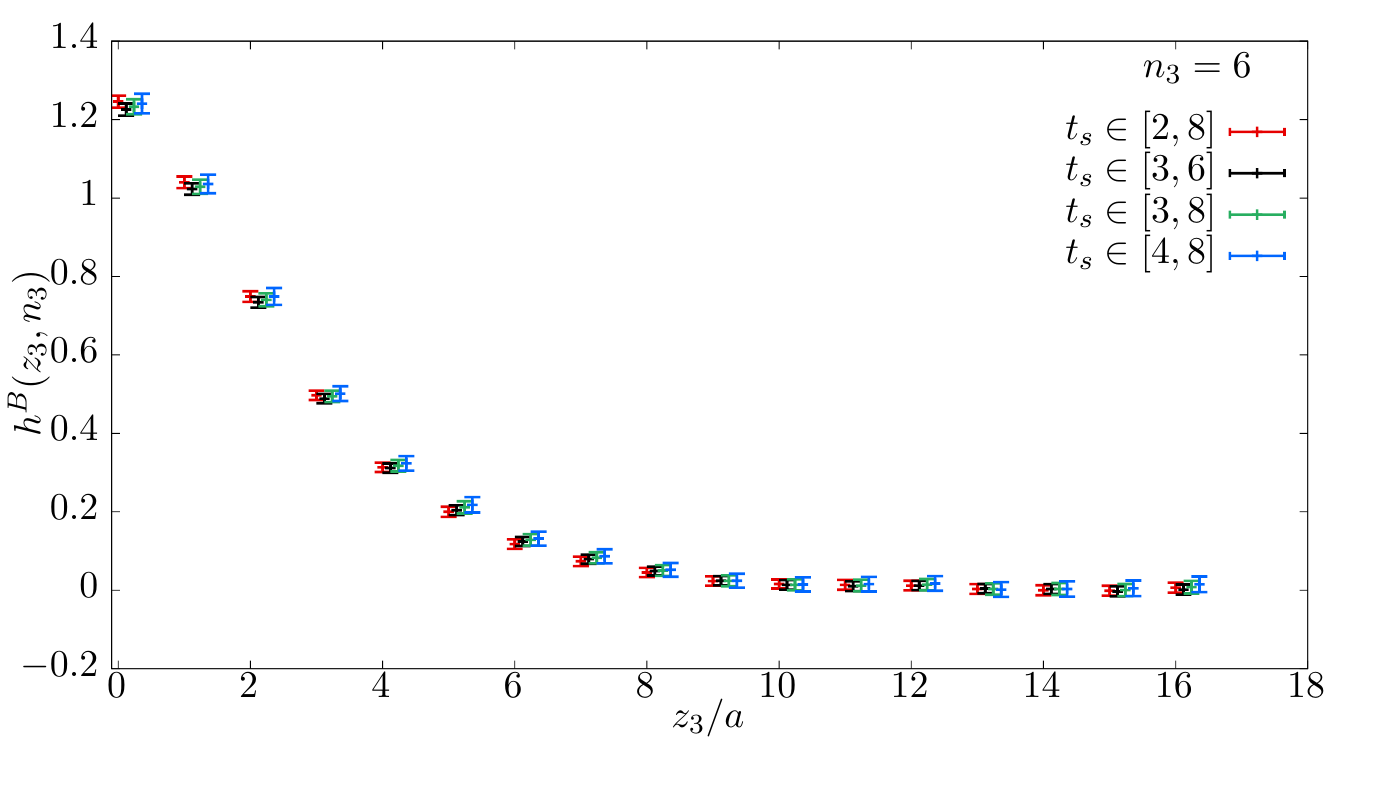}

\includegraphics[scale=0.5]{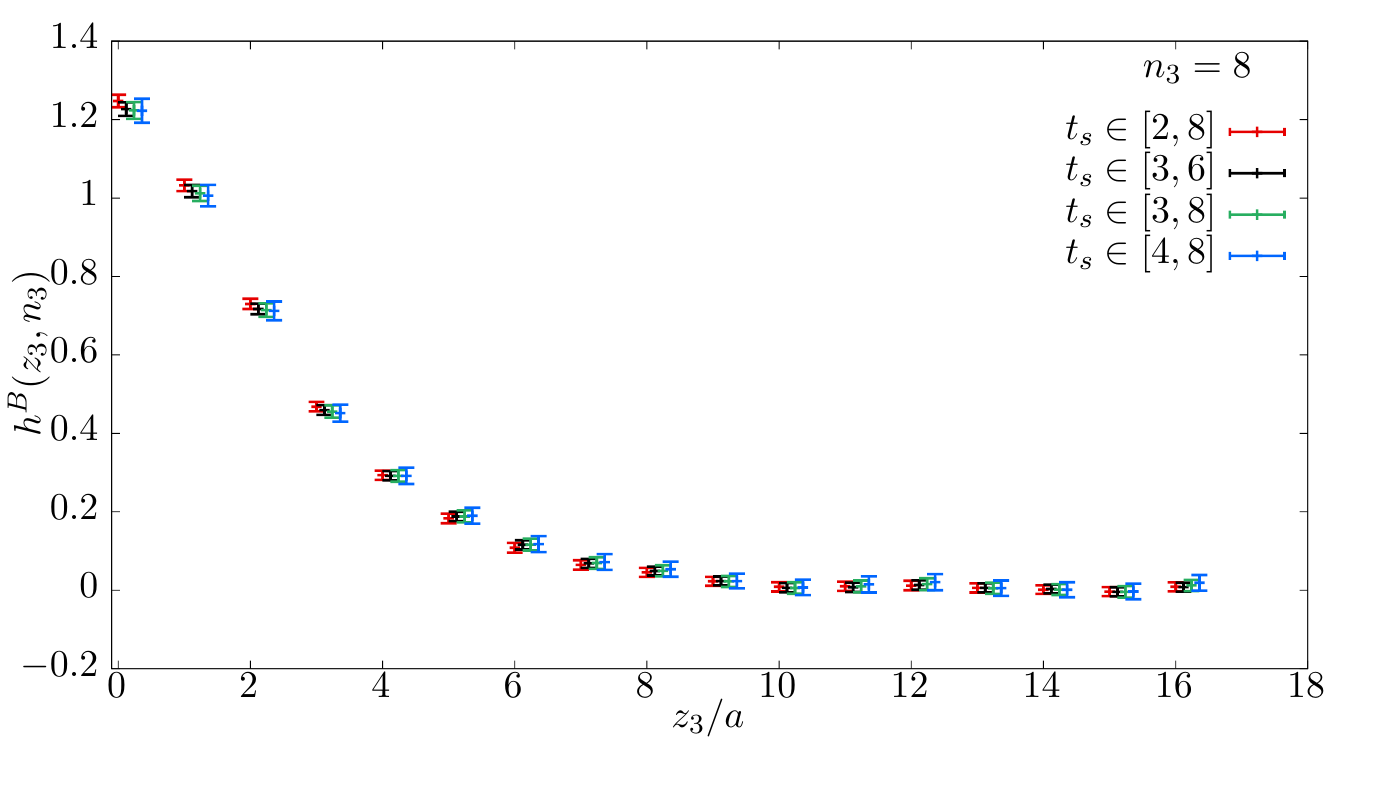}
\includegraphics[scale=0.5]{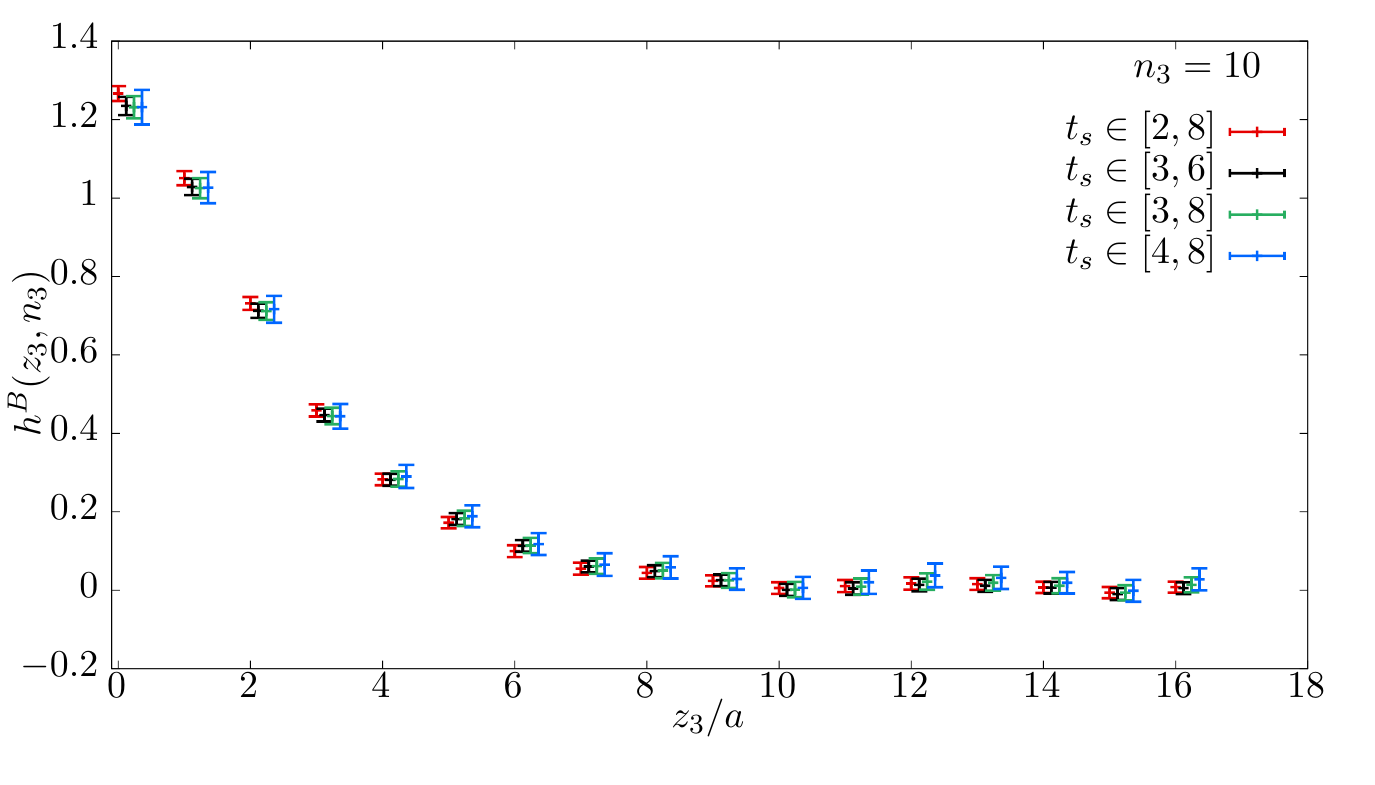}

\includegraphics[scale=0.5]{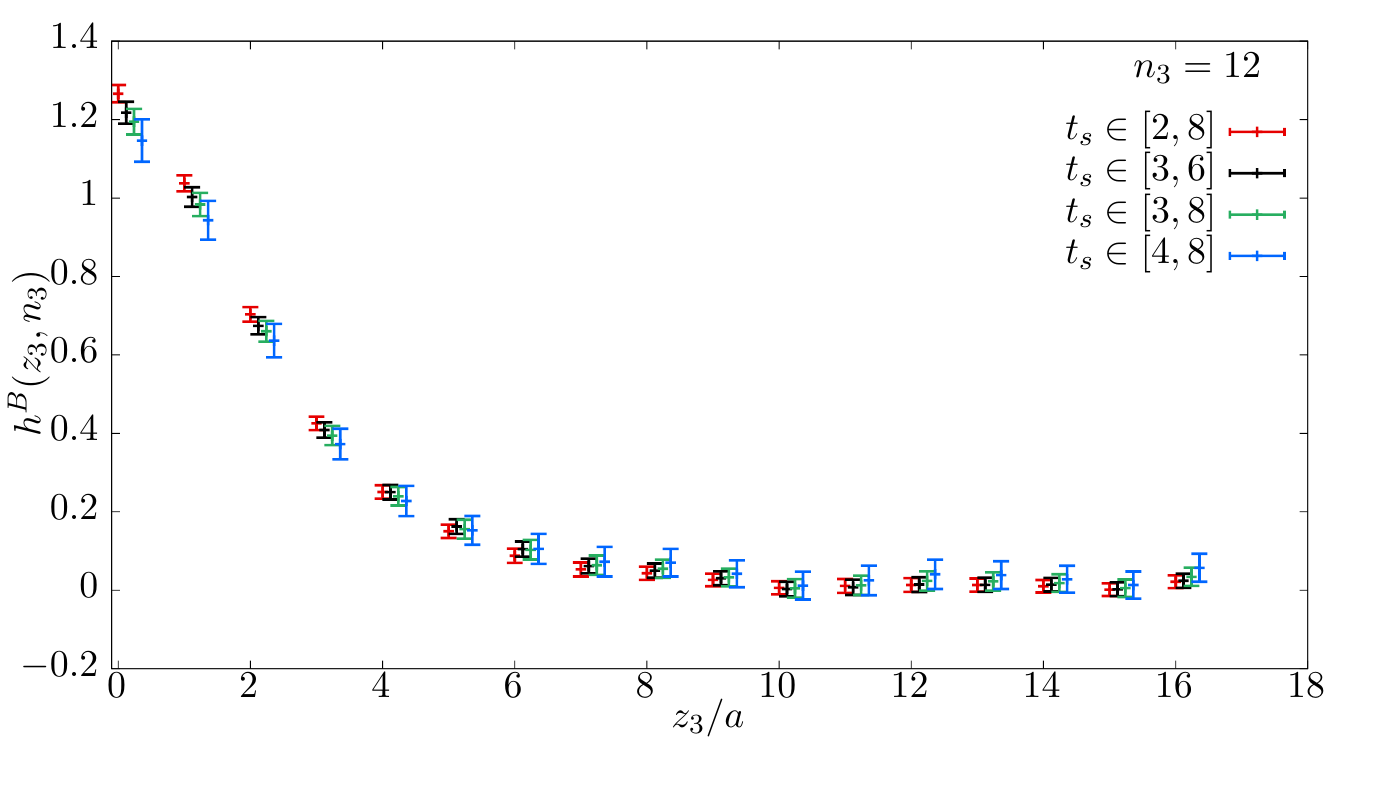}
\includegraphics[scale=0.5]{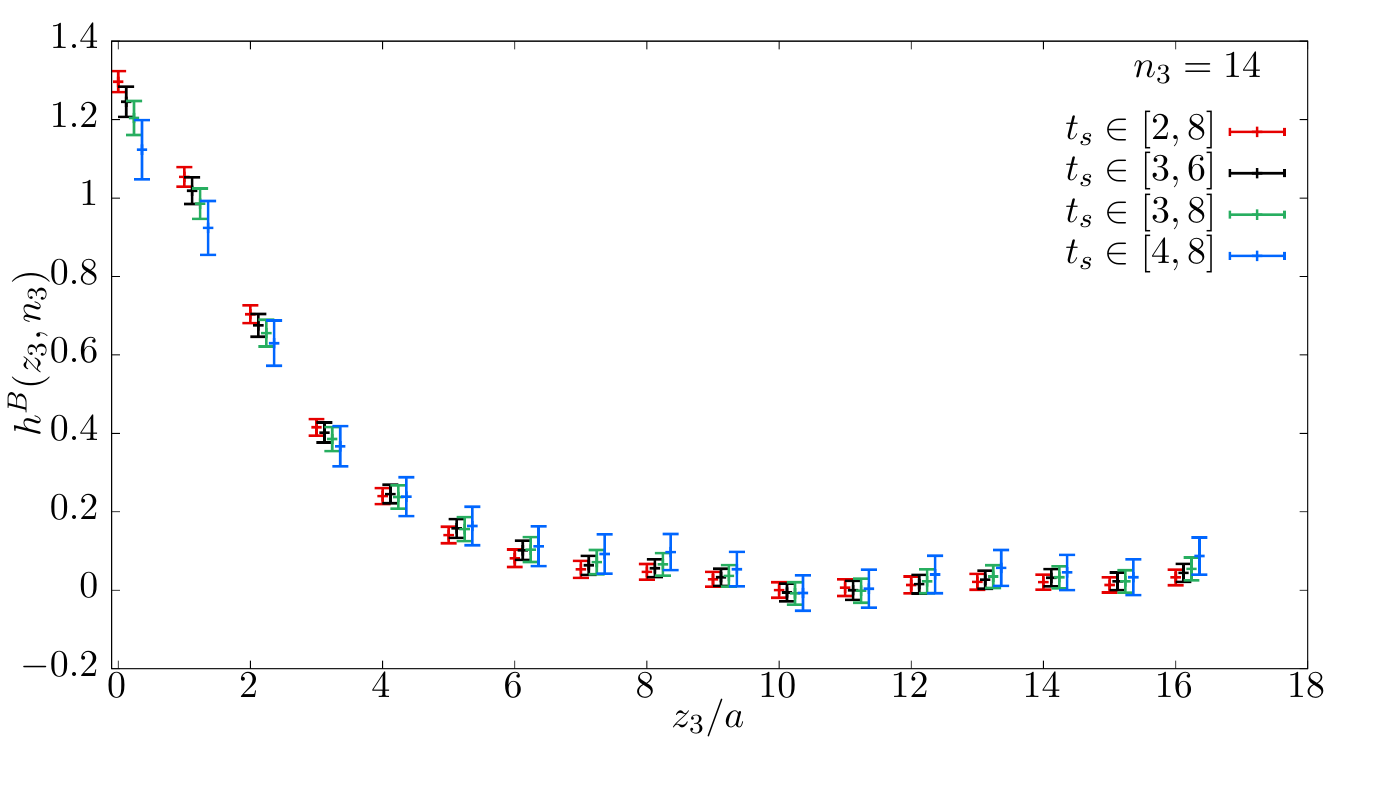}

\includegraphics[scale=0.5]{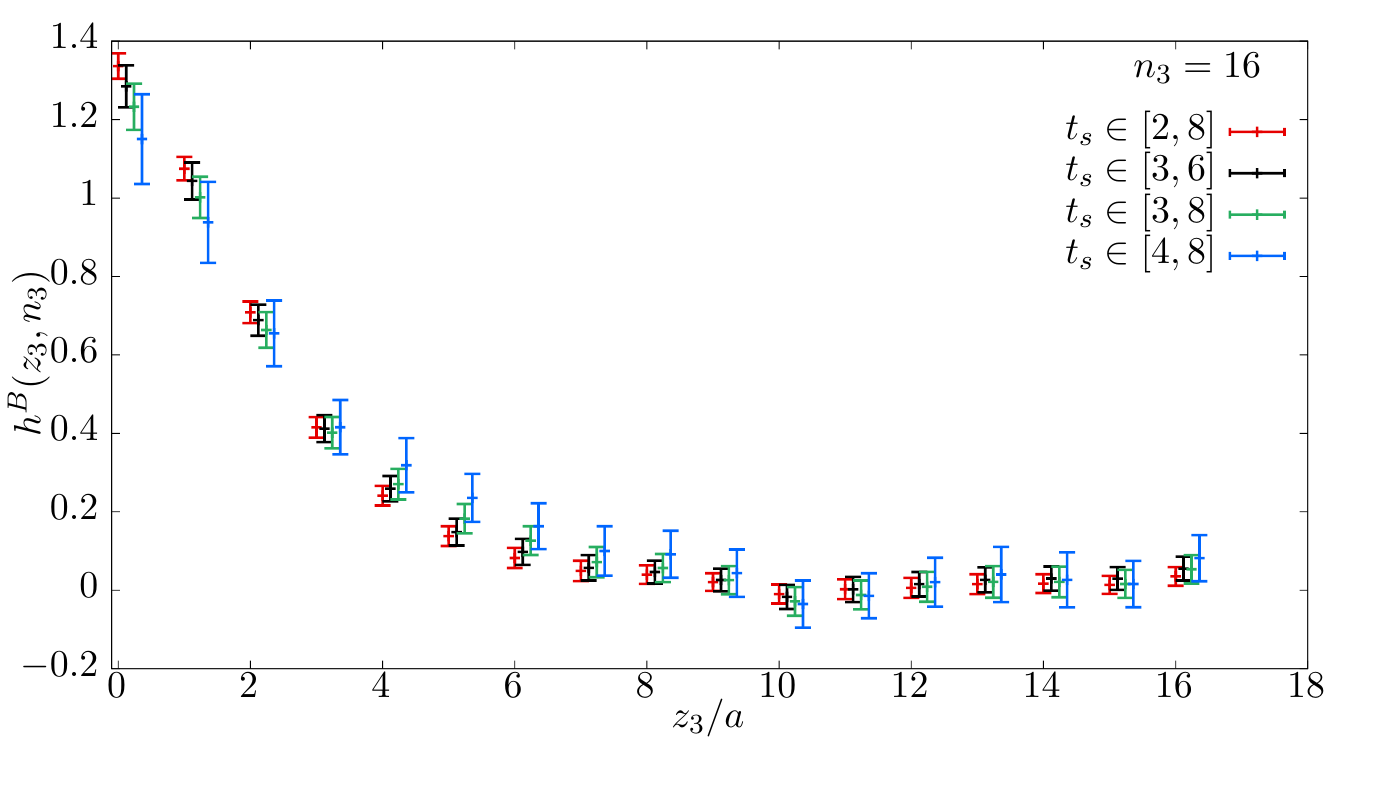}
\caption{
    The real part of the bare quasi-PDF matrix element $h^B(z_3,n_3)$
is shown as a function of $z_3/a$ at different spatial momentum $\propto n_3$ used in this work
as separate panels. The extrapolated results from summation methods 
over different fit ranges in $t_s$ are shown together in the plots.
We used extrapolations from $t_s \in [3a,8a]$ for $n_3\in[0,10]$,
and $t_s\in[3a,6a]$ for $n_3\in[12,16]$ to avoid badly determined points beyond
$t_s \ge 6a$. Using ranges with even
larger minimum $t_s$ was not feasible and forms a limitation of this
work.
}
\eef{rehb}

\bef
\centering
\includegraphics[scale=0.5]{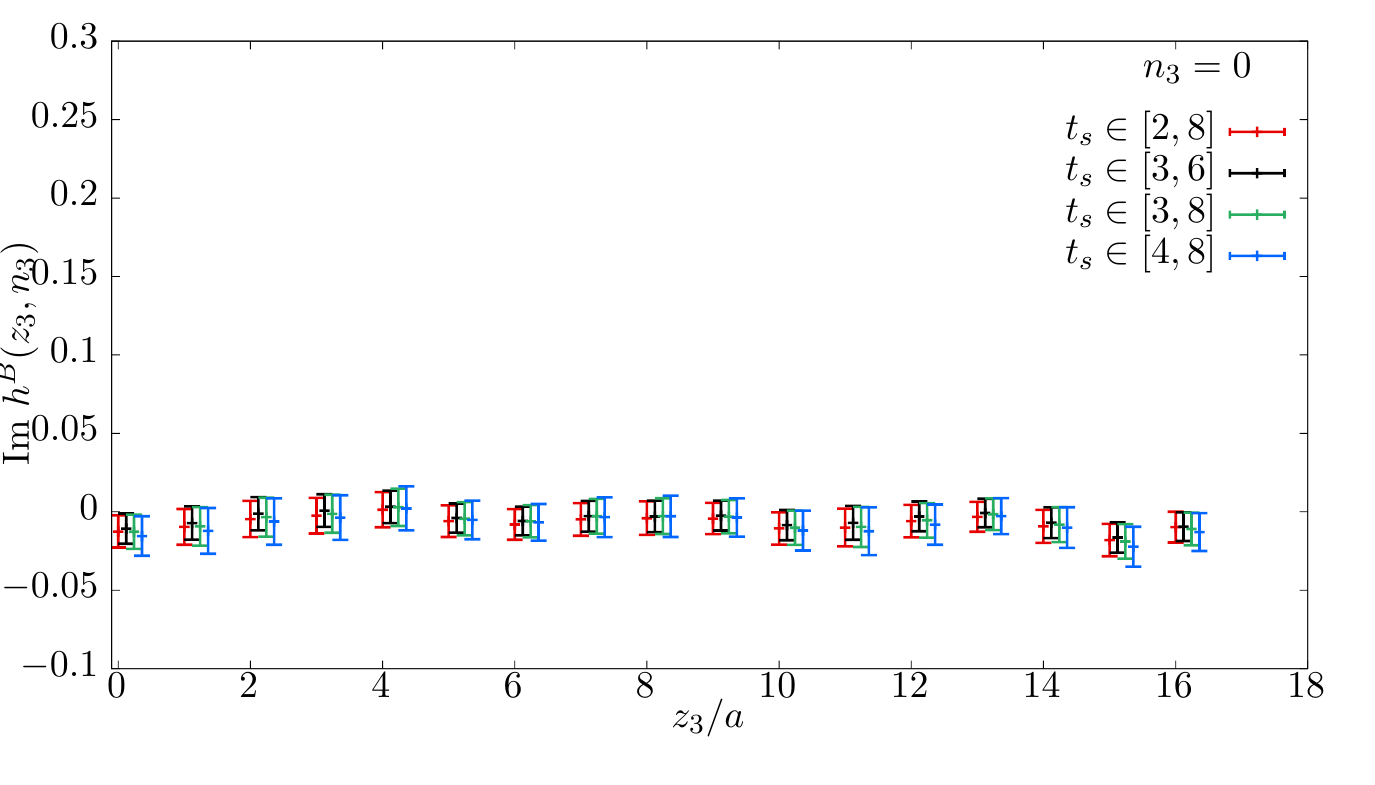}
\includegraphics[scale=0.5]{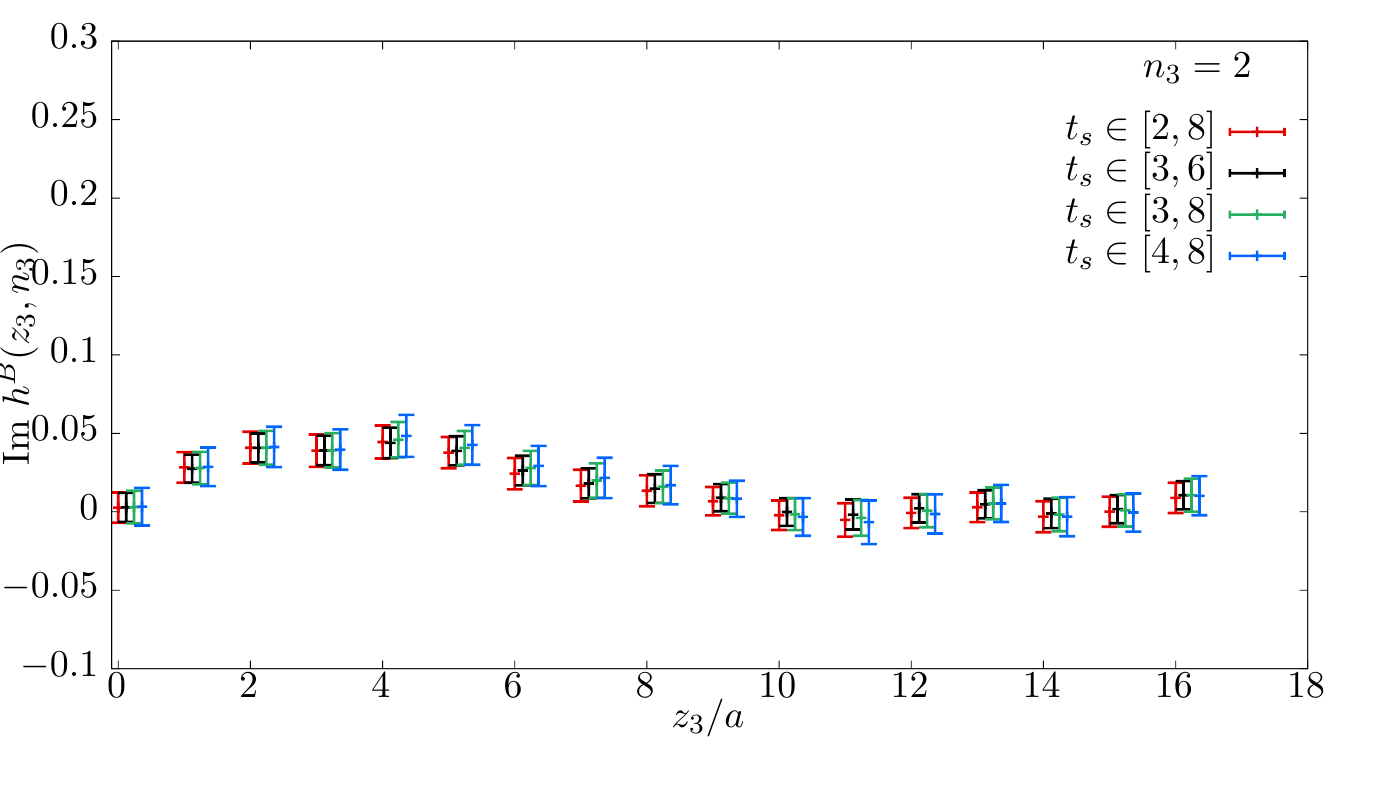}

\includegraphics[scale=0.5]{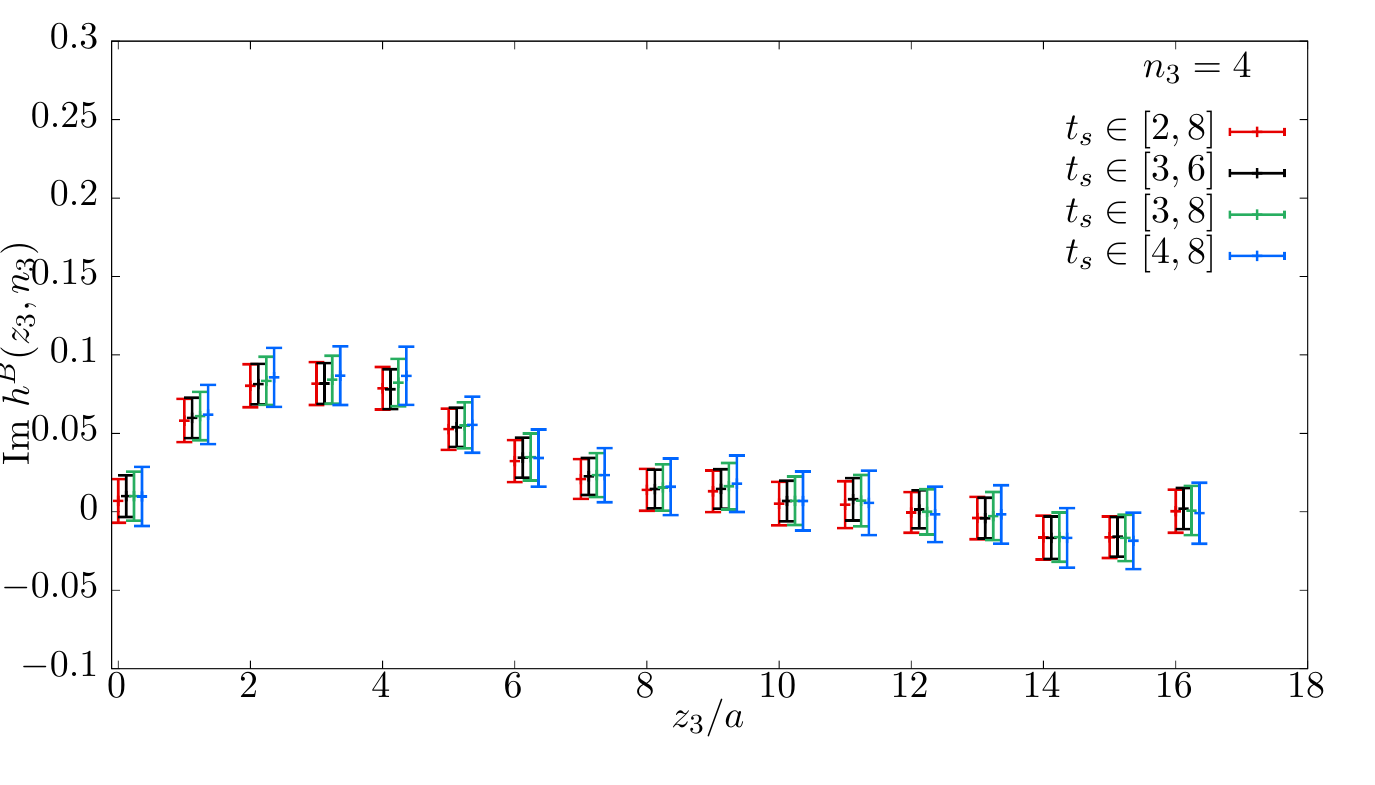}
\includegraphics[scale=0.5]{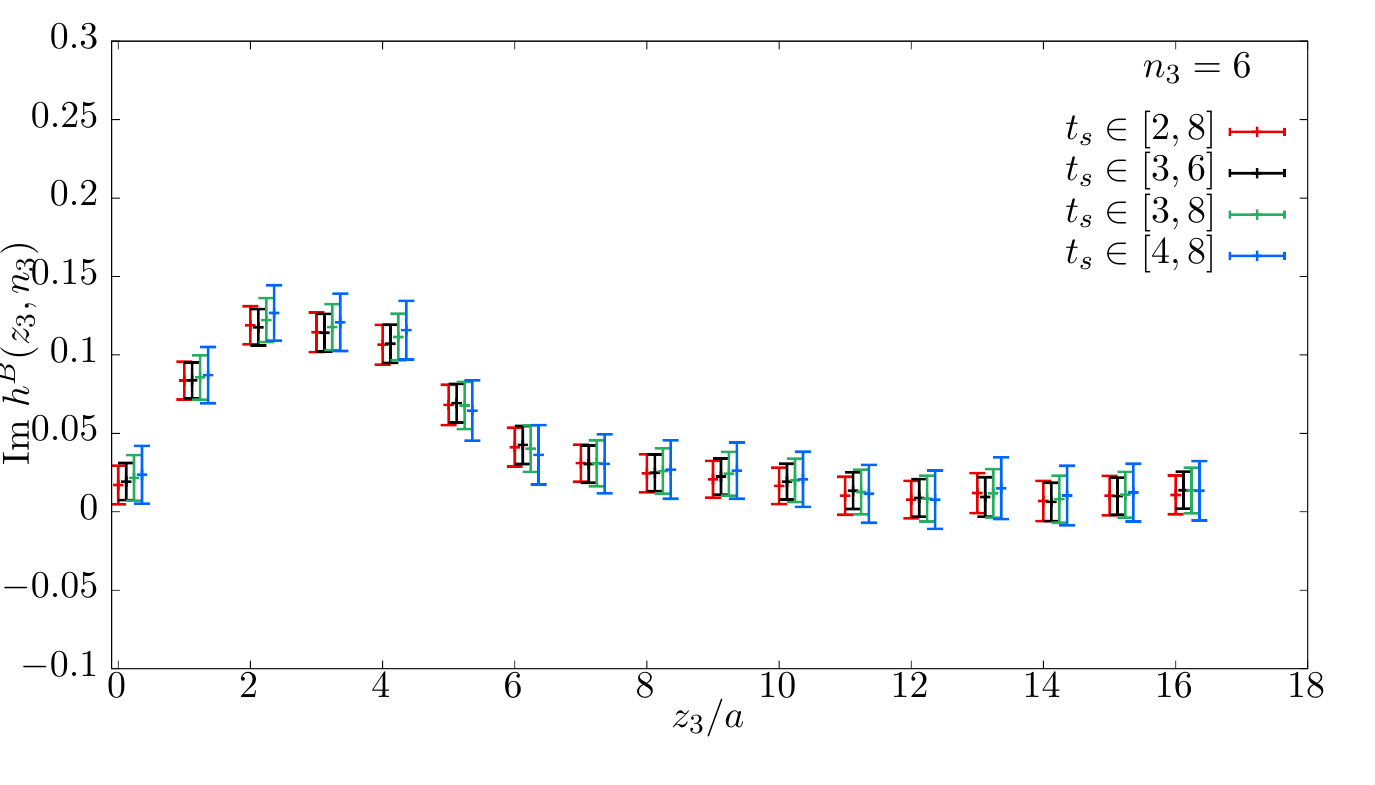}

\includegraphics[scale=0.5]{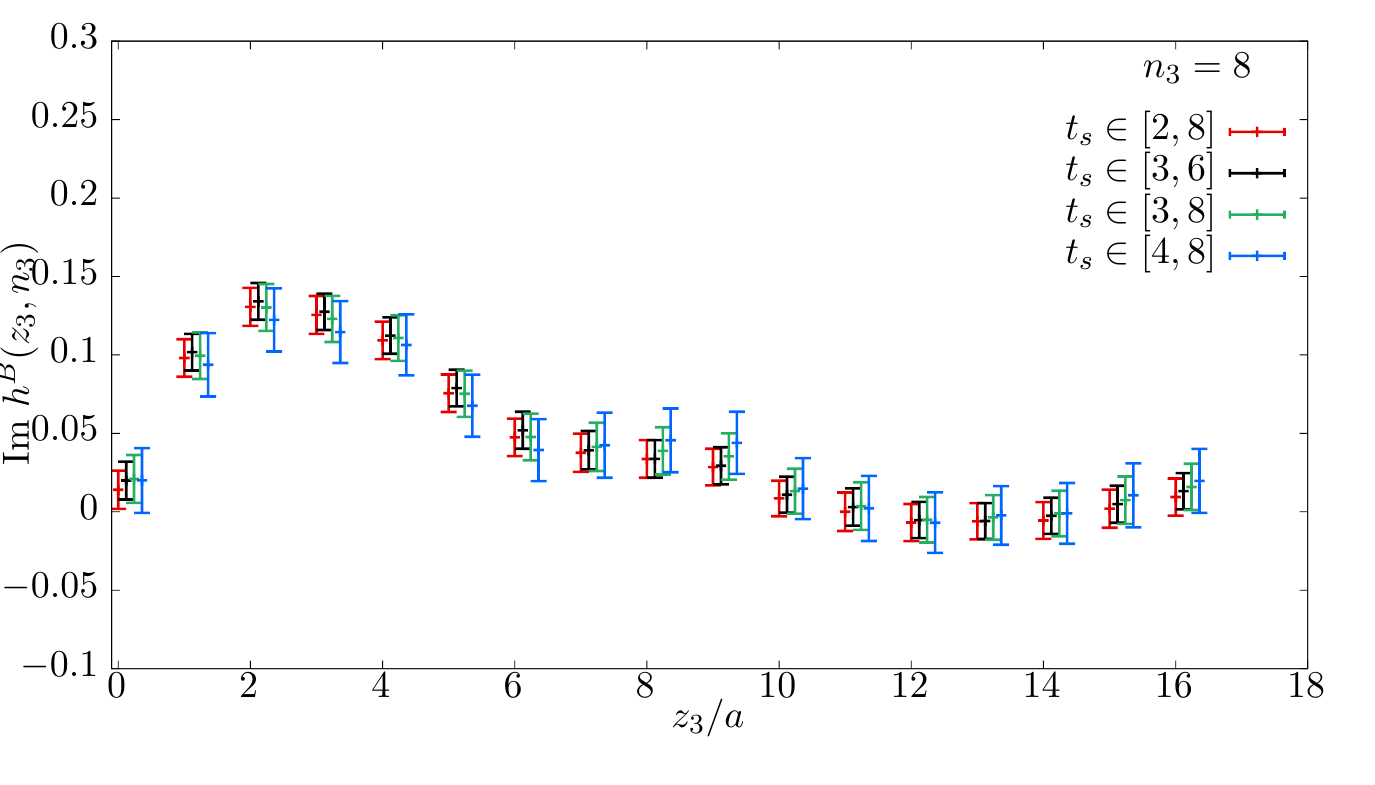}
\includegraphics[scale=0.5]{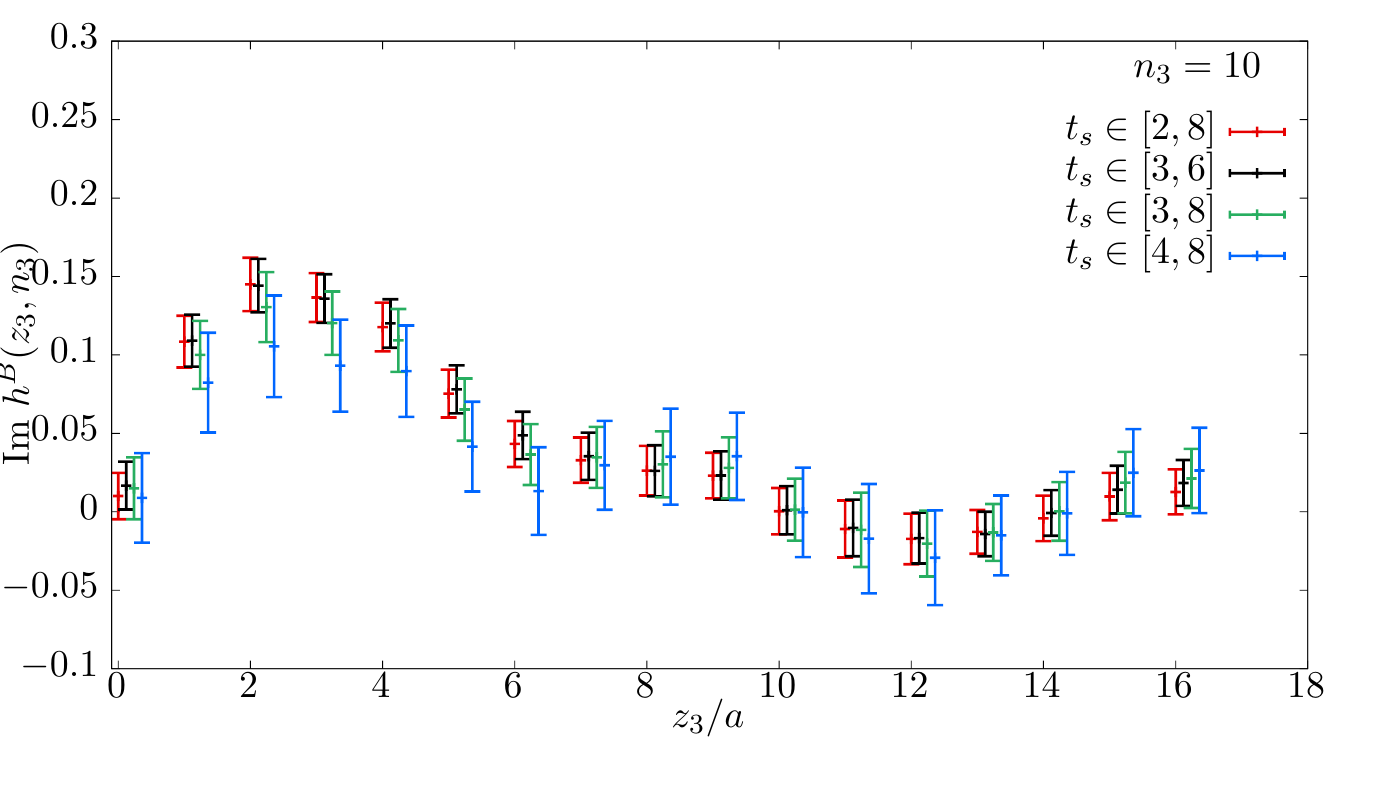}

\includegraphics[scale=0.5]{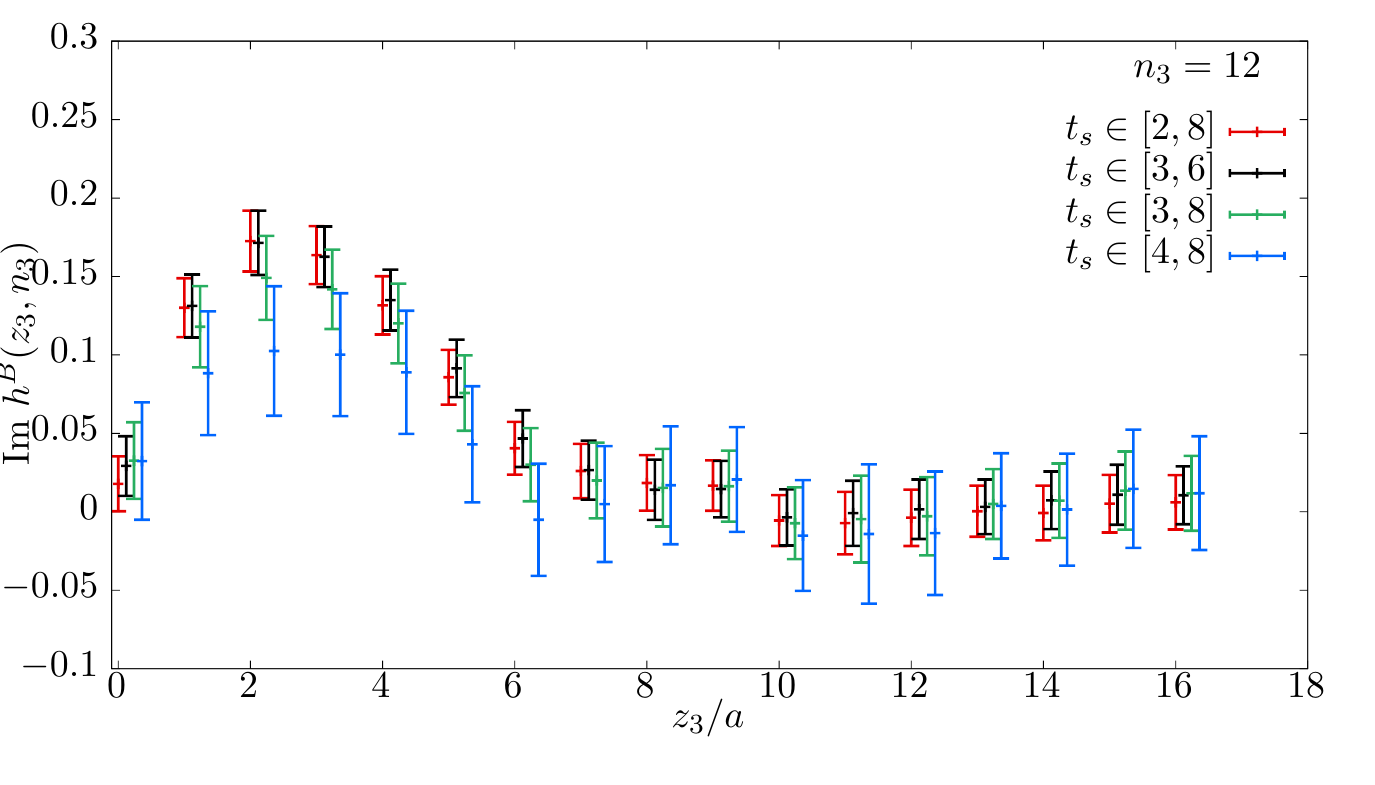}
\includegraphics[scale=0.5]{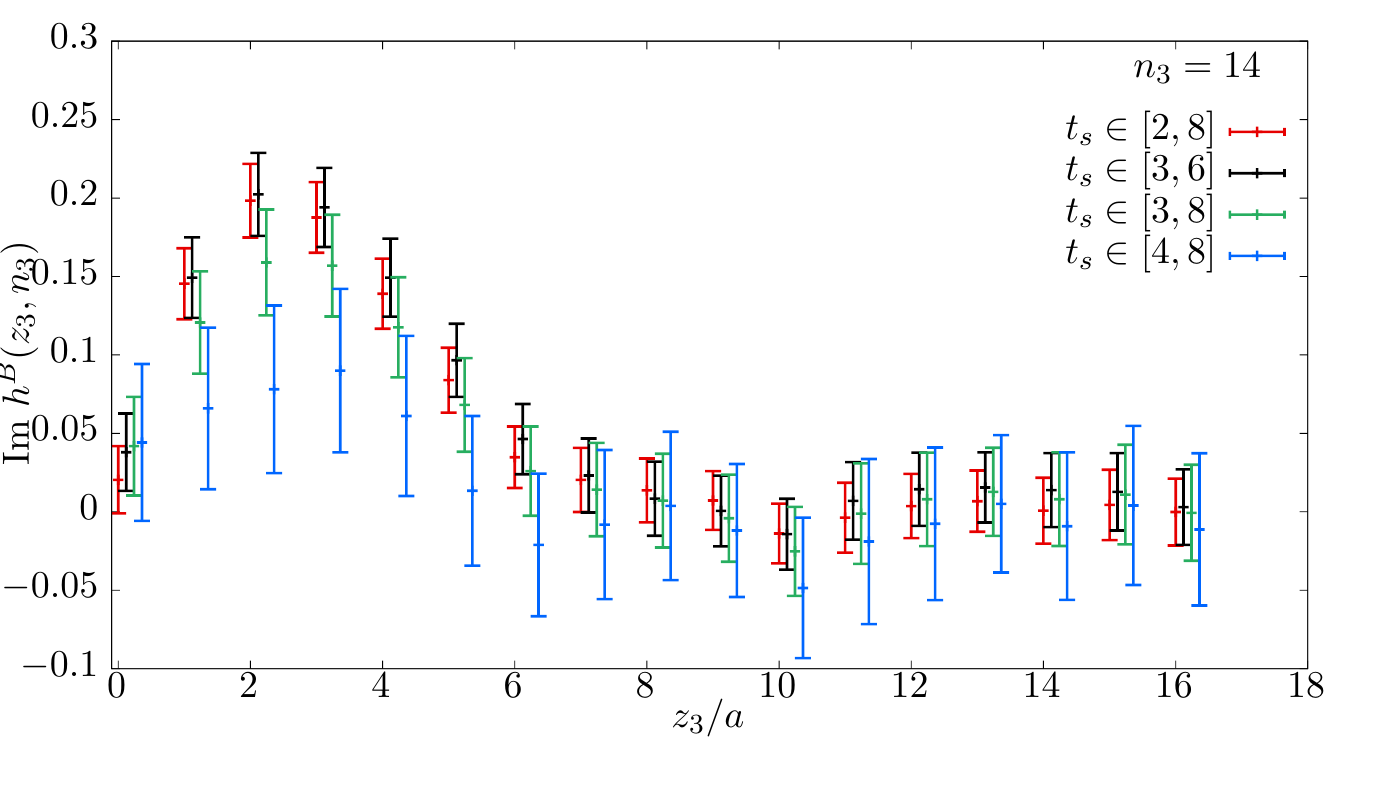}

\includegraphics[scale=0.5]{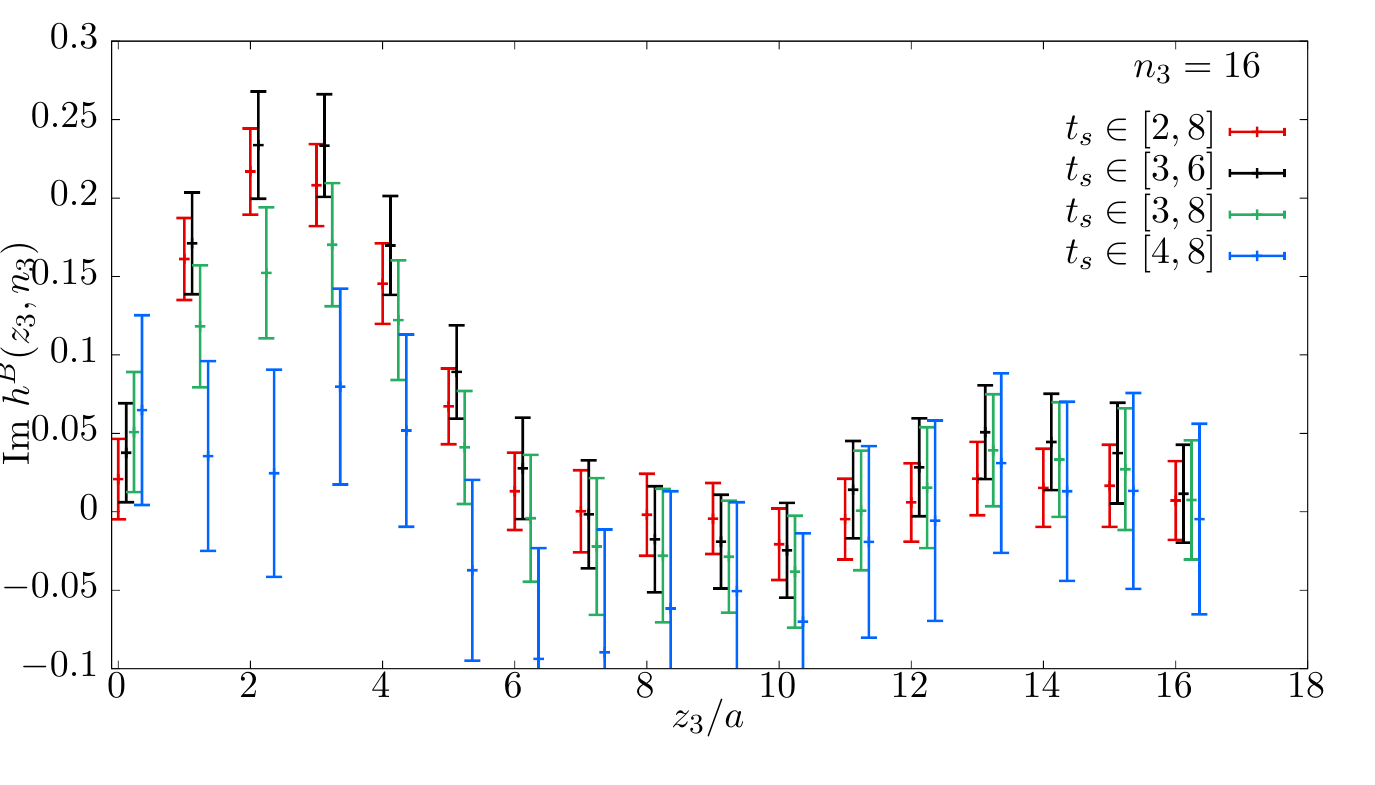}
\caption{The imaginary part of the bare quasi-PDF matrix element $h^B(z_3,n_3)$
is shown as a function of $z_3/a$ at different $n_3$ used in this work
as separate panels. The description is similar to \fgn{rehb}.}
\eef{imhb}

\bef
\centering
\includegraphics[scale=1.0]{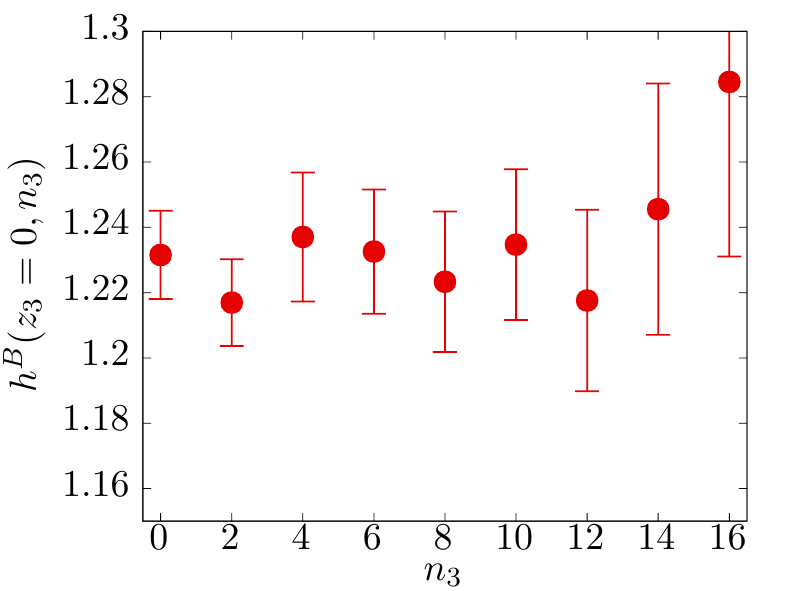}
\caption{A cross-check on the near-constant behavior the extracted bare matrix element 
of local current operator, $h^B(z_3=0,n_3)$ as a function of momentum $n_3$ in lattice units.
The bare matrix element is an estimate of the inverse of the vector current 
renormalization constant $Z_V$.
}
\eef{zv}

\section{Determination of bare pion quasi-PDF matrix elements}
\label{sec:threeptextrapol}

\subsection{Construction of $t_s$ dependent 3-point function}

We used the stochastic estimator in \eqn{3ptstochastic} to determine
the 3-point function in momentum space,
$\tilde{C}_{\rm 3pt}(z,p,q)$. We found it computationally simpler
and cheaper to fix the momentum insertion $q$ and scan the entire
set of $p_0$ for each choice of pion's spatial momentum $P_3$.  We
chose $q_0=0$ as this choice is the summation method as noted in the
main text. Further, we set the spatial part of $q$ also to zero, as we want the forward matrix element in this work 
for the case of PDF; thus $q=0$. We reconstructed the $t_s$ dependence of the
3-point function as,
\beq
C_{\rm 3pt}(z,t_s,P_3) = \sum_{n_0=0}^{L^{\rm eff}_0-1} \tilde{C}_{\rm 3pt}(z,p_0,P_3,q=0) e^{i p_0 \frac{t_s}{a}};\qquad  \tilde{C}_{\rm 3pt}(z,L^{\rm eff}_0 - n_0, n_3,q=0) = -  \tilde{C}_{\rm 3pt}(z,n_0,n_3,q=0).
\eeq{3ptts}
The second identity is simply due to the usage of $\gamma_0$ in the
definition of quasi-PDF operator, which makes the three-point
function to be proportional to $p_0$, and hence, antisymmetric with
respect to $n_0$ and $L^{\rm eff}_0-n_0$.  As is usual, we used
$z=(0,0,0,z_3)$ along the $z$-axis. By using folded Wilson line,
we scanned $z_3/a \in [-16,16]$.  In the end, we only used $|z_3|/a
\le 6$ so as to ensure the applicability of perturbation theory.

Let us make the connection of 3-point function with $q_0=0$ to summation
method obvious. We suppress the arguments for $z$, $p$ in the 2-
and 3-point functions for the sake of brevity, 
and both of them should be at
understood to be at the same $p$ below.  For the sake of argument,
let $C_{\rm 3pt}(t_s,\tau)$ be the 3-point function by Fourier
transforming with respect to both $p_0$ and $q_0$; in that case, $\tau$
is the insertion time of the quasi-PDF operator $\tau$, and it could
be both within and outside the pion source and sink locations. For
the case $\tau \le t_s$, we can do a spectral decomposition to get
\beq
C_{\rm 3pt}(t_s,\tau) = \sum_{i,j=0} A^*_i A_j\langle i|{\cal O}|j\rangle e^{-E_i (t_s-\tau) -E_j \tau};\qquad A_i = \langle 0|\pi^\dagger|E_i\rangle,
\eeq{3ptspect1}
for $\tau\le t_s$. The sum within this region gives, $\sum_{\tau=0}^{t_s}
C_{\rm 3pt}(t_s,\tau) \sim |A_0|^2 \langle 0|{\cal O}|0\rangle t_s
e^{-E_i t_s} + {\rm const.} \times e^{-E_0 t_s}$, up to $O(e^{-E_1
t_s})$ excited state corrections. When the operator is ``outside"
the source and sink, $\tau > t_s$, then the terms which are not
exponentially suppressed with the effective temporal extent, $L_0^{\rm eff}$, are of the form
\beq
C_{\rm 3pt}(t_s,\tau) = \sum_{i,j} \langle 0|{\cal O}|i\rangle e^{-E_i \tau} e^{-E_j t_s}\langle i|\pi|j\rangle\langle j|\pi^\dagger|0\rangle,
\eeq{3ptspect2}
for $\tau > t_s$.  Since the state $j$ cannot be the ground-state
pion, and $i\ne j$, the sum over $\tau> t_s$ cannot have the linear
piece and will contribute simply as yet another exponentially
suppressed excited state contribution. Therefore, we can sum over
$\tau$ for all values from $0$ to $L_0^{\rm eff}$, and the linear piece in $t_s$ gives the
information on the ground state matrix element. The sum,
$\sum_{\tau=0}^{L^{\rm eff}_0} C_{\rm 3pt}(t_s,\tau) = C_{\rm
3pt}(t_s, q_0=0)$.  To cancel off the amplitudes $A_i$, we form the
ratio
\beq
R(t_s) \equiv \frac{C_{\rm 3pt}(t_s,q_0=0)}{C_{\rm 2pt}(t_s)}.
\eeq{rdef}
From the arguments above, we see that
\beq
R(t_s) = \langle E_0| {\cal O} |E_0\rangle t_s + C + {O}(e^{-(E_1-E_0) t_s}).
\eeq{lineardep}
By fitting the linear $t_s$ dependence of $R(t_s)$, we obtained the
bare quasi-PDF matrix element from the slope. 

\subsection{Extraction of ground-state bare quasi-PDF matrix elements}
We fit the functional form
\beq
R(t_s; P_3, z_3) = t_s h^B(z_3, P_3) + C,
\eeq{stfit}
to the summed ratio $R(t_s)$ at different $z_3$ and $P_3$, using
$h^B(z_3,P_3)$ and $C$ as fit parameters over ranges $t_s \in
[t_s^{\rm min}, t_s^{\rm max}]$. At the precision allowed by our
data, we restricted the value of $t_s^{\rm min}=2a,3a,4a$, and
finally used $t_s^{\rm min}=3a$, which in physical units, $t^{\rm
min}_s \sqrt{\sigma} = 0.76$, is on the verge of the typical
nonperturbative mass-gapped scales. By changing $t_s^{\rm min}$ to
$2a$ and $4a$, we checked for the level of consistency as means to
test for the residual presence of excited state contributions. We
used $t_s^{\rm max}$ between $6a$ and $8a$, and avoided points
beyond $8a$ to not use the noisy as well as stochastically not-so
well determined points at even larger $t_s^{\rm max}$.  In the
various panels of \fgn{resumfit} and \fgn{imsumfit}, we show the
data points and  fitted straight-lines to determine ${\rm Re} h^B$
and ${\rm Im} h^B$ respectively. Each column in these figures show
the fits at $z_3 = 0, 2a, 4a, 6a$ at fixed momentum $n_3$.  The
different rows show them at $n_3=0,4,8,12,16$. In each panel, the
points are our lattice determination of $R(t_s)$. We have shown the
fits to \eqn{stfit} for different fit ranges as the bands; the
slopes of these lines are the needed values of $h^B$. Within the
statistical errors, it is clear that the data nicely agrees with a
linear $t_s$ dependence in the ranges of $t_s$ specified above.  At
the smaller $n_3$, where the data quality is better, we see that
the fitted bands from the various ranges agree quite well. At larger
$n_3=12$ to 16, the data quality for $t_s > 6a$ is quite poor, and
the fits that start from $t_s=4a$ and include $t_s > 6a$ data points
behave quite differently. Therefore, for $n_3\ge 12$, we restricted
$t_s^{\rm max}=6a$.

In \fgn{rehb} and \fgn{imhb}, we show the resulting $z_3$ dependence
of the quasi-PDF matrix element $h^B(z_3,P_3)$ from the summation-type
fits over the different $t_s$ ranges. We have slightly displaced
the different estimations for clarity. We see that the estimations
using $t_s \in [3a,8a]$ are quite consistent with those using
$[4a,8a]$ for the momenta $n_3\le 10$. Therefore, we used the
estimated values of $h^B$ from $t_s\in[3a,8a]$ in the main text.
For the higher momenta, as we noted above, we see that estimations
using $t_s^{\rm max}>6a$ are not reliable. Within the larger
statistical errors at the higher momenta $n_3 \ge 12$, we find the
$t_s\in[3a,6a]$ estimates are consistent with the shorter $t_s\in[2a,6a]$
estimates, and also within the larger errors of the $t_s\in[4a,8a]$
estimates which are biased with the poorly determined data beyond
$t_s>6a$ . Therefore, we chose the range containing  $t_s\in [3a,6a]$
for the set of momenta $n_3 \ge 12$.

As a cross-check, we present the values of $h^B(z_3=0,P_3)$ as a
function of $P_3$ in \fgn{zv}. The $z_3=0$ matrix element is nothing
but the pion matrix element of the local vector current operator,
and hence measures the inverse of the vector current renormalization
factor, $Z_V$. If the extraction of matrix elements is done correctly
and there is no momentum dependent lattice corrections, we should
not find any $P_3$ dependence in $Z_V$. Indeed, we find that to be
case in \fgn{zv} up to statistical errors.

\section{Implementation of leading-twist OPE and construction of $\msbar$ ITD}
\label{sec:implementation}
We implemented the leading-twist OPE using the truncated form of \eqn{tw2ope} written explicitly as,
\beqa
&&{\rm Re}{\cal M}(\nu,z_3^2)=1+\sum_{n=1}^{N_{\rm max}} \frac{(-1)^n \nu^{2n}}{(2n)!}C_{2n}(z_3^2 \mu^2) \langle x^{2n}\rangle_{u-\bar u},\cr
&&{\rm Im}{\cal M}(\nu,z_3^2)=\sum_{n=1}^{N_{\rm max}} \frac{-(-1)^n \nu^{2n-1}}{(2n-1)!}C_{2n-1}(z_3^2 \mu^2) \langle x^{2n-1}\rangle_{u+\bar u}.
\eeqa{tw2opetrunc}
We used the truncation as $N_{\rm max}=4$ to fit the data up to
$\nu=3.5$, and we checked that the results do not change well within errors when $N_{\rm max}$ is changed from 3 to 4.  The Wilson coefficients
$C_n$ are the isovector quark coefficients, usually written explicitly
as $C_n^{qq}$.  For the imaginary part, which is not a isovector
quantity, one would have to include the corresponding $C_n^{qq}$
and $C_n^{qg}$ which will cause mixing with quark and gluon
PDFs~\cite{Wang:2017qyg}; here, in the large-$N_c$ limit, the
$C_n^{qg}$ which are proportional to $\alpha_s T_F(N_c)$ are $1/N_c$
suppressed and hence, we have considered only the $C_n^{qq}$ Wilson
coefficients above and in the main text.

Using the above OPE, we performed combined fits to the $z_3$ and $P_3$ dependencies of the 
lattice data. We performed two types of fits in the main text:
\begin{enumerate}
\item Moments fit:  here, we used the Mellin moments $\langle x^n\rangle$ entering the OPE as the free fit parameters. 
    Since, we are assuming no functional form for the $x$-dependence of the PDF, we referred to these types of fits as 
    the model-independent fit analysis. Assuming the positivity of the underlying $u+\bar u$ and $u-\bar u$ PDFs 
        help impose additional constraint~\cite{Gao:2020ito} on their Mellin moments. We implemented such 
        inequalities using a change of variable from moments to $\lambda_i$,
        \beq
        \langle x^{2n}\rangle_{u-\bar u} = \sum_{i=n}^{N_{\rm max}} \sum_{j=i}^{N_{\rm max}} e^{-\lambda_j},
        \eeq{chovar}
        and similarly for odd-moments $\langle x^{2n-1}\rangle_{u+\bar u}$.

\item PDF Ansatz fit: Here, we assumed a global fit analysis inspired
ansatz for the $x$-dependence of the valence PDF,
    $f_{u-\bar u}(x)={\cal N}x^\alpha (1-x)^\beta (1+s x^2)$, with
    $\int_0^1 f_{u-\bar u}(x) dx =1$. In practice,
	it results in Mellin moments $\langle x^{2n}\rangle_{u-\bar
	u}(\alpha,\beta,s)$, that in turn enter \eqn{tw2opetrunc}.
	We fit the parameters $\alpha,\beta$ and $s$ in this manner.
	We imposed a prior $\alpha\in[-0.4,-0.6]$ based on Regge
	intercept expectation for valence PDF. We did not perform
	an equivalent analysis for $u+\bar u$, as it was not clear
	if we should assume it to be a combination of ansatz for
	valence PDF and sea-quark PDF, and what prior to impose on
	small-$x$ behavior of sea-quarks in large-$N_c$ theory.
	Therefore, we avoided such issues here by performing only
	moments fit to $u+\bar u$ case.

\end{enumerate}

In the main text, we constructed the $\msbar$ ITD at $\mu=2$ GeV 
based on the
analysis of pseudo-ITD lattice data above.  These ITDs are defined
as
\beq
{\cal M}^{\msbar}_{u-\bar u}(\nu,\mu)\equiv \int_0^1 f_{u-\bar u}(x,\mu)\cos(x \nu) dx;\quad
{\cal M}^{\msbar}_{u+\bar u}(\nu,\mu)\equiv \int_0^1 f_{u+\bar u}(x,\mu)\sin(x \nu) dx.
\eeq{itddefinition}
In practice, the construction of $\msbar$ ITD using the limited range of $\nu$ is simplified into a truncated series in $\nu$ as
\beqa
&&{\cal M}^{\msbar}_{u-\bar u}(\nu,\mu)=1+\sum_{n=1}^{N_{\rm max}} \frac{(-1)^n \nu^{2n}}{(2n)!} \langle x^{2n}\rangle_{u-\bar u}(\mu),\cr
&&{\cal M}^{\msbar}_{u+\bar u}(\nu,\mu)=\sum_{n=1}^{N_{\rm max}} \frac{-(-1)^n \nu^{2n-1}}{(2n-1)!}\langle x^{2n-1}\rangle_{u+\bar u}(\mu),
\eeqa{itdtrunc}
using the best fit estimates of the Mellin moments from the leading-twist analysis.

\section{Details regarding the perturbative aspects}

\subsection{Coupling constant and Wilson coefficients}
\label{sec:coupling}

We borrowed various existing 
perturbative results computed for general $N_c$, 
and we simply used the large-$N_c$ values of the color factors, 
$C_F(N_c)\to N_c/2$, $C_A(N_c)\to N_c$, and $T_F(N_c)\to 1/2$ in those
expressions.
In the absence of a nonperturbative running of the large-$N_c$ $\msbar$ coupling, we simply 
used the LO 't Hooft coupling in the large-$N_c$ limit,
\beq
\lambda(\mu) \equiv \lim_{N_c\to \infty} \alpha_s(\mu) N_c = \frac{1}{\frac{11}{12\pi}\ln\left(\frac{\mu^2}{\Lambda^2_{\rm \msbar}}\right)}.
\eeq{locoupling}
With $\sqrt{\sigma}=0.44$ GeV to set the scale, we used
$\Lambda_{\msbar}=0.22$. At $\mu=2$ GeV, we get $\lambda({\rm 2
GeV})=0.778$.  For the Wilson coefficients that enter the leading-twist
OPE in \eqn{tw2ope}, we used the 1-loop expressions in
Ref~\cite{Izubuchi:2018srq} with the replacement $C_F(N_c)\alpha_s(\mu)
\to \lambda(\mu)/2 = 0.389$. The Wilson coefficients for the $u+\bar
u$ PDF would differ in the SU(3) QCD due to it being a flavor singlet
quantity. In the large-$N_c$ limit, such differences due to the
mixing terms ($\propto T_F(N_c)$) are sub-leading in $1/N_c$, and
hence, we simply used the non-singlet Wilson coefficients $C_n$ for
odd values of $n$.

\subsection{Large-$N_c$ LO DGLAP evolution}
\label{sec:dglap}
In the main text, we checked whether a universal initial condition
at a low factorization scale $\mu_0$ could explain the observed
differences between SU(3) QCD and in large-$N_c$ theory.  For this,
we performed the DGLAP evolution of large-$N_c$ PDF (or equivalently
its ITD), from scale $\mu$ to a lower-scale $\mu_0$, that is then
used as an initial condition for 3 flavor SU(3) QCD evolution back
to scale $\mu$ using corresponding DGLAP evolution in Mellin space.
That is, taking \beq X^{(N_c)}_n(\mu) = \left[2 \langle x^n\rangle_{u-\bar
u}(\mu), 2 \langle x^n\rangle_{u+\bar u}(\mu), \langle
x^n\rangle_g\right(\mu)], \eeq{xdef} as the array of quark and gluon
moments in $SU(N_c)$ QCD, we evolved them as,
\beq
X^{(N_c)}_n(\mu')=
        \begin{pmatrix} 
            P^{qq,N_c}_{\rm NS}(n,\mu,\mu') & 0 & 0 \\
            0 & P^{qq,N_c}_{\rm S}(n,\mu,\mu') & P^{gq,N_c}_{\rm S}(n,\mu,\mu') \\ 
            0 & P^{qg,N_c}_{\rm S}(n,\mu,\mu') & P^{gg,N_c}_{\rm S}(n,\mu,\mu') 
        \end{pmatrix} \cdot X^{(N_c)}_n(\mu),
\eeq{dglap}
where $P^{ij,N_c}(n,\mu,\mu')$ are the $SU(N_c)$ theory DGLAP factors
from parton species $i$ to species $j$ in Mellin space (e.g.,
textbook such as~\cite{Roberts:1990ww}) that evolve the moments
from scale $\mu$ to $\mu'$.  The subscript S and NS specify singlet
and non-singlet respectively.  In this paper, we used a LO DGLAP
evolution, at which order $P^{qq}_{NS}=P^{qq}_S$.  At LO, the
evolution depends on $\mu$ only via the logarithms
$\frac{\ln(\mu/\Lambda_{\msbar})}{\ln(\mu_0/\Lambda_{\msbar})}$.
Since we were only interested in capturing the qualitative behavior
of $u+\bar u$ ITD in the large-$N_c$ theory and $SU(3)$ theory, we
simply used $\Lambda_{\msbar}=0.22$ GeV in the DGLAP factors of
both the theories. One should note that the ratios of twist-2 operator anomalous dimensions to $\beta$-function coefficient,
$\gamma^{(1)}_n/\beta_0$ have a finite limit when $N_c\to\infty$.
In the $N_c\to\infty$ limit, the cross-term $P^{gq,N_c}_{\rm S}\to
0$, and hence $u+\bar u$ evolves without mixing with the gluon. On
the other hand, the term $ P^{qg,\infty}_{\rm S}$ is non-zero as
gluon radiation from a quark-line is still a leading process in
$N_c$ counting.

In this work, we only computed the quark moments $\langle
x^n\rangle_{u+\bar u}$ in the large-$N_c$ theory, and we did not
explicitly compute the gluon PDF in the large-$N_c$ pion. Therefore,
we deduced the leading moment $\langle x\rangle_g=1-2\langle
x\rangle_{u+\bar u}$ from the momentum sum rule.  
Since we expect the gluon PDF to be contributing
dominantly in the small-$x$ region, we assumed that the next moment
$\langle x^3\rangle_g$ (and all other higher odd moments) can be
neglected. With these inputs from the large-$N_c$ theory, we followed
the chain of evolution,
\beq
X^{(\infty)}_n(\mu)\to X^{(\infty)}_n(\mu_0) \to X^{(3)}_n(\mu). 
\eeq{chain}
Using such an expectation $X^{(3)}_n(\mu)$ for SU(3) QCD moments at 
$\mu=2$ GeV based on the above evolution, we constructed the 
corresponding $\msbar$ ITD by using \eqn{itdtrunc}.

\section{Efficacy of 1-loop large-$N_c$ leading-twist OPE}
\label{sec:efficacy}
\bef
\centering
\includegraphics[scale=0.62]{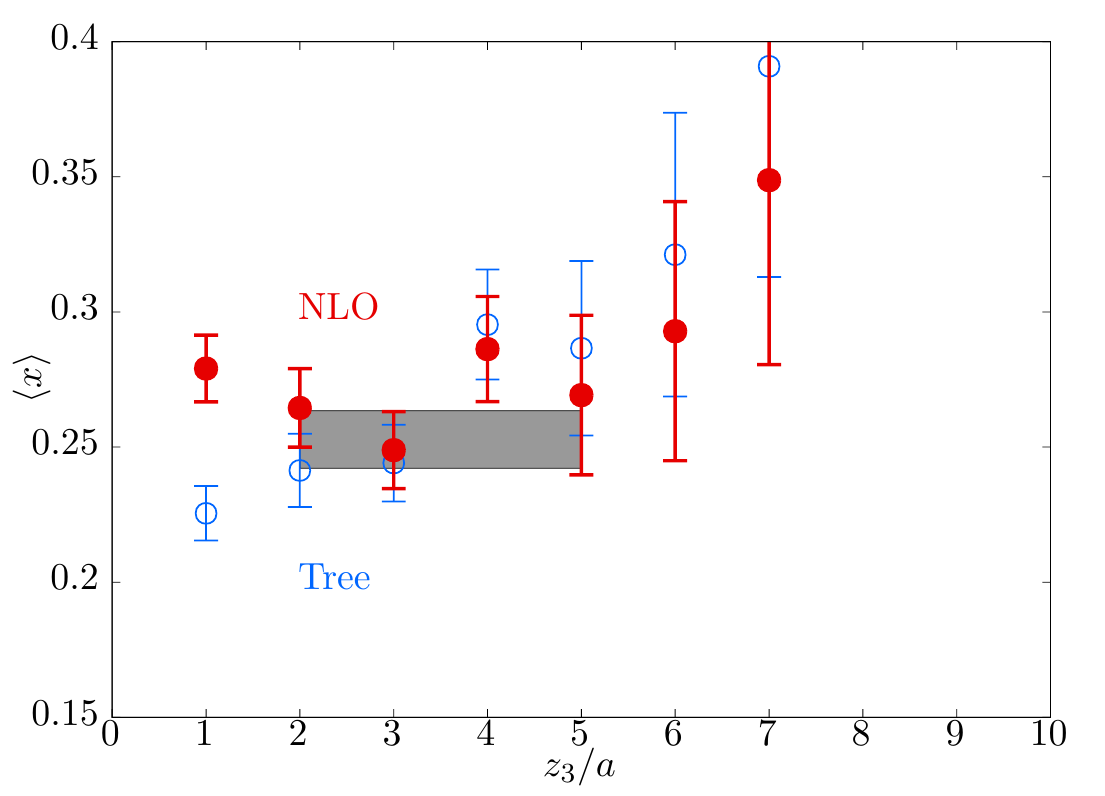}
\includegraphics[scale=0.62]{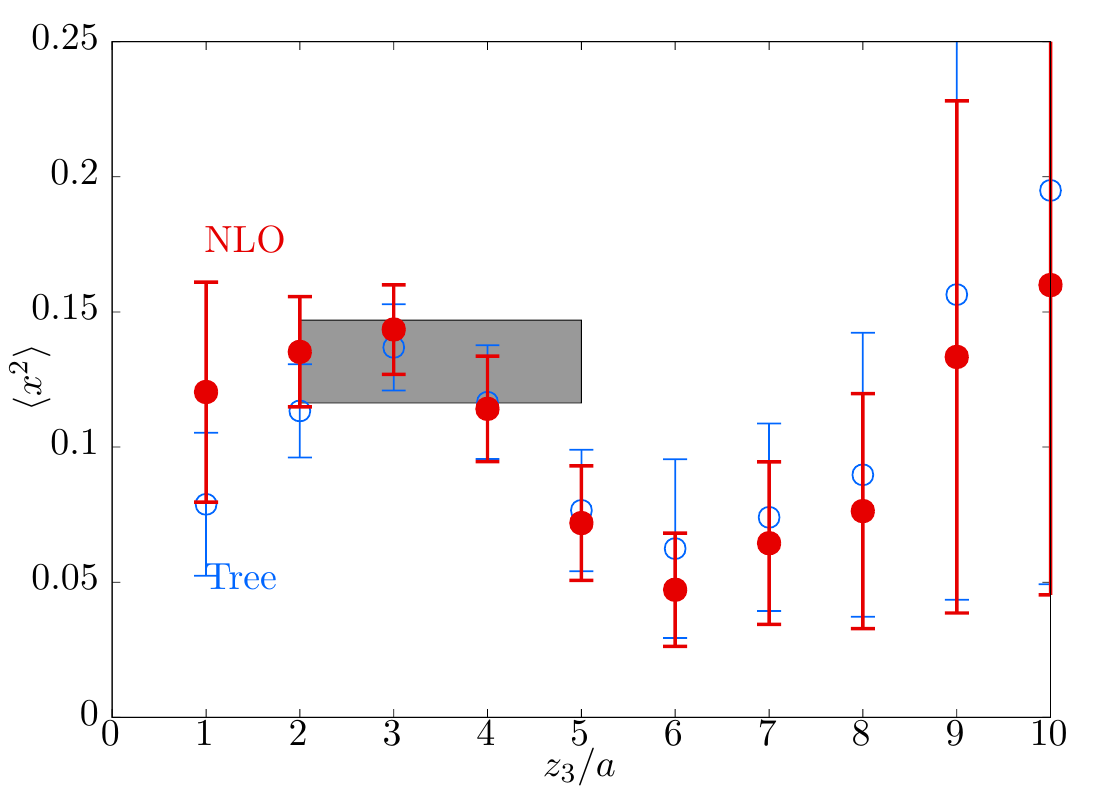}
\caption{
Fixed-$z^2$ analysis of pseudo-ITD ${\cal M}(\nu,z^2$ by 
fitting the 
leading-twist OPE to the $P_3 z_3$ dependence at different values 
of $z_3$ using first few Mellin moments are fit parameters. 
The left and right panels show the resulting $z_3$ dependent 
$\langle x\rangle_{u+\bar u}$ and $\langle x^2\rangle_{u-\bar u}$
moments respectively.
The filled red circles are result of performing such an analysis 
using 1-loop large-$N_c$ Wilson coefficients. The open 
circles are obtained by setting $C_n=1$, that is, to their
tree-level values. Our estimates of those moments based on 
a combined fit to both $z^2$ and $\nu$ dependencies of ${\cal M}$
data in the range $z_3 \in [2a,5a]$ are shown as the gray bands.
}
\eef{opewope}

 The leading-twist OPE can be applied to the lattice data at 
fixed values of $z_3$~\cite{Karpie:2018zaz}, so as to capture the
$\nu$ dependence coming only via variation in the momentum $P_3$. 
Such an application has
been found~\cite{Gao:2020ito,HadStruc:2021qdf} to be a nice diagnostic
of the effectiveness of perturbative as well as leading-twist framework
in a region of $z_3$, and as way to detect corrections to the framework. 
In the left panel of \fgn{opewope}, we show
such a $z_3$ dependent leading non-trivial moment $\langle x\rangle$,
from the analysis of ${\rm Im}{\cal M}(\nu,z_3^2)$. In the right
panel, we show a similar $z_3$ dependence of $\langle x^2\rangle$
from ${\rm Re}{\cal M}(\nu,z_3^2)$. We used $\mu=2$ GeV in the scale
set by $\sqrt{\sigma}=0.44$ GeV as explained above. The red filled
circles are the results using 1-loop Wilson coefficients in the
OPE. If 1-loop is sufficient, and if there are no higher-twist
corrections to the OPE and $z_3$-dependent lattice spacing corrections
to the continuum OPE, then one should observe a plateau in the
moments as a function of $z_3$. In the range $z_3\in[2a,5a]$ that
we used, we see an approximate plateau in the 1-loop results in the
two panels. Our determinations of the two moments via a combined
fits to the entire data in the range of  $z_3\in[2a,5a]$ is shown
as the gray bands, which are consistent with the plateau in the
data. We skipped the $z_3=a$ point to be cautious of avoiding any
lattice corrections at those separations. Given the quality of our
data, we did not add any lattice spacing and higher-twist
corrections by hand to the leading-twist continuum OPE. 
To see the effect of 1-loop
evolution in $z_3$ effected by the Wilson coefficients, we also plot the
results using tree-level (i.e., set $\alpha_s=0$) in the two panels
in \fgn{opewope}. The effect of 1-loop is rather small in comparison
with typical statistical errors, but it is quite pronounced at shorter
$z_3=1a-3a$.

\bibliography{pap.bib}

\begin{thebibliography}{80}%
\makeatletter
\providecommand \@ifxundefined [1]{%
 \@ifx{#1\undefined}
}%
\providecommand \@ifnum [1]{%
 \ifnum #1\expandafter \@firstoftwo
 \else \expandafter \@secondoftwo
 \fi
}%
\providecommand \@ifx [1]{%
 \ifx #1\expandafter \@firstoftwo
 \else \expandafter \@secondoftwo
 \fi
}%
\providecommand \natexlab [1]{#1}%
\providecommand \enquote  [1]{``#1''}%
\providecommand \bibnamefont  [1]{#1}%
\providecommand \bibfnamefont [1]{#1}%
\providecommand \citenamefont [1]{#1}%
\providecommand \href@noop [0]{\@secondoftwo}%
\providecommand \href [0]{\begingroup \@sanitize@url \@href}%
\providecommand \@href[1]{\@@startlink{#1}\@@href}%
\providecommand \@@href[1]{\endgroup#1\@@endlink}%
\providecommand \@sanitize@url [0]{\catcode `\\12\catcode `\$12\catcode
  `\&12\catcode `\#12\catcode `\^12\catcode `\_12\catcode `\%12\relax}%
\providecommand \@@startlink[1]{}%
\providecommand \@@endlink[0]{}%
\providecommand \url  [0]{\begingroup\@sanitize@url \@url }%
\providecommand \@url [1]{\endgroup\@href {#1}{\urlprefix }}%
\providecommand \urlprefix  [0]{URL }%
\providecommand \Eprint [0]{\href }%
\providecommand \doibase [0]{http://dx.doi.org/}%
\providecommand \selectlanguage [0]{\@gobble}%
\providecommand \bibinfo  [0]{\@secondoftwo}%
\providecommand \bibfield  [0]{\@secondoftwo}%
\providecommand \translation [1]{[#1]}%
\providecommand \BibitemOpen [0]{}%
\providecommand \bibitemStop [0]{}%
\providecommand \bibitemNoStop [0]{.\EOS\space}%
\providecommand \EOS [0]{\spacefactor3000\relax}%
\providecommand \BibitemShut  [1]{\csname bibitem#1\endcsname}%
\let\auto@bib@innerbib\@empty
%</preamble>
\bibitem [{\citenamefont {'t~Hooft}(1974{\natexlab{a}})}]{tHooft:1973alw}%
  \BibitemOpen
  \bibfield  {author} {\bibinfo {author} {\bibfnamefont {G.}~\bibnamefont
  {'t~Hooft}},\ }\href {\doibase 10.1016/0550-3213(74)90154-0} {\bibfield
  {journal} {\bibinfo  {journal} {Nucl. Phys. B}\ }\textbf {\bibinfo {volume}
  {72}},\ \bibinfo {pages} {461} (\bibinfo {year}
  {1974}{\natexlab{a}})}\BibitemShut {NoStop}%
\bibitem [{\citenamefont {'t~Hooft}(1974{\natexlab{b}})}]{tHooft:1974pnl}%
  \BibitemOpen
  \bibfield  {author} {\bibinfo {author} {\bibfnamefont {G.}~\bibnamefont
  {'t~Hooft}},\ }\href {\doibase 10.1016/0550-3213(74)90088-1} {\bibfield
  {journal} {\bibinfo  {journal} {Nucl. Phys. B}\ }\textbf {\bibinfo {volume}
  {75}},\ \bibinfo {pages} {461} (\bibinfo {year}
  {1974}{\natexlab{b}})}\BibitemShut {NoStop}%
\bibitem [{\citenamefont {P\'erez}\ \emph {et~al.}(2021)\citenamefont
  {P\'erez}, \citenamefont {Gonz\'alez-Arroyo},\ and\ \citenamefont
  {Okawa}}]{Perez:2020vbn}%
  \BibitemOpen
  \bibfield  {author} {\bibinfo {author} {\bibfnamefont {M.~G.}\ \bibnamefont
  {P\'erez}}, \bibinfo {author} {\bibfnamefont {A.}~\bibnamefont
  {Gonz\'alez-Arroyo}}, \ and\ \bibinfo {author} {\bibfnamefont
  {M.}~\bibnamefont {Okawa}},\ }\href {\doibase 10.1007/JHEP04(2021)230}
  {\bibfield  {journal} {\bibinfo  {journal} {JHEP}\ }\textbf {\bibinfo
  {volume} {04}},\ \bibinfo {pages} {230} (\bibinfo {year} {2021})},\ \Eprint
  {http://arxiv.org/abs/2011.13061} {arXiv:2011.13061 [hep-lat]} \BibitemShut
  {NoStop}%
\bibitem [{\citenamefont {DeGrand}\ and\ \citenamefont
  {Liu}(2016)}]{DeGrand:2016pur}%
  \BibitemOpen
  \bibfield  {author} {\bibinfo {author} {\bibfnamefont {T.}~\bibnamefont
  {DeGrand}}\ and\ \bibinfo {author} {\bibfnamefont {Y.}~\bibnamefont {Liu}},\
  }\href {\doibase 10.1103/PhysRevD.94.034506} {\bibfield  {journal} {\bibinfo
  {journal} {Phys. Rev. D}\ }\textbf {\bibinfo {volume} {94}},\ \bibinfo
  {pages} {034506} (\bibinfo {year} {2016})},\ \bibinfo {note} {[Erratum:
  Phys.Rev.D 95, 019902 (2017)]},\ \Eprint {http://arxiv.org/abs/1606.01277}
  {arXiv:1606.01277 [hep-lat]} \BibitemShut {NoStop}%
\bibitem [{\citenamefont {Hern\'andez}\ \emph {et~al.}(2019)\citenamefont
  {Hern\'andez}, \citenamefont {Pena},\ and\ \citenamefont
  {Romero-L\'opez}}]{Hernandez:2019qed}%
  \BibitemOpen
  \bibfield  {author} {\bibinfo {author} {\bibfnamefont {P.}~\bibnamefont
  {Hern\'andez}}, \bibinfo {author} {\bibfnamefont {C.}~\bibnamefont {Pena}}, \
  and\ \bibinfo {author} {\bibfnamefont {F.}~\bibnamefont {Romero-L\'opez}},\
  }\href {\doibase 10.1140/epjc/s10052-019-7395-y} {\bibfield  {journal}
  {\bibinfo  {journal} {Eur. Phys. J. C}\ }\textbf {\bibinfo {volume} {79}},\
  \bibinfo {pages} {865} (\bibinfo {year} {2019})},\ \Eprint
  {http://arxiv.org/abs/1907.11511} {arXiv:1907.11511 [hep-lat]} \BibitemShut
  {NoStop}%
\bibitem [{\citenamefont {Bali}\ \emph {et~al.}(2013)\citenamefont {Bali},
  \citenamefont {Bursa}, \citenamefont {Castagnini}, \citenamefont {Collins},
  \citenamefont {Del~Debbio}, \citenamefont {Lucini},\ and\ \citenamefont
  {Panero}}]{Bali:2013kia}%
  \BibitemOpen
  \bibfield  {author} {\bibinfo {author} {\bibfnamefont {G.~S.}\ \bibnamefont
  {Bali}}, \bibinfo {author} {\bibfnamefont {F.}~\bibnamefont {Bursa}},
  \bibinfo {author} {\bibfnamefont {L.}~\bibnamefont {Castagnini}}, \bibinfo
  {author} {\bibfnamefont {S.}~\bibnamefont {Collins}}, \bibinfo {author}
  {\bibfnamefont {L.}~\bibnamefont {Del~Debbio}}, \bibinfo {author}
  {\bibfnamefont {B.}~\bibnamefont {Lucini}}, \ and\ \bibinfo {author}
  {\bibfnamefont {M.}~\bibnamefont {Panero}},\ }\href {\doibase
  10.1007/JHEP06(2013)071} {\bibfield  {journal} {\bibinfo  {journal} {JHEP}\
  }\textbf {\bibinfo {volume} {06}},\ \bibinfo {pages} {071} (\bibinfo {year}
  {2013})},\ \Eprint {http://arxiv.org/abs/1304.4437} {arXiv:1304.4437
  [hep-lat]} \BibitemShut {NoStop}%
\bibitem [{\citenamefont {Bali}\ and\ \citenamefont
  {Bursa}(2008)}]{Bali:2008an}%
  \BibitemOpen
  \bibfield  {author} {\bibinfo {author} {\bibfnamefont {G.~S.}\ \bibnamefont
  {Bali}}\ and\ \bibinfo {author} {\bibfnamefont {F.}~\bibnamefont {Bursa}},\
  }\href {\doibase 10.1088/1126-6708/2008/09/110} {\bibfield  {journal}
  {\bibinfo  {journal} {JHEP}\ }\textbf {\bibinfo {volume} {09}},\ \bibinfo
  {pages} {110} (\bibinfo {year} {2008})},\ \Eprint
  {http://arxiv.org/abs/0806.2278} {arXiv:0806.2278 [hep-lat]} \BibitemShut
  {NoStop}%
\bibitem [{\citenamefont {Del~Debbio}\ \emph {et~al.}(2008)\citenamefont
  {Del~Debbio}, \citenamefont {Lucini}, \citenamefont {Patella},\ and\
  \citenamefont {Pica}}]{DelDebbio:2007wk}%
  \BibitemOpen
  \bibfield  {author} {\bibinfo {author} {\bibfnamefont {L.}~\bibnamefont
  {Del~Debbio}}, \bibinfo {author} {\bibfnamefont {B.}~\bibnamefont {Lucini}},
  \bibinfo {author} {\bibfnamefont {A.}~\bibnamefont {Patella}}, \ and\
  \bibinfo {author} {\bibfnamefont {C.}~\bibnamefont {Pica}},\ }\href {\doibase
  10.1088/1126-6708/2008/03/062} {\bibfield  {journal} {\bibinfo  {journal}
  {JHEP}\ }\textbf {\bibinfo {volume} {03}},\ \bibinfo {pages} {062} (\bibinfo
  {year} {2008})},\ \Eprint {http://arxiv.org/abs/0712.3036} {arXiv:0712.3036
  [hep-th]} \BibitemShut {NoStop}%
\bibitem [{\citenamefont {Abdul~Khalek}\ \emph {et~al.}(2022)\citenamefont
  {Abdul~Khalek} \emph {et~al.}}]{AbdulKhalek:2022erw}%
  \BibitemOpen
  \bibfield  {author} {\bibinfo {author} {\bibfnamefont {R.}~\bibnamefont
  {Abdul~Khalek}} \emph {et~al.},\ }in\ \href@noop {} {\emph {\bibinfo
  {booktitle} {{2022 Snowmass Summer Study}}}}\ (\bibinfo {year} {2022})\
  \Eprint {http://arxiv.org/abs/2203.13199} {arXiv:2203.13199 [hep-ph]}
  \BibitemShut {NoStop}%
\bibitem [{\citenamefont {Accardi}\ \emph {et~al.}(2016)\citenamefont {Accardi}
  \emph {et~al.}}]{Accardi:2012qut}%
  \BibitemOpen
  \bibfield  {author} {\bibinfo {author} {\bibfnamefont {A.}~\bibnamefont
  {Accardi}} \emph {et~al.},\ }\href {\doibase 10.1140/epja/i2016-16268-9}
  {\bibfield  {journal} {\bibinfo  {journal} {Eur. Phys. J. A}\ }\textbf
  {\bibinfo {volume} {52}},\ \bibinfo {pages} {268} (\bibinfo {year} {2016})},\
  \Eprint {http://arxiv.org/abs/1212.1701} {arXiv:1212.1701 [nucl-ex]}
  \BibitemShut {NoStop}%
\bibitem [{\citenamefont {Dudek}\ \emph {et~al.}(2012)\citenamefont {Dudek}
  \emph {et~al.}}]{Dudek:2012vr}%
  \BibitemOpen
  \bibfield  {author} {\bibinfo {author} {\bibfnamefont {J.}~\bibnamefont
  {Dudek}} \emph {et~al.},\ }\href {\doibase 10.1140/epja/i2012-12187-1}
  {\bibfield  {journal} {\bibinfo  {journal} {Eur. Phys. J. A}\ }\textbf
  {\bibinfo {volume} {48}},\ \bibinfo {pages} {187} (\bibinfo {year} {2012})},\
  \Eprint {http://arxiv.org/abs/1208.1244} {arXiv:1208.1244 [hep-ex]}
  \BibitemShut {NoStop}%
\bibitem [{\citenamefont {Mueller}(1994)}]{Mueller:1993rr}%
  \BibitemOpen
  \bibfield  {author} {\bibinfo {author} {\bibfnamefont {A.~H.}\ \bibnamefont
  {Mueller}},\ }\href {\doibase 10.1016/0550-3213(94)90116-3} {\bibfield
  {journal} {\bibinfo  {journal} {Nucl. Phys. B}\ }\textbf {\bibinfo {volume}
  {415}},\ \bibinfo {pages} {373} (\bibinfo {year} {1994})}\BibitemShut
  {NoStop}%
\bibitem [{\citenamefont {Poggio}\ \emph {et~al.}(1976)\citenamefont {Poggio},
  \citenamefont {Quinn},\ and\ \citenamefont {Weinberg}}]{Poggio:1975af}%
  \BibitemOpen
  \bibfield  {author} {\bibinfo {author} {\bibfnamefont {E.~C.}\ \bibnamefont
  {Poggio}}, \bibinfo {author} {\bibfnamefont {H.~R.}\ \bibnamefont {Quinn}}, \
  and\ \bibinfo {author} {\bibfnamefont {S.}~\bibnamefont {Weinberg}},\ }\href
  {\doibase 10.1103/PhysRevD.13.1958} {\bibfield  {journal} {\bibinfo
  {journal} {Phys. Rev. D}\ }\textbf {\bibinfo {volume} {13}},\ \bibinfo
  {pages} {1958} (\bibinfo {year} {1976})}\BibitemShut {NoStop}%
\bibitem [{\citenamefont {Shifman}(2000)}]{Shifman:2000jv}%
  \BibitemOpen
  \bibfield  {author} {\bibinfo {author} {\bibfnamefont {M.~A.}\ \bibnamefont
  {Shifman}},\ }in\ \href {\doibase 10.1142/9789812810458_0032} {\emph
  {\bibinfo {booktitle} {{8th ISHFP}}}},\ Vol.~\bibinfo {volume} {3}\ (\bibinfo
   {publisher} {World Scientific},\ \bibinfo {address} {Singapore},\ \bibinfo
  {year} {2000})\ pp.\ \bibinfo {pages} {1447--1494},\ \Eprint
  {http://arxiv.org/abs/hep-ph/0009131} {arXiv:hep-ph/0009131} \BibitemShut
  {NoStop}%
\bibitem [{\citenamefont {Witten}(1979{\natexlab{a}})}]{Witten:1979kh}%
  \BibitemOpen
  \bibfield  {author} {\bibinfo {author} {\bibfnamefont {E.}~\bibnamefont
  {Witten}},\ }\href {\doibase 10.1016/0550-3213(79)90232-3} {\bibfield
  {journal} {\bibinfo  {journal} {Nucl. Phys. B}\ }\textbf {\bibinfo {volume}
  {160}},\ \bibinfo {pages} {57} (\bibinfo {year}
  {1979}{\natexlab{a}})}\BibitemShut {NoStop}%
\bibitem [{\citenamefont {Skyrme}(1962)}]{Skyrme:1962vh}%
  \BibitemOpen
  \bibfield  {author} {\bibinfo {author} {\bibfnamefont {T.~H.~R.}\
  \bibnamefont {Skyrme}},\ }\href {\doibase 10.1016/0029-5582(62)90775-7}
  {\bibfield  {journal} {\bibinfo  {journal} {Nucl. Phys.}\ }\textbf {\bibinfo
  {volume} {31}},\ \bibinfo {pages} {556} (\bibinfo {year} {1962})}\BibitemShut
  {NoStop}%
\bibitem [{\citenamefont {Witten}(1983)}]{Witten:1983tx}%
  \BibitemOpen
  \bibfield  {author} {\bibinfo {author} {\bibfnamefont {E.}~\bibnamefont
  {Witten}},\ }\href {\doibase 10.1016/0550-3213(83)90064-0} {\bibfield
  {journal} {\bibinfo  {journal} {Nucl. Phys. B}\ }\textbf {\bibinfo {volume}
  {223}},\ \bibinfo {pages} {433} (\bibinfo {year} {1983})}\BibitemShut
  {NoStop}%
\bibitem [{\citenamefont {Adkins}\ \emph {et~al.}(1983)\citenamefont {Adkins},
  \citenamefont {Nappi},\ and\ \citenamefont {Witten}}]{Adkins:1983ya}%
  \BibitemOpen
  \bibfield  {author} {\bibinfo {author} {\bibfnamefont {G.~S.}\ \bibnamefont
  {Adkins}}, \bibinfo {author} {\bibfnamefont {C.~R.}\ \bibnamefont {Nappi}}, \
  and\ \bibinfo {author} {\bibfnamefont {E.}~\bibnamefont {Witten}},\ }\href
  {\doibase 10.1016/0550-3213(83)90559-X} {\bibfield  {journal} {\bibinfo
  {journal} {Nucl. Phys. B}\ }\textbf {\bibinfo {volume} {228}},\ \bibinfo
  {pages} {552} (\bibinfo {year} {1983})}\BibitemShut {NoStop}%
\bibitem [{\citenamefont {Diakonov}\ \emph {et~al.}(1988)\citenamefont
  {Diakonov}, \citenamefont {Petrov},\ and\ \citenamefont
  {Pobylitsa}}]{Diakonov:1987ty}%
  \BibitemOpen
  \bibfield  {author} {\bibinfo {author} {\bibfnamefont {D.}~\bibnamefont
  {Diakonov}}, \bibinfo {author} {\bibfnamefont {V.~Y.}\ \bibnamefont
  {Petrov}}, \ and\ \bibinfo {author} {\bibfnamefont {P.~V.}\ \bibnamefont
  {Pobylitsa}},\ }\href {\doibase 10.1016/0550-3213(88)90443-9} {\bibfield
  {journal} {\bibinfo  {journal} {Nucl. Phys. B}\ }\textbf {\bibinfo {volume}
  {306}},\ \bibinfo {pages} {809} (\bibinfo {year} {1988})}\BibitemShut
  {NoStop}%
\bibitem [{\citenamefont {Diakonov}\ \emph {et~al.}(1996)\citenamefont
  {Diakonov}, \citenamefont {Petrov}, \citenamefont {Pobylitsa}, \citenamefont
  {Polyakov},\ and\ \citenamefont {Weiss}}]{Diakonov:1996sr}%
  \BibitemOpen
  \bibfield  {author} {\bibinfo {author} {\bibfnamefont {D.}~\bibnamefont
  {Diakonov}}, \bibinfo {author} {\bibfnamefont {V.}~\bibnamefont {Petrov}},
  \bibinfo {author} {\bibfnamefont {P.}~\bibnamefont {Pobylitsa}}, \bibinfo
  {author} {\bibfnamefont {M.~V.}\ \bibnamefont {Polyakov}}, \ and\ \bibinfo
  {author} {\bibfnamefont {C.}~\bibnamefont {Weiss}},\ }\href {\doibase
  10.1016/S0550-3213(96)00486-5} {\bibfield  {journal} {\bibinfo  {journal}
  {Nucl. Phys. B}\ }\textbf {\bibinfo {volume} {480}},\ \bibinfo {pages} {341}
  (\bibinfo {year} {1996})},\ \Eprint {http://arxiv.org/abs/hep-ph/9606314}
  {arXiv:hep-ph/9606314} \BibitemShut {NoStop}%
\bibitem [{\citenamefont {Diakonov}\ \emph {et~al.}(1997)\citenamefont
  {Diakonov}, \citenamefont {Petrov}, \citenamefont {Pobylitsa}, \citenamefont
  {Polyakov},\ and\ \citenamefont {Weiss}}]{Diakonov:1997vc}%
  \BibitemOpen
  \bibfield  {author} {\bibinfo {author} {\bibfnamefont {D.}~\bibnamefont
  {Diakonov}}, \bibinfo {author} {\bibfnamefont {V.~Y.}\ \bibnamefont
  {Petrov}}, \bibinfo {author} {\bibfnamefont {P.~V.}\ \bibnamefont
  {Pobylitsa}}, \bibinfo {author} {\bibfnamefont {M.~V.}\ \bibnamefont
  {Polyakov}}, \ and\ \bibinfo {author} {\bibfnamefont {C.}~\bibnamefont
  {Weiss}},\ }\href {\doibase 10.1103/PhysRevD.56.4069} {\bibfield  {journal}
  {\bibinfo  {journal} {Phys. Rev. D}\ }\textbf {\bibinfo {volume} {56}},\
  \bibinfo {pages} {4069} (\bibinfo {year} {1997})},\ \Eprint
  {http://arxiv.org/abs/hep-ph/9703420} {arXiv:hep-ph/9703420} \BibitemShut
  {NoStop}%
\bibitem [{\citenamefont {Weigel}\ \emph {et~al.}(1996)\citenamefont {Weigel},
  \citenamefont {Gamberg},\ and\ \citenamefont {Reinhardt}}]{Weigel:1996ef}%
  \BibitemOpen
  \bibfield  {author} {\bibinfo {author} {\bibfnamefont {H.}~\bibnamefont
  {Weigel}}, \bibinfo {author} {\bibfnamefont {L.~P.}\ \bibnamefont {Gamberg}},
  \ and\ \bibinfo {author} {\bibfnamefont {H.}~\bibnamefont {Reinhardt}},\
  }\href {\doibase 10.1142/S021773239600299X} {\bibfield  {journal} {\bibinfo
  {journal} {Mod. Phys. Lett. A}\ }\textbf {\bibinfo {volume} {11}},\ \bibinfo
  {pages} {3021} (\bibinfo {year} {1996})}\BibitemShut {NoStop}%
\bibitem [{\citenamefont {Gamberg}\ \emph {et~al.}(1998)\citenamefont
  {Gamberg}, \citenamefont {Reinhardt},\ and\ \citenamefont
  {Weigel}}]{Gamberg:1998vg}%
  \BibitemOpen
  \bibfield  {author} {\bibinfo {author} {\bibfnamefont {L.~P.}\ \bibnamefont
  {Gamberg}}, \bibinfo {author} {\bibfnamefont {H.}~\bibnamefont {Reinhardt}},
  \ and\ \bibinfo {author} {\bibfnamefont {H.}~\bibnamefont {Weigel}},\ }\href
  {\doibase 10.1103/PhysRevD.58.054014} {\bibfield  {journal} {\bibinfo
  {journal} {Phys. Rev. D}\ }\textbf {\bibinfo {volume} {58}},\ \bibinfo
  {pages} {054014} (\bibinfo {year} {1998})},\ \Eprint
  {http://arxiv.org/abs/hep-ph/9801379} {arXiv:hep-ph/9801379} \BibitemShut
  {NoStop}%
\bibitem [{\citenamefont {Aguilar}\ \emph {et~al.}(2019)\citenamefont {Aguilar}
  \emph {et~al.}}]{Aguilar:2019teb}%
  \BibitemOpen
  \bibfield  {author} {\bibinfo {author} {\bibfnamefont {A.~C.}\ \bibnamefont
  {Aguilar}} \emph {et~al.},\ }\href {\doibase 10.1140/epja/i2019-12885-0}
  {\bibfield  {journal} {\bibinfo  {journal} {Eur. Phys. J. A}\ }\textbf
  {\bibinfo {volume} {55}},\ \bibinfo {pages} {190} (\bibinfo {year} {2019})},\
  \Eprint {http://arxiv.org/abs/1907.08218} {arXiv:1907.08218 [nucl-ex]}
  \BibitemShut {NoStop}%
\bibitem [{\citenamefont {Adams}\ \emph {et~al.}(2018)\citenamefont {Adams}
  \emph {et~al.}}]{Adams:2018pwt}%
  \BibitemOpen
  \bibfield  {author} {\bibinfo {author} {\bibfnamefont {B.}~\bibnamefont
  {Adams}} \emph {et~al.},\ }\href@noop {} {\  (\bibinfo {year} {2018})},\
  \Eprint {http://arxiv.org/abs/1808.00848} {arXiv:1808.00848 [hep-ex]}
  \BibitemShut {NoStop}%
\bibitem [{\citenamefont {Roberts}\ \emph {et~al.}(2021)\citenamefont
  {Roberts}, \citenamefont {Richards}, \citenamefont {Horn},\ and\
  \citenamefont {Chang}}]{Roberts:2021nhw}%
  \BibitemOpen
  \bibfield  {author} {\bibinfo {author} {\bibfnamefont {C.~D.}\ \bibnamefont
  {Roberts}}, \bibinfo {author} {\bibfnamefont {D.~G.}\ \bibnamefont
  {Richards}}, \bibinfo {author} {\bibfnamefont {T.}~\bibnamefont {Horn}}, \
  and\ \bibinfo {author} {\bibfnamefont {L.}~\bibnamefont {Chang}},\ }\href
  {\doibase 10.1016/j.ppnp.2021.103883} {\bibfield  {journal} {\bibinfo
  {journal} {Prog. Part. Nucl. Phys.}\ }\textbf {\bibinfo {volume} {120}},\
  \bibinfo {pages} {103883} (\bibinfo {year} {2021})},\ \Eprint
  {http://arxiv.org/abs/2102.01765} {arXiv:2102.01765 [hep-ph]} \BibitemShut
  {NoStop}%
\bibitem [{\citenamefont {Sufian}\ \emph {et~al.}(2019)\citenamefont {Sufian},
  \citenamefont {Karpie}, \citenamefont {Egerer}, \citenamefont {Orginos},
  \citenamefont {Qiu},\ and\ \citenamefont {Richards}}]{Sufian:2019bol}%
  \BibitemOpen
  \bibfield  {author} {\bibinfo {author} {\bibfnamefont {R.~S.}\ \bibnamefont
  {Sufian}}, \bibinfo {author} {\bibfnamefont {J.}~\bibnamefont {Karpie}},
  \bibinfo {author} {\bibfnamefont {C.}~\bibnamefont {Egerer}}, \bibinfo
  {author} {\bibfnamefont {K.}~\bibnamefont {Orginos}}, \bibinfo {author}
  {\bibfnamefont {J.-W.}\ \bibnamefont {Qiu}}, \ and\ \bibinfo {author}
  {\bibfnamefont {D.~G.}\ \bibnamefont {Richards}},\ }\href {\doibase
  10.1103/PhysRevD.99.074507} {\bibfield  {journal} {\bibinfo  {journal} {Phys.
  Rev. D}\ }\textbf {\bibinfo {volume} {99}},\ \bibinfo {pages} {074507}
  (\bibinfo {year} {2019})},\ \Eprint {http://arxiv.org/abs/1901.03921}
  {arXiv:1901.03921 [hep-lat]} \BibitemShut {NoStop}%
\bibitem [{\citenamefont {Sufian}\ \emph {et~al.}(2020)\citenamefont {Sufian},
  \citenamefont {Egerer}, \citenamefont {Karpie}, \citenamefont {Edwards},
  \citenamefont {Jo\'o}, \citenamefont {Ma}, \citenamefont {Orginos},
  \citenamefont {Qiu},\ and\ \citenamefont {Richards}}]{Sufian:2020vzb}%
  \BibitemOpen
  \bibfield  {author} {\bibinfo {author} {\bibfnamefont {R.~S.}\ \bibnamefont
  {Sufian}}, \bibinfo {author} {\bibfnamefont {C.}~\bibnamefont {Egerer}},
  \bibinfo {author} {\bibfnamefont {J.}~\bibnamefont {Karpie}}, \bibinfo
  {author} {\bibfnamefont {R.~G.}\ \bibnamefont {Edwards}}, \bibinfo {author}
  {\bibfnamefont {B.}~\bibnamefont {Jo\'o}}, \bibinfo {author} {\bibfnamefont
  {Y.-Q.}\ \bibnamefont {Ma}}, \bibinfo {author} {\bibfnamefont
  {K.}~\bibnamefont {Orginos}}, \bibinfo {author} {\bibfnamefont {J.-W.}\
  \bibnamefont {Qiu}}, \ and\ \bibinfo {author} {\bibfnamefont {D.~G.}\
  \bibnamefont {Richards}},\ }\href {\doibase 10.1103/PhysRevD.102.054508}
  {\bibfield  {journal} {\bibinfo  {journal} {Phys. Rev. D}\ }\textbf {\bibinfo
  {volume} {102}},\ \bibinfo {pages} {054508} (\bibinfo {year} {2020})},\
  \Eprint {http://arxiv.org/abs/2001.04960} {arXiv:2001.04960 [hep-lat]}
  \BibitemShut {NoStop}%
\bibitem [{\citenamefont {Izubuchi}\ \emph {et~al.}(2019)\citenamefont
  {Izubuchi}, \citenamefont {Jin}, \citenamefont {Kallidonis}, \citenamefont
  {Karthik}, \citenamefont {Mukherjee}, \citenamefont {Petreczky},
  \citenamefont {Shugert},\ and\ \citenamefont {Syritsyn}}]{Izubuchi:2019lyk}%
  \BibitemOpen
  \bibfield  {author} {\bibinfo {author} {\bibfnamefont {T.}~\bibnamefont
  {Izubuchi}}, \bibinfo {author} {\bibfnamefont {L.}~\bibnamefont {Jin}},
  \bibinfo {author} {\bibfnamefont {C.}~\bibnamefont {Kallidonis}}, \bibinfo
  {author} {\bibfnamefont {N.}~\bibnamefont {Karthik}}, \bibinfo {author}
  {\bibfnamefont {S.}~\bibnamefont {Mukherjee}}, \bibinfo {author}
  {\bibfnamefont {P.}~\bibnamefont {Petreczky}}, \bibinfo {author}
  {\bibfnamefont {C.}~\bibnamefont {Shugert}}, \ and\ \bibinfo {author}
  {\bibfnamefont {S.}~\bibnamefont {Syritsyn}},\ }\href {\doibase
  10.1103/PhysRevD.100.034516} {\bibfield  {journal} {\bibinfo  {journal}
  {Phys. Rev. D}\ }\textbf {\bibinfo {volume} {100}},\ \bibinfo {pages}
  {034516} (\bibinfo {year} {2019})},\ \Eprint
  {http://arxiv.org/abs/1905.06349} {arXiv:1905.06349 [hep-lat]} \BibitemShut
  {NoStop}%
\bibitem [{\citenamefont {Gao}\ \emph {et~al.}(2020)\citenamefont {Gao},
  \citenamefont {Jin}, \citenamefont {Kallidonis}, \citenamefont {Karthik},
  \citenamefont {Mukherjee}, \citenamefont {Petreczky}, \citenamefont
  {Shugert}, \citenamefont {Syritsyn},\ and\ \citenamefont
  {Zhao}}]{Gao:2020ito}%
  \BibitemOpen
  \bibfield  {author} {\bibinfo {author} {\bibfnamefont {X.}~\bibnamefont
  {Gao}}, \bibinfo {author} {\bibfnamefont {L.}~\bibnamefont {Jin}}, \bibinfo
  {author} {\bibfnamefont {C.}~\bibnamefont {Kallidonis}}, \bibinfo {author}
  {\bibfnamefont {N.}~\bibnamefont {Karthik}}, \bibinfo {author} {\bibfnamefont
  {S.}~\bibnamefont {Mukherjee}}, \bibinfo {author} {\bibfnamefont
  {P.}~\bibnamefont {Petreczky}}, \bibinfo {author} {\bibfnamefont
  {C.}~\bibnamefont {Shugert}}, \bibinfo {author} {\bibfnamefont
  {S.}~\bibnamefont {Syritsyn}}, \ and\ \bibinfo {author} {\bibfnamefont
  {Y.}~\bibnamefont {Zhao}},\ }\href {\doibase 10.1103/PhysRevD.102.094513}
  {\bibfield  {journal} {\bibinfo  {journal} {Phys. Rev. D}\ }\textbf {\bibinfo
  {volume} {102}},\ \bibinfo {pages} {094513} (\bibinfo {year} {2020})},\
  \Eprint {http://arxiv.org/abs/2007.06590} {arXiv:2007.06590 [hep-lat]}
  \BibitemShut {NoStop}%
\bibitem [{\citenamefont {Gao}\ \emph {et~al.}(2021)\citenamefont {Gao},
  \citenamefont {Karthik}, \citenamefont {Mukherjee}, \citenamefont
  {Petreczky}, \citenamefont {Syritsyn},\ and\ \citenamefont
  {Zhao}}]{Gao:2021hvs}%
  \BibitemOpen
  \bibfield  {author} {\bibinfo {author} {\bibfnamefont {X.}~\bibnamefont
  {Gao}}, \bibinfo {author} {\bibfnamefont {N.}~\bibnamefont {Karthik}},
  \bibinfo {author} {\bibfnamefont {S.}~\bibnamefont {Mukherjee}}, \bibinfo
  {author} {\bibfnamefont {P.}~\bibnamefont {Petreczky}}, \bibinfo {author}
  {\bibfnamefont {S.}~\bibnamefont {Syritsyn}}, \ and\ \bibinfo {author}
  {\bibfnamefont {Y.}~\bibnamefont {Zhao}},\ }\href {\doibase
  10.1103/PhysRevD.103.094510} {\bibfield  {journal} {\bibinfo  {journal}
  {Phys. Rev. D}\ }\textbf {\bibinfo {volume} {103}},\ \bibinfo {pages}
  {094510} (\bibinfo {year} {2021})},\ \Eprint
  {http://arxiv.org/abs/2101.11632} {arXiv:2101.11632 [hep-lat]} \BibitemShut
  {NoStop}%
\bibitem [{\citenamefont {Karthik}(2021)}]{Karthik:2021qwz}%
  \BibitemOpen
  \bibfield  {author} {\bibinfo {author} {\bibfnamefont {N.}~\bibnamefont
  {Karthik}},\ }\href {\doibase 10.1103/PhysRevD.103.074512} {\bibfield
  {journal} {\bibinfo  {journal} {Phys. Rev. D}\ }\textbf {\bibinfo {volume}
  {103}},\ \bibinfo {pages} {074512} (\bibinfo {year} {2021})},\ \Eprint
  {http://arxiv.org/abs/2101.02224} {arXiv:2101.02224 [hep-lat]} \BibitemShut
  {NoStop}%
\bibitem [{\citenamefont {Lin}\ \emph {et~al.}(2021)\citenamefont {Lin},
  \citenamefont {Chen}, \citenamefont {Fan}, \citenamefont {Zhang},\ and\
  \citenamefont {Zhang}}]{Lin:2020ssv}%
  \BibitemOpen
  \bibfield  {author} {\bibinfo {author} {\bibfnamefont {H.-W.}\ \bibnamefont
  {Lin}}, \bibinfo {author} {\bibfnamefont {J.-W.}\ \bibnamefont {Chen}},
  \bibinfo {author} {\bibfnamefont {Z.}~\bibnamefont {Fan}}, \bibinfo {author}
  {\bibfnamefont {J.-H.}\ \bibnamefont {Zhang}}, \ and\ \bibinfo {author}
  {\bibfnamefont {R.}~\bibnamefont {Zhang}},\ }\href {\doibase
  10.1103/PhysRevD.103.014516} {\bibfield  {journal} {\bibinfo  {journal}
  {Phys. Rev. D}\ }\textbf {\bibinfo {volume} {103}},\ \bibinfo {pages}
  {014516} (\bibinfo {year} {2021})},\ \Eprint
  {http://arxiv.org/abs/2003.14128} {arXiv:2003.14128 [hep-lat]} \BibitemShut
  {NoStop}%
\bibitem [{\citenamefont {Gao}\ \emph {et~al.}(2022)\citenamefont {Gao},
  \citenamefont {Hanlon}, \citenamefont {Mukherjee}, \citenamefont {Petreczky},
  \citenamefont {Scior}, \citenamefont {Syritsyn},\ and\ \citenamefont
  {Zhao}}]{Gao:2021dbh}%
  \BibitemOpen
  \bibfield  {author} {\bibinfo {author} {\bibfnamefont {X.}~\bibnamefont
  {Gao}}, \bibinfo {author} {\bibfnamefont {A.~D.}\ \bibnamefont {Hanlon}},
  \bibinfo {author} {\bibfnamefont {S.}~\bibnamefont {Mukherjee}}, \bibinfo
  {author} {\bibfnamefont {P.}~\bibnamefont {Petreczky}}, \bibinfo {author}
  {\bibfnamefont {P.}~\bibnamefont {Scior}}, \bibinfo {author} {\bibfnamefont
  {S.}~\bibnamefont {Syritsyn}}, \ and\ \bibinfo {author} {\bibfnamefont
  {Y.}~\bibnamefont {Zhao}},\ }\href {\doibase 10.1103/PhysRevLett.128.142003}
  {\bibfield  {journal} {\bibinfo  {journal} {Phys. Rev. Lett.}\ }\textbf
  {\bibinfo {volume} {128}},\ \bibinfo {pages} {142003} (\bibinfo {year}
  {2022})},\ \Eprint {http://arxiv.org/abs/2112.02208} {arXiv:2112.02208
  [hep-lat]} \BibitemShut {NoStop}%
\bibitem [{\citenamefont {Detmold}\ \emph {et~al.}(2022)\citenamefont
  {Detmold}, \citenamefont {Grebe}, \citenamefont {Kanamori}, \citenamefont
  {Lin}, \citenamefont {Mondal}, \citenamefont {Perry},\ and\ \citenamefont
  {Zhao}}]{Detmold:2021qln}%
  \BibitemOpen
  \bibfield  {author} {\bibinfo {author} {\bibfnamefont {W.}~\bibnamefont
  {Detmold}}, \bibinfo {author} {\bibfnamefont {A.~V.}\ \bibnamefont {Grebe}},
  \bibinfo {author} {\bibfnamefont {I.}~\bibnamefont {Kanamori}}, \bibinfo
  {author} {\bibfnamefont {C.~J.~D.}\ \bibnamefont {Lin}}, \bibinfo {author}
  {\bibfnamefont {S.}~\bibnamefont {Mondal}}, \bibinfo {author} {\bibfnamefont
  {R.~J.}\ \bibnamefont {Perry}}, \ and\ \bibinfo {author} {\bibfnamefont
  {Y.}~\bibnamefont {Zhao}} (\bibinfo {collaboration} {HOPE}),\ }\href
  {\doibase 10.1103/PhysRevD.105.034506} {\bibfield  {journal} {\bibinfo
  {journal} {Phys. Rev. D}\ }\textbf {\bibinfo {volume} {105}},\ \bibinfo
  {pages} {034506} (\bibinfo {year} {2022})},\ \Eprint
  {http://arxiv.org/abs/2109.15241} {arXiv:2109.15241 [hep-lat]} \BibitemShut
  {NoStop}%
\bibitem [{\citenamefont {Hua}\ \emph {et~al.}(2022)\citenamefont {Hua} \emph
  {et~al.}}]{Hua:2022kcm}%
  \BibitemOpen
  \bibfield  {author} {\bibinfo {author} {\bibfnamefont {J.}~\bibnamefont
  {Hua}} \emph {et~al.},\ }\href@noop {} {\  (\bibinfo {year} {2022})},\
  \Eprint {http://arxiv.org/abs/2201.09173} {arXiv:2201.09173 [hep-lat]}
  \BibitemShut {NoStop}%
\bibitem [{\citenamefont {Ruiz~Arriola}\ and\ \citenamefont
  {Broniowski}(2006)}]{RuizArriola:2006jge}%
  \BibitemOpen
  \bibfield  {author} {\bibinfo {author} {\bibfnamefont {E.}~\bibnamefont
  {Ruiz~Arriola}}\ and\ \bibinfo {author} {\bibfnamefont {W.}~\bibnamefont
  {Broniowski}},\ }\href {\doibase 10.1103/PhysRevD.74.034008} {\bibfield
  {journal} {\bibinfo  {journal} {Phys. Rev. D}\ }\textbf {\bibinfo {volume}
  {74}},\ \bibinfo {pages} {034008} (\bibinfo {year} {2006})},\ \Eprint
  {http://arxiv.org/abs/hep-ph/0605318} {arXiv:hep-ph/0605318} \BibitemShut
  {NoStop}%
\bibitem [{\citenamefont {Braun}\ and\ \citenamefont
  {M\"uller}(2008)}]{Braun:2007wv}%
  \BibitemOpen
  \bibfield  {author} {\bibinfo {author} {\bibfnamefont {V.}~\bibnamefont
  {Braun}}\ and\ \bibinfo {author} {\bibfnamefont {D.}~\bibnamefont
  {M\"uller}},\ }\href {\doibase 10.1140/epjc/s10052-008-0608-4} {\bibfield
  {journal} {\bibinfo  {journal} {Eur. Phys. J. C}\ }\textbf {\bibinfo {volume}
  {55}},\ \bibinfo {pages} {349} (\bibinfo {year} {2008})},\ \Eprint
  {http://arxiv.org/abs/0709.1348} {arXiv:0709.1348 [hep-ph]} \BibitemShut
  {NoStop}%
\bibitem [{\citenamefont {Ji}(2013)}]{Ji:2013dva}%
  \BibitemOpen
  \bibfield  {author} {\bibinfo {author} {\bibfnamefont {X.}~\bibnamefont
  {Ji}},\ }\href {\doibase 10.1103/PhysRevLett.110.262002} {\bibfield
  {journal} {\bibinfo  {journal} {Phys. Rev. Lett.}\ }\textbf {\bibinfo
  {volume} {110}},\ \bibinfo {pages} {262002} (\bibinfo {year} {2013})},\
  \Eprint {http://arxiv.org/abs/1305.1539} {arXiv:1305.1539 [hep-ph]}
  \BibitemShut {NoStop}%
\bibitem [{\citenamefont {Radyushkin}(2017)}]{Radyushkin:2017cyf}%
  \BibitemOpen
  \bibfield  {author} {\bibinfo {author} {\bibfnamefont {A.~V.}\ \bibnamefont
  {Radyushkin}},\ }\href {\doibase 10.1103/PhysRevD.96.034025} {\bibfield
  {journal} {\bibinfo  {journal} {Phys. Rev. D}\ }\textbf {\bibinfo {volume}
  {96}},\ \bibinfo {pages} {034025} (\bibinfo {year} {2017})},\ \Eprint
  {http://arxiv.org/abs/1705.01488} {arXiv:1705.01488 [hep-ph]} \BibitemShut
  {NoStop}%
\bibitem [{\citenamefont {Ma}\ and\ \citenamefont {Qiu}(2018)}]{Ma:2014jla}%
  \BibitemOpen
  \bibfield  {author} {\bibinfo {author} {\bibfnamefont {Y.-Q.}\ \bibnamefont
  {Ma}}\ and\ \bibinfo {author} {\bibfnamefont {J.-W.}\ \bibnamefont {Qiu}},\
  }\href {\doibase 10.1103/PhysRevD.98.074021} {\bibfield  {journal} {\bibinfo
  {journal} {Phys. Rev. D}\ }\textbf {\bibinfo {volume} {98}},\ \bibinfo
  {pages} {074021} (\bibinfo {year} {2018})},\ \Eprint
  {http://arxiv.org/abs/1404.6860} {arXiv:1404.6860 [hep-ph]} \BibitemShut
  {NoStop}%
\bibitem [{\citenamefont {Gross}\ and\ \citenamefont
  {Witten}(1980)}]{Gross:1980he}%
  \BibitemOpen
  \bibfield  {author} {\bibinfo {author} {\bibfnamefont {D.~J.}\ \bibnamefont
  {Gross}}\ and\ \bibinfo {author} {\bibfnamefont {E.}~\bibnamefont {Witten}},\
  }\href {\doibase 10.1103/PhysRevD.21.446} {\bibfield  {journal} {\bibinfo
  {journal} {Phys. Rev. D}\ }\textbf {\bibinfo {volume} {21}},\ \bibinfo
  {pages} {446} (\bibinfo {year} {1980})}\BibitemShut {NoStop}%
\bibitem [{\citenamefont {Eguchi}\ and\ \citenamefont
  {Kawai}(1982)}]{Eguchi:1982nm}%
  \BibitemOpen
  \bibfield  {author} {\bibinfo {author} {\bibfnamefont {T.}~\bibnamefont
  {Eguchi}}\ and\ \bibinfo {author} {\bibfnamefont {H.}~\bibnamefont {Kawai}},\
  }\href {\doibase 10.1103/PhysRevLett.48.1063} {\bibfield  {journal} {\bibinfo
   {journal} {Phys. Rev. Lett.}\ }\textbf {\bibinfo {volume} {48}},\ \bibinfo
  {pages} {1063} (\bibinfo {year} {1982})}\BibitemShut {NoStop}%
\bibitem [{\citenamefont {Narayanan}(2009)}]{Narayanan:2009xh}%
  \BibitemOpen
  \bibfield  {author} {\bibinfo {author} {\bibfnamefont {R.}~\bibnamefont
  {Narayanan}},\ }\href@noop {} {\bibfield  {journal} {\bibinfo  {journal}
  {Acta Phys. Polon. B}\ }\textbf {\bibinfo {volume} {40}},\ \bibinfo {pages}
  {3231} (\bibinfo {year} {2009})},\ \Eprint {http://arxiv.org/abs/0910.3711}
  {arXiv:0910.3711 [hep-lat]} \BibitemShut {NoStop}%
\bibitem [{\citenamefont {Narayanan}\ and\ \citenamefont
  {Neuberger}(2007)}]{Narayanan:2007fb}%
  \BibitemOpen
  \bibfield  {author} {\bibinfo {author} {\bibfnamefont {R.}~\bibnamefont
  {Narayanan}}\ and\ \bibinfo {author} {\bibfnamefont {H.}~\bibnamefont
  {Neuberger}},\ }\href {\doibase 10.22323/1.042.0020} {\bibfield  {journal}
  {\bibinfo  {journal} {PoS}\ }\textbf {\bibinfo {volume} {LATTICE2007}},\
  \bibinfo {pages} {020} (\bibinfo {year} {2007})},\ \Eprint
  {http://arxiv.org/abs/0710.0098} {arXiv:0710.0098 [hep-lat]} \BibitemShut
  {NoStop}%
\bibitem [{\citenamefont {Narayanan}\ \emph {et~al.}(2007)\citenamefont
  {Narayanan}, \citenamefont {Neuberger},\ and\ \citenamefont
  {Reynoso}}]{Narayanan:2007ug}%
  \BibitemOpen
  \bibfield  {author} {\bibinfo {author} {\bibfnamefont {R.}~\bibnamefont
  {Narayanan}}, \bibinfo {author} {\bibfnamefont {H.}~\bibnamefont
  {Neuberger}}, \ and\ \bibinfo {author} {\bibfnamefont {F.}~\bibnamefont
  {Reynoso}},\ }\href {\doibase 10.1016/j.physletb.2007.06.016} {\bibfield
  {journal} {\bibinfo  {journal} {Phys. Lett. B}\ }\textbf {\bibinfo {volume}
  {651}},\ \bibinfo {pages} {246} (\bibinfo {year} {2007})},\ \Eprint
  {http://arxiv.org/abs/0704.2591} {arXiv:0704.2591 [hep-lat]} \BibitemShut
  {NoStop}%
\bibitem [{\citenamefont {Narayanan}\ and\ \citenamefont
  {Neuberger}(2003)}]{Narayanan:2003fc}%
  \BibitemOpen
  \bibfield  {author} {\bibinfo {author} {\bibfnamefont {R.}~\bibnamefont
  {Narayanan}}\ and\ \bibinfo {author} {\bibfnamefont {H.}~\bibnamefont
  {Neuberger}},\ }\href {\doibase 10.1103/PhysRevLett.91.081601} {\bibfield
  {journal} {\bibinfo  {journal} {Phys. Rev. Lett.}\ }\textbf {\bibinfo
  {volume} {91}},\ \bibinfo {pages} {081601} (\bibinfo {year} {2003})},\
  \Eprint {http://arxiv.org/abs/hep-lat/0303023} {arXiv:hep-lat/0303023}
  \BibitemShut {NoStop}%
\bibitem [{\citenamefont {Kiskis}\ \emph {et~al.}(2003)\citenamefont {Kiskis},
  \citenamefont {Narayanan},\ and\ \citenamefont {Neuberger}}]{Kiskis:2003rd}%
  \BibitemOpen
  \bibfield  {author} {\bibinfo {author} {\bibfnamefont {J.}~\bibnamefont
  {Kiskis}}, \bibinfo {author} {\bibfnamefont {R.}~\bibnamefont {Narayanan}}, \
  and\ \bibinfo {author} {\bibfnamefont {H.}~\bibnamefont {Neuberger}},\ }\href
  {\doibase 10.1016/j.physletb.2003.08.070} {\bibfield  {journal} {\bibinfo
  {journal} {Phys. Lett. B}\ }\textbf {\bibinfo {volume} {574}},\ \bibinfo
  {pages} {65} (\bibinfo {year} {2003})},\ \Eprint
  {http://arxiv.org/abs/hep-lat/0308033} {arXiv:hep-lat/0308033} \BibitemShut
  {NoStop}%
\bibitem [{\citenamefont {Izubuchi}\ \emph {et~al.}(2018)\citenamefont
  {Izubuchi}, \citenamefont {Ji}, \citenamefont {Jin}, \citenamefont
  {Stewart},\ and\ \citenamefont {Zhao}}]{Izubuchi:2018srq}%
  \BibitemOpen
  \bibfield  {author} {\bibinfo {author} {\bibfnamefont {T.}~\bibnamefont
  {Izubuchi}}, \bibinfo {author} {\bibfnamefont {X.}~\bibnamefont {Ji}},
  \bibinfo {author} {\bibfnamefont {L.}~\bibnamefont {Jin}}, \bibinfo {author}
  {\bibfnamefont {I.~W.}\ \bibnamefont {Stewart}}, \ and\ \bibinfo {author}
  {\bibfnamefont {Y.}~\bibnamefont {Zhao}},\ }\href {\doibase
  10.1103/PhysRevD.98.056004} {\bibfield  {journal} {\bibinfo  {journal} {Phys.
  Rev. D}\ }\textbf {\bibinfo {volume} {98}},\ \bibinfo {pages} {056004}
  (\bibinfo {year} {2018})},\ \Eprint {http://arxiv.org/abs/1801.03917}
  {arXiv:1801.03917 [hep-ph]} \BibitemShut {NoStop}%
\bibitem [{\citenamefont {Maiani}\ \emph {et~al.}(1987)\citenamefont {Maiani},
  \citenamefont {Martinelli}, \citenamefont {Paciello},\ and\ \citenamefont
  {Taglienti}}]{Maiani:1987by}%
  \BibitemOpen
  \bibfield  {author} {\bibinfo {author} {\bibfnamefont {L.}~\bibnamefont
  {Maiani}}, \bibinfo {author} {\bibfnamefont {G.}~\bibnamefont {Martinelli}},
  \bibinfo {author} {\bibfnamefont {M.~L.}\ \bibnamefont {Paciello}}, \ and\
  \bibinfo {author} {\bibfnamefont {B.}~\bibnamefont {Taglienti}},\ }\href
  {\doibase 10.1016/0550-3213(87)90078-2} {\bibfield  {journal} {\bibinfo
  {journal} {Nucl. Phys. B}\ }\textbf {\bibinfo {volume} {293}},\ \bibinfo
  {pages} {420} (\bibinfo {year} {1987})}\BibitemShut {NoStop}%
\bibitem [{\citenamefont {Kiskis}\ and\ \citenamefont
  {Narayanan}(2009)}]{Kiskis:2009rf}%
  \BibitemOpen
  \bibfield  {author} {\bibinfo {author} {\bibfnamefont {J.}~\bibnamefont
  {Kiskis}}\ and\ \bibinfo {author} {\bibfnamefont {R.}~\bibnamefont
  {Narayanan}},\ }\href {\doibase 10.1016/j.physletb.2009.10.043} {\bibfield
  {journal} {\bibinfo  {journal} {Phys. Lett. B}\ }\textbf {\bibinfo {volume}
  {681}},\ \bibinfo {pages} {372} (\bibinfo {year} {2009})},\ \Eprint
  {http://arxiv.org/abs/0908.1451} {arXiv:0908.1451 [hep-lat]} \BibitemShut
  {NoStop}%
\bibitem [{\citenamefont {Lucini}\ \emph {et~al.}(2004)\citenamefont {Lucini},
  \citenamefont {Teper},\ and\ \citenamefont {Wenger}}]{Lucini:2003zr}%
  \BibitemOpen
  \bibfield  {author} {\bibinfo {author} {\bibfnamefont {B.}~\bibnamefont
  {Lucini}}, \bibinfo {author} {\bibfnamefont {M.}~\bibnamefont {Teper}}, \
  and\ \bibinfo {author} {\bibfnamefont {U.}~\bibnamefont {Wenger}},\ }\href
  {\doibase 10.1088/1126-6708/2004/01/061} {\bibfield  {journal} {\bibinfo
  {journal} {JHEP}\ }\textbf {\bibinfo {volume} {01}},\ \bibinfo {pages} {061}
  (\bibinfo {year} {2004})},\ \Eprint {http://arxiv.org/abs/hep-lat/0307017}
  {arXiv:hep-lat/0307017} \BibitemShut {NoStop}%
\bibitem [{\citenamefont {Ishikawa}\ \emph {et~al.}(2017)\citenamefont
  {Ishikawa}, \citenamefont {Ma}, \citenamefont {Qiu},\ and\ \citenamefont
  {Yoshida}}]{Ishikawa:2017faj}%
  \BibitemOpen
  \bibfield  {author} {\bibinfo {author} {\bibfnamefont {T.}~\bibnamefont
  {Ishikawa}}, \bibinfo {author} {\bibfnamefont {Y.-Q.}\ \bibnamefont {Ma}},
  \bibinfo {author} {\bibfnamefont {J.-W.}\ \bibnamefont {Qiu}}, \ and\
  \bibinfo {author} {\bibfnamefont {S.}~\bibnamefont {Yoshida}},\ }\href
  {\doibase 10.1103/PhysRevD.96.094019} {\bibfield  {journal} {\bibinfo
  {journal} {Phys. Rev. D}\ }\textbf {\bibinfo {volume} {96}},\ \bibinfo
  {pages} {094019} (\bibinfo {year} {2017})},\ \Eprint
  {http://arxiv.org/abs/1707.03107} {arXiv:1707.03107 [hep-ph]} \BibitemShut
  {NoStop}%
\bibitem [{\citenamefont {Ji}\ \emph {et~al.}(2018)\citenamefont {Ji},
  \citenamefont {Zhang},\ and\ \citenamefont {Zhao}}]{Ji:2017oey}%
  \BibitemOpen
  \bibfield  {author} {\bibinfo {author} {\bibfnamefont {X.}~\bibnamefont
  {Ji}}, \bibinfo {author} {\bibfnamefont {J.-H.}\ \bibnamefont {Zhang}}, \
  and\ \bibinfo {author} {\bibfnamefont {Y.}~\bibnamefont {Zhao}},\ }\href
  {\doibase 10.1103/PhysRevLett.120.112001} {\bibfield  {journal} {\bibinfo
  {journal} {Phys. Rev. Lett.}\ }\textbf {\bibinfo {volume} {120}},\ \bibinfo
  {pages} {112001} (\bibinfo {year} {2018})},\ \Eprint
  {http://arxiv.org/abs/1706.08962} {arXiv:1706.08962 [hep-ph]} \BibitemShut
  {NoStop}%
\bibitem [{\citenamefont {Orginos}\ \emph {et~al.}(2017)\citenamefont
  {Orginos}, \citenamefont {Radyushkin}, \citenamefont {Karpie},\ and\
  \citenamefont {Zafeiropoulos}}]{Orginos:2017kos}%
  \BibitemOpen
  \bibfield  {author} {\bibinfo {author} {\bibfnamefont {K.}~\bibnamefont
  {Orginos}}, \bibinfo {author} {\bibfnamefont {A.}~\bibnamefont {Radyushkin}},
  \bibinfo {author} {\bibfnamefont {J.}~\bibnamefont {Karpie}}, \ and\ \bibinfo
  {author} {\bibfnamefont {S.}~\bibnamefont {Zafeiropoulos}},\ }\href {\doibase
  10.1103/PhysRevD.96.094503} {\bibfield  {journal} {\bibinfo  {journal} {Phys.
  Rev. D}\ }\textbf {\bibinfo {volume} {96}},\ \bibinfo {pages} {094503}
  (\bibinfo {year} {2017})},\ \Eprint {http://arxiv.org/abs/1706.05373}
  {arXiv:1706.05373 [hep-ph]} \BibitemShut {NoStop}%
\bibitem [{\citenamefont {Allton}\ \emph {et~al.}(2008)\citenamefont {Allton},
  \citenamefont {Teper},\ and\ \citenamefont {Trivini}}]{Allton:2008ty}%
  \BibitemOpen
  \bibfield  {author} {\bibinfo {author} {\bibfnamefont {C.}~\bibnamefont
  {Allton}}, \bibinfo {author} {\bibfnamefont {M.}~\bibnamefont {Teper}}, \
  and\ \bibinfo {author} {\bibfnamefont {A.}~\bibnamefont {Trivini}},\ }\href
  {\doibase 10.1088/1126-6708/2008/07/021} {\bibfield  {journal} {\bibinfo
  {journal} {JHEP}\ }\textbf {\bibinfo {volume} {07}},\ \bibinfo {pages} {021}
  (\bibinfo {year} {2008})},\ \Eprint {http://arxiv.org/abs/0803.1092}
  {arXiv:0803.1092 [hep-lat]} \BibitemShut {NoStop}%
\bibitem [{\citenamefont {Datta}\ and\ \citenamefont
  {Gupta}(2009)}]{Datta:2009ef}%
  \BibitemOpen
  \bibfield  {author} {\bibinfo {author} {\bibfnamefont {S.}~\bibnamefont
  {Datta}}\ and\ \bibinfo {author} {\bibfnamefont {S.}~\bibnamefont {Gupta}},\
  }\href {\doibase 10.22323/1.091.0178} {\bibfield  {journal} {\bibinfo
  {journal} {PoS}\ }\textbf {\bibinfo {volume} {LAT2009}},\ \bibinfo {pages}
  {178} (\bibinfo {year} {2009})},\ \Eprint {http://arxiv.org/abs/0910.2889}
  {arXiv:0910.2889 [hep-lat]} \BibitemShut {NoStop}%
\bibitem [{\citenamefont {Barry}\ \emph {et~al.}(2021)\citenamefont {Barry},
  \citenamefont {Ji}, \citenamefont {Sato},\ and\ \citenamefont
  {Melnitchouk}}]{Barry:2021osv}%
  \BibitemOpen
  \bibfield  {author} {\bibinfo {author} {\bibfnamefont {P.~C.}\ \bibnamefont
  {Barry}}, \bibinfo {author} {\bibfnamefont {C.-R.}\ \bibnamefont {Ji}},
  \bibinfo {author} {\bibfnamefont {N.}~\bibnamefont {Sato}}, \ and\ \bibinfo
  {author} {\bibfnamefont {W.}~\bibnamefont {Melnitchouk}} (\bibinfo
  {collaboration} {Jefferson Lab Angular Momentum (JAM)}),\ }\href {\doibase
  10.1103/PhysRevLett.127.232001} {\bibfield  {journal} {\bibinfo  {journal}
  {Phys. Rev. Lett.}\ }\textbf {\bibinfo {volume} {127}},\ \bibinfo {pages}
  {232001} (\bibinfo {year} {2021})},\ \Eprint
  {http://arxiv.org/abs/2108.05822} {arXiv:2108.05822 [hep-ph]} \BibitemShut
  {NoStop}%
\bibitem [{\citenamefont {Barry}\ \emph {et~al.}(2018)\citenamefont {Barry},
  \citenamefont {Sato}, \citenamefont {Melnitchouk},\ and\ \citenamefont
  {Ji}}]{Barry:2018ort}%
  \BibitemOpen
  \bibfield  {author} {\bibinfo {author} {\bibfnamefont {P.~C.}\ \bibnamefont
  {Barry}}, \bibinfo {author} {\bibfnamefont {N.}~\bibnamefont {Sato}},
  \bibinfo {author} {\bibfnamefont {W.}~\bibnamefont {Melnitchouk}}, \ and\
  \bibinfo {author} {\bibfnamefont {C.-R.}\ \bibnamefont {Ji}},\ }\href
  {\doibase 10.1103/PhysRevLett.121.152001} {\bibfield  {journal} {\bibinfo
  {journal} {Phys. Rev. Lett.}\ }\textbf {\bibinfo {volume} {121}},\ \bibinfo
  {pages} {152001} (\bibinfo {year} {2018})},\ \Eprint
  {http://arxiv.org/abs/1804.01965} {arXiv:1804.01965 [hep-ph]} \BibitemShut
  {NoStop}%
\bibitem [{\citenamefont {Teper}(1997)}]{Teper:1997am}%
  \BibitemOpen
  \bibfield  {author} {\bibinfo {author} {\bibfnamefont {M.~J.}\ \bibnamefont
  {Teper}},\ }in\ \href@noop {} {\emph {\bibinfo {booktitle} {{NATO Advanced
  Study Institute on Confinement, Duality and Nonperturbative Aspects of
  QCD}}}}\ (\bibinfo {year} {1997})\ pp.\ \bibinfo {pages} {43--74},\ \Eprint
  {http://arxiv.org/abs/hep-lat/9711011} {arXiv:hep-lat/9711011} \BibitemShut
  {NoStop}%
\bibitem [{\citenamefont {Gonz\'alez-Arroyo}\ and\ \citenamefont
  {Okawa}(2016)}]{Gonzalez-Arroyo:2015bya}%
  \BibitemOpen
  \bibfield  {author} {\bibinfo {author} {\bibfnamefont {A.}~\bibnamefont
  {Gonz\'alez-Arroyo}}\ and\ \bibinfo {author} {\bibfnamefont {M.}~\bibnamefont
  {Okawa}},\ }\href {\doibase 10.1016/j.physletb.2016.02.001} {\bibfield
  {journal} {\bibinfo  {journal} {Phys. Lett. B}\ }\textbf {\bibinfo {volume}
  {755}},\ \bibinfo {pages} {132} (\bibinfo {year} {2016})},\ \Eprint
  {http://arxiv.org/abs/1510.05428} {arXiv:1510.05428 [hep-lat]} \BibitemShut
  {NoStop}%
\bibitem [{\citenamefont {Bars}\ and\ \citenamefont
  {Green}(1978)}]{Bars:1977ud}%
  \BibitemOpen
  \bibfield  {author} {\bibinfo {author} {\bibfnamefont {I.}~\bibnamefont
  {Bars}}\ and\ \bibinfo {author} {\bibfnamefont {M.~B.}\ \bibnamefont
  {Green}},\ }\href {\doibase 10.1103/PhysRevD.17.537} {\bibfield  {journal}
  {\bibinfo  {journal} {Phys. Rev. D}\ }\textbf {\bibinfo {volume} {17}},\
  \bibinfo {pages} {537} (\bibinfo {year} {1978})}\BibitemShut {NoStop}%
\bibitem [{\citenamefont {Jia}\ \emph {et~al.}(2017)\citenamefont {Jia},
  \citenamefont {Liang}, \citenamefont {Li},\ and\ \citenamefont
  {Xiong}}]{Jia:2017uul}%
  \BibitemOpen
  \bibfield  {author} {\bibinfo {author} {\bibfnamefont {Y.}~\bibnamefont
  {Jia}}, \bibinfo {author} {\bibfnamefont {S.}~\bibnamefont {Liang}}, \bibinfo
  {author} {\bibfnamefont {L.}~\bibnamefont {Li}}, \ and\ \bibinfo {author}
  {\bibfnamefont {X.}~\bibnamefont {Xiong}},\ }\href {\doibase
  10.1007/JHEP11(2017)151} {\bibfield  {journal} {\bibinfo  {journal} {JHEP}\
  }\textbf {\bibinfo {volume} {11}},\ \bibinfo {pages} {151} (\bibinfo {year}
  {2017})},\ \Eprint {http://arxiv.org/abs/1708.09379} {arXiv:1708.09379
  [hep-ph]} \BibitemShut {NoStop}%
\bibitem [{\citenamefont {Jia}\ \emph {et~al.}(2018)\citenamefont {Jia},
  \citenamefont {Liang}, \citenamefont {Xiong},\ and\ \citenamefont
  {Yu}}]{Jia:2018qee}%
  \BibitemOpen
  \bibfield  {author} {\bibinfo {author} {\bibfnamefont {Y.}~\bibnamefont
  {Jia}}, \bibinfo {author} {\bibfnamefont {S.}~\bibnamefont {Liang}}, \bibinfo
  {author} {\bibfnamefont {X.}~\bibnamefont {Xiong}}, \ and\ \bibinfo {author}
  {\bibfnamefont {R.}~\bibnamefont {Yu}},\ }\href {\doibase
  10.1103/PhysRevD.98.054011} {\bibfield  {journal} {\bibinfo  {journal} {Phys.
  Rev. D}\ }\textbf {\bibinfo {volume} {98}},\ \bibinfo {pages} {054011}
  (\bibinfo {year} {2018})},\ \Eprint {http://arxiv.org/abs/1804.04644}
  {arXiv:1804.04644 [hep-th]} \BibitemShut {NoStop}%
\bibitem [{\citenamefont {Burkardt}(2000)}]{Burkardt:2000uu}%
  \BibitemOpen
  \bibfield  {author} {\bibinfo {author} {\bibfnamefont {M.}~\bibnamefont
  {Burkardt}},\ }\href {\doibase 10.1103/PhysRevD.62.094003} {\bibfield
  {journal} {\bibinfo  {journal} {Phys. Rev. D}\ }\textbf {\bibinfo {volume}
  {62}},\ \bibinfo {pages} {094003} (\bibinfo {year} {2000})},\ \Eprint
  {http://arxiv.org/abs/hep-ph/0005209} {arXiv:hep-ph/0005209} \BibitemShut
  {NoStop}%
\bibitem [{\citenamefont {Witten}(1979{\natexlab{b}})}]{Witten:1978bc}%
  \BibitemOpen
  \bibfield  {author} {\bibinfo {author} {\bibfnamefont {E.}~\bibnamefont
  {Witten}},\ }\href {\doibase 10.1016/0550-3213(79)90243-8} {\bibfield
  {journal} {\bibinfo  {journal} {Nucl. Phys. B}\ }\textbf {\bibinfo {volume}
  {149}},\ \bibinfo {pages} {285} (\bibinfo {year}
  {1979}{\natexlab{b}})}\BibitemShut {NoStop}%
\bibitem [{\citenamefont {Teper}(1980)}]{Teper:1979tq}%
  \BibitemOpen
  \bibfield  {author} {\bibinfo {author} {\bibfnamefont {M.~J.}\ \bibnamefont
  {Teper}},\ }\href {\doibase 10.1007/BF01421781} {\bibfield  {journal}
  {\bibinfo  {journal} {Z. Phys. C}\ }\textbf {\bibinfo {volume} {5}},\
  \bibinfo {pages} {233} (\bibinfo {year} {1980})}\BibitemShut {NoStop}%
\bibitem [{\citenamefont {Lucini}\ and\ \citenamefont
  {Teper}(2001)}]{Lucini:2001ej}%
  \BibitemOpen
  \bibfield  {author} {\bibinfo {author} {\bibfnamefont {B.}~\bibnamefont
  {Lucini}}\ and\ \bibinfo {author} {\bibfnamefont {M.}~\bibnamefont {Teper}},\
  }\href {\doibase 10.1088/1126-6708/2001/06/050} {\bibfield  {journal}
  {\bibinfo  {journal} {JHEP}\ }\textbf {\bibinfo {volume} {06}},\ \bibinfo
  {pages} {050} (\bibinfo {year} {2001})},\ \Eprint
  {http://arxiv.org/abs/hep-lat/0103027} {arXiv:hep-lat/0103027} \BibitemShut
  {NoStop}%
\bibitem [{\citenamefont {Nason}\ and\ \citenamefont
  {Palassini}(1995)}]{Nason:1994xw}%
  \BibitemOpen
  \bibfield  {author} {\bibinfo {author} {\bibfnamefont {P.}~\bibnamefont
  {Nason}}\ and\ \bibinfo {author} {\bibfnamefont {M.}~\bibnamefont
  {Palassini}},\ }\href {\doibase 10.1016/0550-3213(95)00082-4} {\bibfield
  {journal} {\bibinfo  {journal} {Nucl. Phys. B}\ }\textbf {\bibinfo {volume}
  {444}},\ \bibinfo {pages} {310} (\bibinfo {year} {1995})},\ \Eprint
  {http://arxiv.org/abs/hep-ph/9411246} {arXiv:hep-ph/9411246} \BibitemShut
  {NoStop}%
\bibitem [{\citenamefont {Shifman}\ \emph {et~al.}(1979)\citenamefont
  {Shifman}, \citenamefont {Vainshtein},\ and\ \citenamefont
  {Zakharov}}]{Shifman:1978bx}%
  \BibitemOpen
  \bibfield  {author} {\bibinfo {author} {\bibfnamefont {M.~A.}\ \bibnamefont
  {Shifman}}, \bibinfo {author} {\bibfnamefont {A.~I.}\ \bibnamefont
  {Vainshtein}}, \ and\ \bibinfo {author} {\bibfnamefont {V.~I.}\ \bibnamefont
  {Zakharov}},\ }\href {\doibase 10.1016/0550-3213(79)90022-1} {\bibfield
  {journal} {\bibinfo  {journal} {Nucl. Phys. B}\ }\textbf {\bibinfo {volume}
  {147}},\ \bibinfo {pages} {385} (\bibinfo {year} {1979})}\BibitemShut
  {NoStop}%
\bibitem [{\citenamefont {Andrei}\ and\ \citenamefont
  {Gross}(1978)}]{Andrei:1978xg}%
  \BibitemOpen
  \bibfield  {author} {\bibinfo {author} {\bibfnamefont {N.}~\bibnamefont
  {Andrei}}\ and\ \bibinfo {author} {\bibfnamefont {D.~J.}\ \bibnamefont
  {Gross}},\ }\href {\doibase 10.1103/PhysRevD.18.468} {\bibfield  {journal}
  {\bibinfo  {journal} {Phys. Rev. D}\ }\textbf {\bibinfo {volume} {18}},\
  \bibinfo {pages} {468} (\bibinfo {year} {1978})}\BibitemShut {NoStop}%
\bibitem [{\citenamefont {Holl}\ \emph {et~al.}(2004)\citenamefont {Holl},
  \citenamefont {Krassnigg},\ and\ \citenamefont {Roberts}}]{Holl:2004fr}%
  \BibitemOpen
  \bibfield  {author} {\bibinfo {author} {\bibfnamefont {A.}~\bibnamefont
  {Holl}}, \bibinfo {author} {\bibfnamefont {A.}~\bibnamefont {Krassnigg}}, \
  and\ \bibinfo {author} {\bibfnamefont {C.~D.}\ \bibnamefont {Roberts}},\
  }\href {\doibase 10.1103/PhysRevC.70.042203} {\bibfield  {journal} {\bibinfo
  {journal} {Phys. Rev. C}\ }\textbf {\bibinfo {volume} {70}},\ \bibinfo
  {pages} {042203} (\bibinfo {year} {2004})},\ \Eprint
  {http://arxiv.org/abs/nucl-th/0406030} {arXiv:nucl-th/0406030} \BibitemShut
  {NoStop}%
\bibitem [{\citenamefont {Morningstar}\ and\ \citenamefont
  {Peardon}(2004)}]{Morningstar:2003gk}%
  \BibitemOpen
  \bibfield  {author} {\bibinfo {author} {\bibfnamefont {C.}~\bibnamefont
  {Morningstar}}\ and\ \bibinfo {author} {\bibfnamefont {M.~J.}\ \bibnamefont
  {Peardon}},\ }\href {\doibase 10.1103/PhysRevD.69.054501} {\bibfield
  {journal} {\bibinfo  {journal} {Phys. Rev. D}\ }\textbf {\bibinfo {volume}
  {69}},\ \bibinfo {pages} {054501} (\bibinfo {year} {2004})},\ \Eprint
  {http://arxiv.org/abs/hep-lat/0311018} {arXiv:hep-lat/0311018} \BibitemShut
  {NoStop}%
\bibitem [{\citenamefont {Frommer}\ \emph {et~al.}(1994)\citenamefont
  {Frommer}, \citenamefont {Hannemann}, \citenamefont {Nockel}, \citenamefont
  {Lippert},\ and\ \citenamefont {Schilling}}]{Frommer:1994vn}%
  \BibitemOpen
  \bibfield  {author} {\bibinfo {author} {\bibfnamefont {A.}~\bibnamefont
  {Frommer}}, \bibinfo {author} {\bibfnamefont {V.}~\bibnamefont {Hannemann}},
  \bibinfo {author} {\bibfnamefont {B.}~\bibnamefont {Nockel}}, \bibinfo
  {author} {\bibfnamefont {T.}~\bibnamefont {Lippert}}, \ and\ \bibinfo
  {author} {\bibfnamefont {K.}~\bibnamefont {Schilling}},\ }\href {\doibase
  10.1142/S012918319400115X} {\bibfield  {journal} {\bibinfo  {journal} {Int.
  J. Mod. Phys. C}\ }\textbf {\bibinfo {volume} {5}},\ \bibinfo {pages} {1073}
  (\bibinfo {year} {1994})},\ \Eprint {http://arxiv.org/abs/hep-lat/9404013}
  {arXiv:hep-lat/9404013} \BibitemShut {NoStop}%
\bibitem [{\citenamefont {Gusken}\ \emph {et~al.}(1989)\citenamefont {Gusken},
  \citenamefont {Low}, \citenamefont {Mutter}, \citenamefont {Sommer},
  \citenamefont {Patel},\ and\ \citenamefont {Schilling}}]{Gusken:1989ad}%
  \BibitemOpen
  \bibfield  {author} {\bibinfo {author} {\bibfnamefont {S.}~\bibnamefont
  {Gusken}}, \bibinfo {author} {\bibfnamefont {U.}~\bibnamefont {Low}},
  \bibinfo {author} {\bibfnamefont {K.~H.}\ \bibnamefont {Mutter}}, \bibinfo
  {author} {\bibfnamefont {R.}~\bibnamefont {Sommer}}, \bibinfo {author}
  {\bibfnamefont {A.}~\bibnamefont {Patel}}, \ and\ \bibinfo {author}
  {\bibfnamefont {K.}~\bibnamefont {Schilling}},\ }\href {\doibase
  10.1016/S0370-2693(89)80034-6} {\bibfield  {journal} {\bibinfo  {journal}
  {Phys. Lett. B}\ }\textbf {\bibinfo {volume} {227}},\ \bibinfo {pages} {266}
  (\bibinfo {year} {1989})}\BibitemShut {NoStop}%
\bibitem [{\citenamefont {Bali}\ \emph {et~al.}(2016)\citenamefont {Bali},
  \citenamefont {Lang}, \citenamefont {Musch},\ and\ \citenamefont
  {Sch\"afer}}]{Bali:2016lva}%
  \BibitemOpen
  \bibfield  {author} {\bibinfo {author} {\bibfnamefont {G.~S.}\ \bibnamefont
  {Bali}}, \bibinfo {author} {\bibfnamefont {B.}~\bibnamefont {Lang}}, \bibinfo
  {author} {\bibfnamefont {B.~U.}\ \bibnamefont {Musch}}, \ and\ \bibinfo
  {author} {\bibfnamefont {A.}~\bibnamefont {Sch\"afer}},\ }\href {\doibase
  10.1103/PhysRevD.93.094515} {\bibfield  {journal} {\bibinfo  {journal} {Phys.
  Rev. D}\ }\textbf {\bibinfo {volume} {93}},\ \bibinfo {pages} {094515}
  (\bibinfo {year} {2016})},\ \Eprint {http://arxiv.org/abs/1602.05525}
  {arXiv:1602.05525 [hep-lat]} \BibitemShut {NoStop}%
\bibitem [{\citenamefont {Wang}\ \emph {et~al.}(2018)\citenamefont {Wang},
  \citenamefont {Zhao},\ and\ \citenamefont {Zhu}}]{Wang:2017qyg}%
  \BibitemOpen
  \bibfield  {author} {\bibinfo {author} {\bibfnamefont {W.}~\bibnamefont
  {Wang}}, \bibinfo {author} {\bibfnamefont {S.}~\bibnamefont {Zhao}}, \ and\
  \bibinfo {author} {\bibfnamefont {R.}~\bibnamefont {Zhu}},\ }\href {\doibase
  10.1140/epjc/s10052-018-5617-3} {\bibfield  {journal} {\bibinfo  {journal}
  {Eur. Phys. J. C}\ }\textbf {\bibinfo {volume} {78}},\ \bibinfo {pages} {147}
  (\bibinfo {year} {2018})},\ \Eprint {http://arxiv.org/abs/1708.02458}
  {arXiv:1708.02458 [hep-ph]} \BibitemShut {NoStop}%
\bibitem [{\citenamefont {Roberts}(1994)}]{Roberts:1990ww}%
  \BibitemOpen
  \bibfield  {author} {\bibinfo {author} {\bibfnamefont {R.~G.}\ \bibnamefont
  {Roberts}},\ }\href {\doibase 10.1017/CBO9780511564062} {\emph {\bibinfo
  {title} {{The Structure of the proton: Deep inelastic scattering}}}},\
  Cambridge Monographs on Mathematical Physics\ (\bibinfo  {publisher}
  {Cambridge University Press},\ \bibinfo {year} {1994})\BibitemShut {NoStop}%
\bibitem [{\citenamefont {Karpie}\ \emph {et~al.}(2018)\citenamefont {Karpie},
  \citenamefont {Orginos},\ and\ \citenamefont
  {Zafeiropoulos}}]{Karpie:2018zaz}%
  \BibitemOpen
  \bibfield  {author} {\bibinfo {author} {\bibfnamefont {J.}~\bibnamefont
  {Karpie}}, \bibinfo {author} {\bibfnamefont {K.}~\bibnamefont {Orginos}}, \
  and\ \bibinfo {author} {\bibfnamefont {S.}~\bibnamefont {Zafeiropoulos}},\
  }\href {\doibase 10.1007/JHEP11(2018)178} {\bibfield  {journal} {\bibinfo
  {journal} {JHEP}\ }\textbf {\bibinfo {volume} {11}},\ \bibinfo {pages} {178}
  (\bibinfo {year} {2018})},\ \Eprint {http://arxiv.org/abs/1807.10933}
  {arXiv:1807.10933 [hep-lat]} \BibitemShut {NoStop}%
\bibitem [{\citenamefont {Egerer}\ \emph {et~al.}(2022)\citenamefont {Egerer}
  \emph {et~al.}}]{HadStruc:2021qdf}%
  \BibitemOpen
  \bibfield  {author} {\bibinfo {author} {\bibfnamefont {C.}~\bibnamefont
  {Egerer}} \emph {et~al.} (\bibinfo {collaboration} {HadStruc}),\ }\href
  {\doibase 10.1103/PhysRevD.105.034507} {\bibfield  {journal} {\bibinfo
  {journal} {Phys. Rev. D}\ }\textbf {\bibinfo {volume} {105}},\ \bibinfo
  {pages} {034507} (\bibinfo {year} {2022})},\ \Eprint
  {http://arxiv.org/abs/2111.01808} {arXiv:2111.01808 [hep-lat]} \BibitemShut
  {NoStop}%
\end{thebibliography}%

\end{document}